\documentclass[longauth,desactivate]{aa}  

\pdfoutput=1

\usepackage{amsmath}
\usepackage{amssymb}
\usepackage{blindtext}
\usepackage{bm}
\usepackage{color}
\usepackage{graphicx}
\usepackage{hyperref}
\usepackage{longtable}
\usepackage{multicol}
\usepackage{multirow}
\usepackage{natbib}
\usepackage{orcidlink}
\usepackage{placeins}
\usepackage{txfonts}
\usepackage{wasysym}
\usepackage[capitalize,nameinlink]{cleveref}

\creflabelformat{equation}{#2#1#3}
\crefrangelabelformat{equation}{#3#1#4 to #5#2#6}

\hypersetup{colorlinks,
  linkcolor=red,
  filecolor=red,
  urlcolor=magenta,
  citecolor=blue}

\usepackage{txfonts}
\usepackage{xspace}

\newcommand{\redmapper}{redMaPPer}
\newcommand{\Msun}{\mathrm{M}_\odot}
\newcommand{\mgii}{Mg~{\sc ii}}

\usepackage{enumitem}

\begin{document}

\title{
CHANCES, the Chilean Cluster Galaxy Evolution Survey: Selection and initial characterisation of clusters and superclusters
}
\titlerunning{The CHANCES cluster and supercluster sample}

\author{
Cristóbal~Sifón\orcidlink{0000-0002-8149-1352}\inst{1}
\and Alexis~Finoguenov\orcidlink{0000-0002-4606-5403}\inst{2}
\and Christopher~P.~Haines\orcidlink{0000-0002-8814-8960}\inst{3,33}
\and Yara~Jaffé\orcidlink{0000-0003-2150-1130}\inst{4,33}
\and B.~M.~Amrutha\orcidlink{0000-0002-2732-7519}\inst{3}
\and Ricardo~Demarco\orcidlink{0000-0003-3921-2177}\inst{5}
\and E.~V.~R.~Lima\orcidlink{0000-0002-6268-8600}\inst{6}
\and Ciria~Lima-Dias\orcidlink{0009-0006-0373-8168}\inst{7,8}
\and Hugo~M\'endez-Hern\'andez\orcidlink{0000-0003-3057-9677}\inst{7,8,33}
\and Paola~Merluzzi\orcidlink{0000-0003-3966-2397}\inst{9}
\and Antonela~Monachesi\orcidlink{0000-0003-2325-9616}\inst{7}
\and Gabriel~S.~M.~Teixeira\orcidlink{0000-0002-1594-208X}\inst{10}
\and Nicolas~Tejos\orcidlink{0000-0002-1883-4252}\inst{1}
\and F.~Almeida-Fernandes\orcidlink{0000-0002-8048-8717}\inst{6,11}
\and Pablo~Araya-Araya\orcidlink{0000-0003-2860-5717}\inst{6}
\and Maria~Argudo-Fernández\orcidlink{0000-0002-0789-2326}\inst{12,13}
\and Raúl~Baier-Soto\orcidlink{0009-0008-4255-1309}\inst{4,33}
\and Lawrence~E.~Bilton\orcidlink{0000-0002-4780-129X}\inst{14,4,1,15}
\and C.~R.~Bom\orcidlink{0000-0003-4383-2969}\inst{10}
\and Juan~Pablo~Calder\'on\orcidlink{0000-0003-2931-2932}\inst{16,17,18}
\and Letizia~P.~Cassar\`a\orcidlink{0000-0001-5760-089X}\inst{19}
\and Johan~Comparat\orcidlink{0000-0001-9200-1497}\inst{20}
\and H.~M.~Courtois\orcidlink{0000-0003-0509-1776}\inst{21}
\and Giuseppe~D'Ago\orcidlink{0000-0001-9697-7331}\inst{22}
\and Alexandra~Dupuy\orcidlink{0000-0001-9005-2792}\inst{23}
\and Alexander~Fritz\inst{24}
\and Rodrigo~F.~Haack\orcidlink{0009-0005-6830-1832}\inst{17,18}
\and Fabio~R.~Herpich\orcidlink{0000-0001-7907-7884}\inst{25}
\and E.~Ibar\orcidlink{0009-0008-9801-2224}\inst{14,33}
\and Ulrike~Kuchner\orcidlink{0000-0002-0035-5202}\inst{26}
\and Ivan~Lacerna\orcidlink{0000-0002-7802-7356}\inst{3}
\and Amanda~R.~Lopes\orcidlink{0000-0002-6164-5051}\inst{17,18}
\and Sebastian~Lopez\orcidlink{0000-0003-0389-0902}\inst{27}
\and Elismar~Lösch\orcidlink{0000-0003-2561-0756}\inst{6}
\and Sean~McGee\orcidlink{0000-0003-3255-3139}\inst{28}
\and C.~Mendes~de~Oliveira\orcidlink{0000-0002-5267-9065}\inst{6}
\and Lorenzo~Morelli\orcidlink{0000-0001-6890-3503}\inst{3}
\and Alessia~Moretti\orcidlink{0000-0002-1688-482X}\inst{29}
\and Diego~Pallero\orcidlink{0000-0002-1577-7475}\inst{4,33}
\and Franco~Piraino-Cerda\orcidlink{0009-0008-0197-3337}\inst{4,33}
\and Emanuela~Pompei\orcidlink{0000-0001-7578-8160}\inst{30}
\and U.~Rescigno\orcidlink{0000-0002-9280-0173}\inst{3}
\and Anal\'ia~V.~Smith~Castelli\orcidlink{0009-0007-2396-0003}\inst{17,18}
\and Rory~Smith\orcidlink{0000-0001-5303-6830}\inst{4,33}
\and Laerte~Sodr\'e~Jr\orcidlink{0000-0002-3876-268X}\inst{6}
\and Elmo~Tempel\orcidlink{0000-0002-5249-7018}\inst{31,32}
}

\institute{
Instituto de Física, Pontificia Universidad Católica de Valparaíso, Casilla 4059, Valparaíso, Chile
\and Department of Physics, University of Helsinki, Gustaf Hällströmin katu 2, 00560 Helsinki, Finland
\and Instituto de Astronomía y Ciencias Planetarias (INCT), Universidad de Atacama, Copayapu 485, Copiapó, Chile
\and Departamento de Física, Universidad Técnica Federico Santa María, Avenida España 1680, Valparaíso, Chile
\and Institute of Astrophysics, Facultad de Ciencias Exactas, Universidad Andr\'es Bello, Sede Concepci\'on, Talcahuano, Chile
\and Departamento de Astronomia, Instituto de Astronomia, Geofísica e Ciências Atmosféricas, Universidade de São Paulo, Rua do Matão 1226, Cidade Universitária, São Paulo, 05508-090, Brazil
\and Departamento de Astronom\'ia, Universidad de La Serena, Avda. Ra\'ul Bitr\'an 1305, La Serena, Chile
\and Instituto Multidisciplinario de Investigaci\'on y Postgrado, Ra\'ul Bitr\'an 1305, La Serena, Chile
\and INAF - Osservatorio Astronomico di Capodimonte, Salita Moiariello 16 80131 Napoli Italy
\and Centro Brasileiro de Pesquisas F\'isicas, Rua Dr. Xavier Sigaud 150, 22290-180 Rio de Janeiro, RJ, Brazil
\and Observat\'orio do Valongo, Ladeira Pedro Ant\^onio, 43, Sa\'ude, Rio de Janeiro, 20080-090, BR
\and Departamento de Física Teórica y del Cosmos, Edificio Mecenas, Campus Fuentenueva, Universidad de Granada, E-18071, Granada, Spain
\and Instituto Universitario Carlos I de F\'isica Te\'orica y Computacional, Universidad de Granada, 18071 Granada, Spain
\and Instituto de F\'isica y Astronom\'ia, Universidad de Valpara\'iso, Avda. Gran Breta\~na 1111, Valpara\'iso, Chile
\and Centre of Excellence for Data Science, Artificial Intelligence \& Modelling, The University of Hull, Cottingham Road, Kingston-Upon-Hull, HU6 7RX, UK
\and Consejo Nacional de Investigaciones Científicas y Técnicas, Ciudad Autónoma de Buenos Aires, Argentina
\and Instituto de Astrofísica de La Plata, CONICET-UNLP, Paseo del Bosque s/n, B1900FWA, Argentina
\and Facultad de Ciencias Astronómicas y Geofísicas, Universidad Nacional de La Plata, Paseo del Bosque s/n, B1900FWA, Argentina
\and INAF-IASF Milano, Via Alfonso Corti 12, 20133, Milano, Italy
\and Max-Planck-Institut f\"{u}r extraterrestrische Physik (MPE), Gie{\ss}enbachstra{\ss}e 1, D-85748 Garching bei M\"unchen
\and Universit\'e Claude Bernard Lyon 1, IUF, IP2I Lyon, 4 rue Enrico Fermi, 69622 Villeurbanne, France
\and Institute of Astronomy, University of Cambridge, Madingley Road, Cambridge CB3 0HA, United Kingdom
\and Korea Institute for Advanced Study, 85, Hoegi-ro, Dongdaemun-gu, Seoul 02455, Republic of Korea
\and Kuffner Observatory, Johann-Staud-Straße 10, 1160 Wien, Austria
\and CASU, Institute of Astronomy, Cambridge, UK
\and School of Physics, Astronomy, University of Nottingham, Nottingham NG7 2RD, UK
\and Universidad de Chile
\and School of Physics and Astronomy, University of Birmingham, Birmingham, B15 2TT, UK
\and INAF-Padova Astronomical Observatory, Vicolo dell'Osservatorio 5, I-35122 Padova, Italy
\and ESO
\and Tartu Observatory, University of Tartu, Observatooriumi 1, Tõravere 61602, Estonia
\and Estonian Academy of Sciences, Kohtu 6, 10130 Tallinn, Estonia
\and Millennium Nucleus for Galaxies (MINGAL)
}

\abstract{
CHANCES, the CHileAN Cluster galaxy Evolution Survey, will study the evolution of galaxies in and around 100 massive galaxy clusters from the local Universe out to $z=0.45$, and two superclusters at $z\sim0.05$ that contain roughly 25 Abell clusters each.
CHANCES will use the new 4MOST Spectroscopic Survey Facility on the VISTA 4m telescope to obtain spectra for ${\sim}$500,000 galaxies with magnitudes $r_\mathrm{AB} < 20.4$, providing comprehensive spectroscopic coverage of each cluster out to $5r_{200}$. 
Its wide and deep scope will trace massive and dwarf galaxies from the surrounding filaments and groups to the cores of galaxy clusters. This will enable the study of galaxy preprocessing and of the role of the evolving environment on galaxy evolution.
In this paper, we present and characterise the sample of clusters and superclusters to be targeted by CHANCES.
We used literature catalogues based on X-ray emission and the Sunyaev-Zel'dovich effect to define the cluster sample in a homogeneous way, with attention to cluster mass and redshift, as well as the availability of ancillary data.
We calibrated literature mass estimates from various surveys against each other and provide an initial mass estimate for each cluster, which we used to define the radial extent of the 4MOST coverage.
We also present an initial assessment of the structure surrounding these clusters based on the redMaPPer red-sequence algorithm as a preview of some of the science CHANCES will enable.
}


\maketitle

\section{Introduction}

Understanding what drives the evolution of galaxies and determines whether they end up as star-forming spirals or quiescent early-type galaxies remains a fundamental task within astrophysics.  While most isolated galaxies remain gas-rich star-forming spirals to the present day, the bulk of galaxies within massive clusters has lost their gas and transformed into quiescent early-type galaxies \citep{Wilman2009}.
Internal energetic mechanisms and external environmental processes are both expected to play major roles in the transformation of galaxies. The former may include gravitational instabilities, supernovae, stellar winds, or feedback from the central supermassive black hole. The latter can include gravitational processes, such as tidal stripping by the cluster halo and galaxy-galaxy interactions \citep[e.g.][]{Natarajan2002,Gnedin2003,Smith2016,Tollet2017,Smith2022}, and hydrodynamical processes due to the hot intracluster medium (ICM), such as gas heating or ram-pressure stripping \citep[e.g.][]{Gunn1972,Treu2003,Ebeling2014,Brown2017,Quilis2017,Kulier2023} (for a review of these various effects, we refer to \cite{Boselli2022}, and for a review that focused on environment-related scenarios, we refer to \cite{Alberts2022}).

It is now well established that a significant fraction of this transformation takes place not inside the clusters themselves, but in the surrounding filamentary large-scale structure out to at least $5r_{200}$. This is evidenced by the reduced level of star formation and the lower fraction of star-forming galaxies at large cluster-centric radii (2--5$r_{200}$) compared to coeval field galaxy populations \citep{Hou2014,Haines2015,Lopes2024,deVos2024}. This shortfall of star-forming galaxies at large cluster-centric radii was interpreted as evidence that star-forming galaxies are at least partially quenched within galaxy groups \citep{Zabludoff1996} before they become part of the clusters. This interpretation is supported by recent studies identifying galaxy groups in the infall regions of clusters. For instance, the fraction of star-forming galaxies within these groups is lower than was seen among other galaxies at the same cluster-centric distance \citep{Bianconi2018,Lopes2024} and lower than for galaxies in groups that are not associated with more massive clusters \citep[e.g.][]{Montaguth2024}. This cumulative effect of processes outside of the cluster is known as preprocessing of galaxies \citep{Fujita2004} and is also a topic of increased investigation in hydrodynamical simulations \citep[e.g.][]{Bahe2013,Bahe2019,Pallero2022,Sifon2024}.

Beyond groups and clusters, intermediate-density environments of the cosmic web (filaments and sheets) may play an equally important role in galaxy evolution. The cosmic web dynamically impacts about half of all galaxies that fall into clusters \citep{Cautun2014,Kuchner2022} and might cause gas accretion, secondary infall, and disk re-formation, but also gas stripping and shock heating, which suppress star formation \citep{Martinez2016,Hasan2023}. Galaxies in filament cores are redder, more massive, and tend to be elliptical, and they have lower star formation rates and higher metallicities \citep{Donnan2022}. Our model of galaxy quenching is consequently changing to account for the physical processes and the time spent in filaments, walls, and groups. We currently do not know whether these observations are solely a consequence of the relations with local density (e.g.\ morphology-density and star formation-density relations) or if the physical processes in cosmic filaments cause the effects we observe \citep{OKane2024,Raj2024}.

The cosmic evolution of star formation in and around clusters provides another key piece to constrain the timescales of galaxy transformation \citep{Haines2013,Stroe2017,Kesebonye2023}. The relevance of preprocessing for the evolution of present-day cluster galaxies is a consequence of both the late assembly of massive clusters (they have accreted half their mass and galaxy populations since $z\sim0.5$) and the fact that a significant fraction of this accreted material is in the form of galaxy groups \citep{McGee2009}. \cite{Haines2018} showed that massive clusters contain a wealth of X-ray galaxy groups at distances $\sim r_{200}$, whose accretion can explain half of the expected mass-growth rate of clusters at late epochs.   

CHANCES\footnote{\url{https://chances.uda.cl/}}, the CHileAN Cluster galaxy Evolution Survey \citep{Haines2023}, is a 4MOST Community Survey \citep{deJong2019} designed to uncover the relation between the evolution of galaxies and hierarchical structure formation as it occurs through deep and wide multi-object spectroscopy of galaxy clusters and their surroundings. During its five-year survey, CHANCES will target $\sim$500,000 cluster galaxies out to $5r_{200}$. This is approximately the distance at which environmental effects acting on infalling galaxies are expected to be sufficiently strong to start removing the extended hot gas atmospheres of galaxies and cutting off their gas supply \citep{Bahe2013}. It is also well beyond the maximum distance of 2--$3\,r_{200}$ that can be reached by back-splash galaxies \citep{Mamon2004,Kuchner2022,Pizzardo2024}. A distance of $5r_{200}$ also corresponds approximately to the turn-around radius within which matter has detached from the Hubble-Lema\^itre flow and is gravitationally bounded to collapse \citep[e.g.][]{Bertschinger1985,Rines2006}. In combination with other multi-wavelength surveys, CHANCES will capture all relevant environments in and around massive clusters, including filaments and groups, to determine the prevalence of preprocessing. This will provide a comprehensive view of the evolution of galaxies and the growth of massive clusters over the past 5~Gyr. A complementary survey, the WEAVE Wide-Field Cluster Survey \citep{Jin2024}, will be performed in the northern hemisphere with the WEAVE multi-fibre spectrograph on the \textit{William Herschel} Telescope, characterising environmentally driven galaxy evolution with a breadth and depth comparable to those of CHANCES.

In this paper, we present the cluster selection for CHANCES (Sect. \ref{s:design}). We then discuss a uniform calibration of cluster masses based on literature estimates that we used to define the radial extent of the 4MOST coverage (Sect. \ref{s:masses}). We also provide an initial evaluation of the infalling structures surrounding these clusters using a photometric cluster catalogue (Sect. \ref{s:environment}). We conclude with a summary and final remarks (Sect. \ref{s:conclusions}). The CHANCES target selection and observational setup will be described in more detail in Haines et al.\ (in prep.) and other forthcoming papers.

Throughout this paper, $M_\Delta$ (where $\Delta\in\{200,500\}$) refers to the mass within $r_\Delta$, corresponding to the radius enclosing a density $\Delta$ times the critical matter density of the Universe at each redshift. We assumed a flat $\Lambda$CDM cosmology with cosmological parameters equal to the central values inferred by \cite{PlanckCollaboration2020}, the most relevant of which are the current expansion rate, $H_0=67.4\,\mathrm{km\,s^{-1}Mpc^{-1}}$, and the present-day matter density parameter, $\Omega_\mathrm{m}=0.315$.

\section{Survey design}
\label{s:design}

The 4MOST Spectroscopic Survey Facility is a multi-fibre spectrograph about to be installed on the VISTA 4m telescope at Paranal Observatory in Chile. It is equipped with 2,436 science fibres, 1,624 of which feed two low-resolution spectrographs with $R\equiv\lambda/\Delta\lambda=6,\!500$, and 812 fibres feed a single high-resolution spectrograph with $R=20,\!000$. Fibres can be positioned across a field of view with a diameter of 2.5 degrees \citep{deJong2016,deJong2022}. 4MOST will simultaneously carry out 18 public spectroscopic surveys over its first five years of science operations, sharing the focal plane among the surveys in every observation to maximise efficiency \citep{deJong2019,Tempel2020,Tempel2020a}.

With the 4MOST survey structure, we devised a strategy to study the evolution of galaxies in and around clusters by combining a unique breadth and depth. In order to efficiently span both stellar mass and lookback times, we split the CHANCES cluster survey into a Low-z subsurvey that includes clusters and superclusters at $z<0.07$ and down to $M_{200}\sim10^{14}\,\Msun$ (Sect. \ref{s:lowz}), and an Evolution subsurvey that consists of massive ($M_{200}>5\times10^{14}\,\Msun$) clusters over $0.07 < z < 0.45$ (Sect. \ref{s:evolution}). We refer to these two collectively as the CHANCES cluster surveys. Combined, they make CHANCES a survey with continuous coverage of massive clusters from the present day to $z=0.45$, over the past 4.8 Gyr of cosmic time. The redshift limit of 0.45 was chosen as a compromise between maximising the redshift extent and minimising the stellar mass limit that can be achieved in the available observing time, while taking advantage of available spectroscopy (Haines et al.\ in prep.). In addition, CHANCES includes a novel survey of the circumgalactic medium (CGM) of cluster galaxies (CHANCES-CGM) that we briefly describe in Sect. \ref{s:cgm}. Because of the 4MOST observing strategy, we restrict CHANCES targets to declinations $-80^\circ<\delta<+5^\circ$. We also exclude clusters with Galactic latitudes $|b|<20^\circ$.

By covering clusters out to $5r_{200}$, CHANCES will observe, in addition to the nominal cluster list, roughly ten times as many infalling systems with masses reaching about an order of magnitude lower than the main system (see Sect. \ref{s:infalling}). Thus, CHANCES will provide an unprecedented view of environment-driven galaxy evolution not only as a function of stellar mass, but also of the masses of the main clusters and infalling groups. At a given cluster mass, we will also control for the large-scale cluster environment by assessing the impact of residing in a main cluster or in (or near) an infalling system such as a group or filament.

The primary requirement for a cluster to be included in CHANCES is the availability of photometry from the Dark Energy Camera Legacy Survey \citep[DECaLS,][]{Dey2019} data release 10 (LSDR10), which ensures accurate astrometry and photometry over our entire survey. 
We make one exception with the Antlia cluster (Abell~S636, $z=0.0087$, which is part of the Low-z subsurvey; Lima-Dias et al.\ in prep.), which is not within the LSDR10 coverage. This extremely nearby system has full coverage from the S-PLUS survey, which provides photometry in 12 narrow bands across the optical wavelength range \citep{MendesdeOliveira2019}. Target selection in the Antlia field relies on photometric redshifts based on S-PLUS photometry \citep{Lima2022}. 

For both the Low-z (including the superclusters) and Evolution subsurveys, we will target galaxies down to an $r$-band magnitude of 20.4. We estimate stellar masses for galaxies in Low-z clusters using the $g-i$ colour-stellar mass relation of the \cite{Taylor2011} relation. For galaxies in Evolution clusters, we do this by fitting the spectral energy distributions from the \cite{Bruzual2003} library and assuming a \cite{Chabrier2003} initial mass function using the \textsc{fast++} library\footnote{\url{https://github.com/cschreib/fastpp}} \citep{Kriek2009}. This translates into stellar mass limits of $m_\star=10^{8.5}\,\Msun$ and $m_\star=10^{10}\,\Msun$, respectively.\footnote{We verified that the \cite{Taylor2011} relation applied to Evolution targets gives consistent results.}
The selection of target galaxies is based on photometric redshifts from LSDR10, and for a subsample of the Low-z subsurvey for which S-PLUS data are available, it is based on S-PLUS photometric redshifts, with stellar masses derived from the same photometry. Of the 500,000 spectroscopic targets within the CHANCES cluster survey, approximately 70\% will be devoted to the Low-z subsurvey, with 70\% of these corresponding to individual clusters and the remaining 30\% to the superclusters. This means that we will target approximately 4,500 galaxies per cluster in the Low-z cluster sample and approximately 2,500 galaxies per Evolution cluster. All CHANCES targets will be observed with the low-resolution spectrographs. The details will be presented by Haines et~al.\ (in prep.) and Méndez-Hernández et~al.\ (in prep). 

\begin{figure*}
    \centering
    \includegraphics[width=0.9\linewidth]{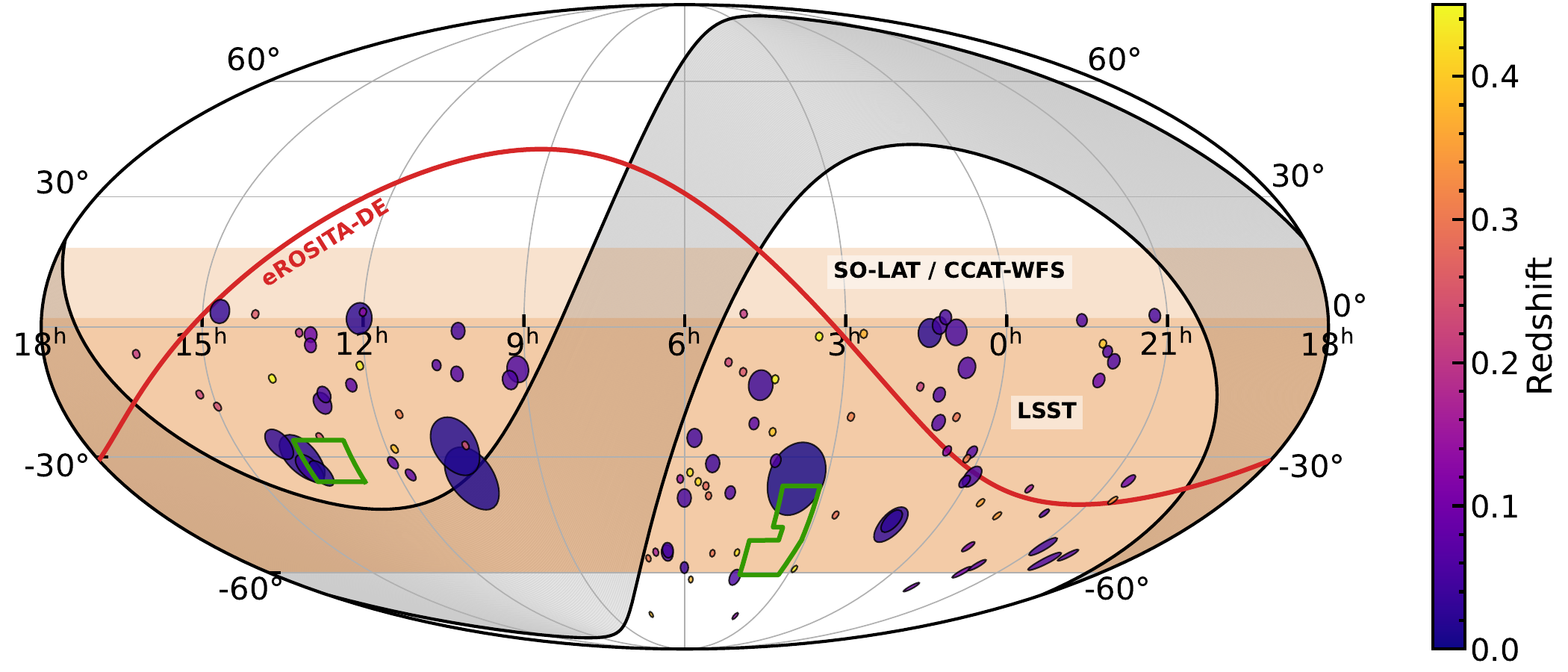}
    \caption{Sky distribution of CHANCES clusters in equatorial coordinates in a Mollweide projection. The radius of each circle corresponds to the maximum between $5r_{200}$ and 1 deg (the latter set for visibility) and is colour-coded by redshift. The supercluster regions are shown with green polygons. The region below the red curve corresponds to the eROSITA-DE survey, the dark orange region shows the approximate LSST survey area, and the light orange region shows SO-LAT and CCAT-WFS (for simplicity, we draw both as delimited by $-60^\circ\leq\delta\leq+18^\circ$). The grey band bounded by black lines shows Galactic latitudes $|b|\leq20^\circ$.}
    \label{f:sky}
\end{figure*}

Figure \ref{f:sky} shows the CHANCES footprint on the sky, which covers over 1,600 sq.\ deg.\ in all.\footnote{This only refers to the cluster subsurveys described in Sect. \ref{s:lowz} and \ref{s:evolution}, that is, it does not include the CHANCES-CGM subsurvey that is briefly described in Sect. \ref{s:cgm}.} For visibility purposes, we set each circle to have at least a radius of 1 deg. In practice, this means that essentially all of the Evolution clusters are enlarged to this size while all Low-z cluster footprints retain their original extent of $5r_{200}$. We describe the $r_{200}$ estimates in Sect. \ref{s:masses}. We also show several multi-wavelength surveys of interest: the Simons Observatory Large-Aperture Telescope (SO-LAT) survey \citep[$-60^\circ\lesssim\delta\lesssim+20^\circ$]{Ade2019}, the Cerro Chajnantor Atacama Telescope (CCAT) Wide-Field Survey \citep[WFS,][$-61^\circ\leq\delta\leq+18^\circ$]{CPC2023}, and the Vera C.~Rubin Observatory Legacy Survey of Space and Time \citep[LSST,][$-60^\circ\lesssim\delta\lesssim+2^\circ$]{Ivezic2019}, as well as the German half of the eROSITA all-sky survey, which covers Galactic longitudes $l>180^\circ$ \citep{Merloni2024}. In addition to its main science goals, CHANCES will therefore also enable unique cluster and galaxy science in synergy with a rich array of multi-wavelength data. 

\subsection{CHANCES Low-z subsurvey}
\label{s:lowz}

The CHANCES Low-z subsurvey is designed to target 50 clusters at $z \leq 0.07$. In addition, it will cover large regions within the Shapley and Horologium-Reticulum superclusters that each contain a large number of clusters within a rich cosmic web.

\subsubsection{Galaxy cluster sample}

The Low-z subsurvey includes clusters selected from the All-Sky X-ray Extended Source (AXES) catalogue \citep{Damsted2024,Khalil2024}, the Wide-field Nearby Galaxy cluster Survey \citep[WINGS,][]{Fasano2006}, and a number of clusters selected individually based on their known properties or available datasets. We describe each subset below. 

AXES is a reanalysis of the ROSAT All-Sky Survey (RASS) data in which X-ray emission from extended sources was detected on scales of 12$-$24 arcmin via wavelet decomposition \citep{Damsted2024}. It is complete in the extragalactic area down to a flux of $1 \times 10^{-12}$ ergs s$^{-1}$ cm$^{-2}$. AXES sources have been validated against a group catalogue based on Sloan Digital Sky Survey (SDSS) data \citep{Tempel2017} by \cite{Damsted2024}, against the Two-Micron All Sky Survey (2MASS) Redshift Survey (2MRS) spectroscopic galaxy group catalogue \citep{Tempel2018} by \cite{Khalil2024}, and against a cluster catalogue constructed with the \redmapper\ cluster red-sequence finder applied to LSDR10 by Finoguenov et al.\ (in prep). We use the latter two  cross-matched catalogues in this work and refer to them as AXES-2MRS and AXES-LEGACY, respectively. 
The scales probed by AXES correspond to about $r_{500}-r_{200}$ at $z<0.04$, which offers the unique advantage of using the scales at which cluster behaviour is best understood. However, beyond $z=0.04$, these angular scales extend beyond $r_{200}$, and AXES sources become affected by blending. In addition, the feedback processes on low-mass clusters lead to baryonic lifting, which results in additional flux compared to estimates based on $L_{500}$ (the X-ray luminosity within $r_{500}$). This means that the scales used in the AXES analyses are best suited to cluster characterisation at $z<0.04$ \citep{Khalil2024, Damsted2024}. Conversely, below $z=0.01$, the scales used by AXES only probe cluster cores, which changes the selection function.\footnote{The lowest-redshift cluster in AXES-2MRS has $z=0.0015$ and $M_{200}=1.1\times10^{13}\,\Msun$.} In any case, there are few massive clusters at these low redshifts, and most of them have good auxiliary information. In addition, to ensure low contamination, we only considered AXES clusters with at least five member galaxies (spectroscopic in the case of AXES-2MRS and photometric in the case of AXES-LEGACY). Therefore, we used the subset of AXES-2MRS with at least five spectroscopic member galaxies as our parent sample at $z<0.04$.

WINGS \citep{Fasano2006}, together with its extension OmegaWINGS \citep{Gullieuszik2015}, is a multi-wavelength imaging and spectroscopic survey of 77 galaxy clusters at $0.04\leq z\leq0.07$ (perfectly complementing the redshift coverage of AXES-2MRS) that were selected from cluster catalogues constructed from the RASS. We included WINGS as one of the CHANCES Low-z parent catalogues because of the wealth of optical spectroscopy that is already available in the central regions of these clusters \citep{Cava2009,Moretti2014,Moretti2017}. The 4MOST observations will complement the available data to reach the CHANCES specifications.

With these two parent samples, we designed the Low-z subsurvey to cover an order of magnitude in mass at redshifts $z<0.07$, sampling known clusters with an essentially uniform distribution in mass at $M_{200}\gtrsim10^{14}\,\Msun$, plus a few well-known lower-mass systems. The Low-z cluster sample is listed in \Cref{t:lowz}.

\subsubsection{Superclusters}
\label{s:design_superclusters}

The CHANCES Low-z subsurvey also includes wide regions covering two well-known superclusters: the Shapley supercluster at $z=0.048$ \citep[e.g.][]{Reisenegger2000,Proust2006,Merluzzi2015,Haines2018} and the Horologium-Reticulum supercluster at $z=0.060$ \citep[e.g.][]{Lucey1983,Fleenor2005,Fleenor2006}. The exact regions were chosen as a compromise between the number of included member clusters and the available fibre hours for the entire survey. Spectroscopic targets are selected using the same magnitude limit as the Low-z cluster subsurvey: an $r$-band magnitude lower than 20.4 across the entire supercluster regions, independently of local density. This ensures a complete census of the structure within them (Haines et al.\ in prep.).

The CHANCES coverage of the Shapley supercluster corresponds to a rectangle bounded by right ascensions $192^\circ\leq\alpha\leq207^\circ$ and declinations $-36^\circ\leq\delta\leq-26^\circ$, containing 128 deg$^2$, plus the remaining area within $5r_{200}$ of Abell~3571 at the same redshift. The coverage of the Horologium-Reticulum supercluster is enclosed by the following polygon vertices\footnote{These are given as [$\alpha$, $\delta$] in degrees.}: \hbox{[46.5, -51]}; \hbox{[49, -60.7]}; \hbox{[66, -60.7]}; \hbox{[66, -51]}; \hbox{[55, -51]}; \hbox{[55, -47.5]}; \hbox{[58.5, -47.5]}; \hbox{[58.5, -37]}; \hbox{[46.5, -37]}. This polygon covers an area of 225 deg$^2$. The supercluster region is supplemented by the $5r_{200}$ area around Abell~3266, which marginally overlaps the supercluster in the south-eastern corner, also at the supercluster redshift. Other Low-z clusters overlapping the superclusters in the sky but not in redshift are Abell~3565 and Abell~3574 with Shapley and Fornax with Horologium-Reticulum. 
In the regions in common with the latter clusters, we include in our target catalogues galaxies with photometric redshifts that match any of the overlapping systems. However, for faint galaxies approaching our magnitude limit, the photometric redshift errors are larger than the redshift difference between the clusters and superclusters. This means in practice that faint targets are common to both systems.
This overlap will, in any case, ensure a high completeness for the clusters and superclusters (Méndez-Hernández et al.\ in prep.).
The supercluster survey regions are shown in Fig. \ref{f:sky}.

\subsection{CHANCES Evolution subsurvey}
\label{s:evolution}

As part of the CHANCES Evolution subsurvey, we will target 50 of the most massive galaxy clusters distributed evenly over \hbox{$0.07<z<0.45$}. The Evolution sample is primarily selected from the second Planck catalogue of Sunyaev–Zel'dovich (SZ) sources \citep[PSZ2;][]{PlanckCollaboration2016}, which provides a homogeneous all-sky sample of massive clusters over this redshift range and is not noticeably biased by dynamical state \citep{PlanckCollaboration2016,AndradeSantos2017,Rossetti2017}. 
We selected the 10 most massive PSZ2 clusters with available DESI Legacy Imaging Survey DR10 $grz$ imaging and photometric redshifts out to 5$r_{200}$ in each of five linearly spaced intervals over \hbox{$0.07<z<0.45$}. This binned selection ensures continuity with the Low-z subsurvey: There are 8 Evolution clusters at \hbox{$0.07<z<0.10$}. It also translates to a minimum mass $M_{200} = 7 \times 10^{14} M_\odot$ at $z > 0.2$, while at lower redshifts, the mass limit is progressively reduced to account for the smaller available volume. 
This selection is similar to the selection that was used for cluster cosmology by \cite{PlanckCollaboration2016a}, and the larger extent of more massive clusters allows an efficient use 4MOST at $z>0.3$.
Most of the CHANCES Evolution cluster sample is covered by the CHEX-MATE XMM Heritage programme \citep{CMC2021}, which provides high-quality X-ray data that are suitable for characterising the ICM and mass distributions of each CHANCES cluster.

The only cluster in the Evolution subsurvey that is not matched to any PSZ2 source is MACS~J0329.7$-$0211 at $z=0.45$. We include it in CHANCES to take advantage of the extensive available spectroscopy from the CLASH-VLT survey \citep{Girardi2024}. The Evolution cluster sample is listed in \Cref{t:evolution}.

\subsection{CHANCES CGM subsurvey}
\label{s:cgm}

In addition to the cluster samples described above, CHANCES will offer a unique view of the CGM in and around clusters by observing $\sim$50,000 galaxies around $\sim$10,000 background quasars behind roughly 4,000 unique galaxy clusters. Quasars are selected either from optical spectroscopy \citep{Lyke2020, Anand2021} or from X-ray imaging \citep{Merloni2019}. We require that they lie within 6~Mpc in projection from foreground clusters at $0.35 < z < 0.7$, where the \mgii\ line falls in the wavelength range covered by 4MOST. This experimental setup builds up from the setup that was pioneered by \cite{Lopez2008}. Clusters for the CGM subsurvey are selected from a catalogue constructed by applying the \redmapper\ algorithm \citep{Rykoff2014} to the LSDR10 data (\citealt{Kluge2024}; see Sect. \ref{s:environment}). The main goal of the CGM subsurvey is to establish the origin of intervening \mgii\ absorbers and their relation to galaxies as a function of galaxy type and both local and large-scale environment. In this manner, we will construct a detailed view of the CGM transformations that occur in dense environments and their impact on galaxy evolution, which will provide a novel complement to the CHANCES cluster survey. More details will be given in a forthcoming paper.

\section{Cluster mass estimates}
\label{s:masses}

\begin{figure*}
    \centering
    \includegraphics[width=0.33\linewidth]{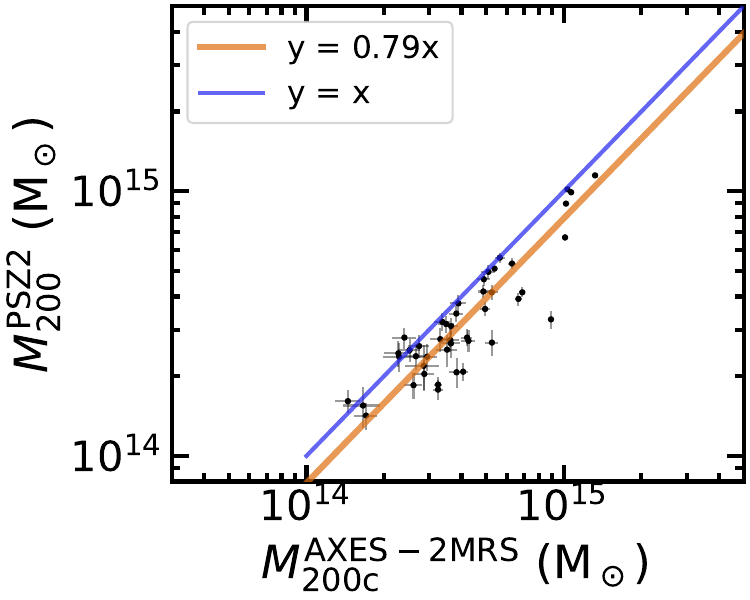}
    \includegraphics[width=0.33\linewidth]{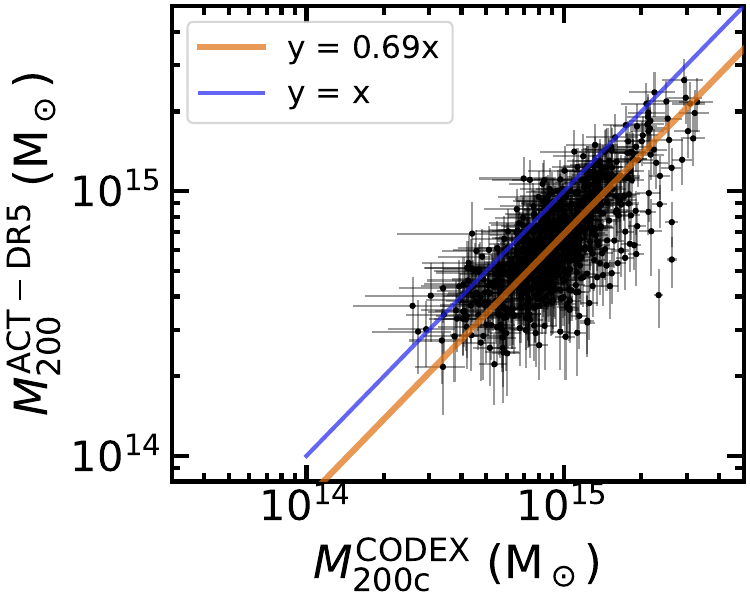}
    \includegraphics[width=0.33\linewidth]{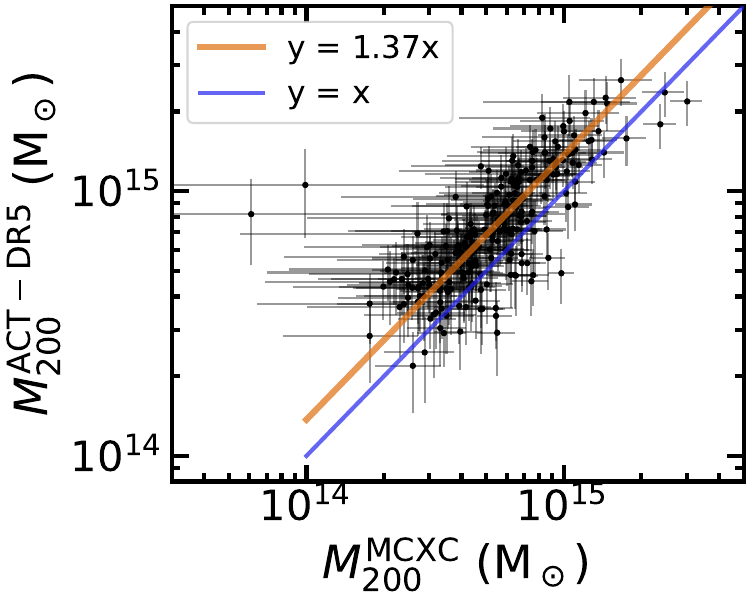}
    \caption{Mass comparison examples. From left to right, we compare the mass estimates of PSZ2 to AXES-2MRS, ACT-DR5 to CODEX, and ACT-DR5 to MCXC. The points with error bars show all clusters in common between each pair of catalogues, regardless of whether they are part of CHANCES. Thin blue lines show $y=x$, and the thick orange lines show the maximum-likelihood estimates given in the legends. The extremely small uncertainties in PSZ2 and AXES-2MRS only consider statistical uncertainties. We account for this fact by including an intrinsic scatter term in the fits. These fits correspond to the adopted normalisations discussed in Sect. \ref{s:masses} and listed in \Cref{t:sources_lowz,t:sources_evol}.}
    \label{f:mass_comparison}
\end{figure*}

\begin{figure*}[]
    \centering
    \includegraphics[width=0.48\linewidth]{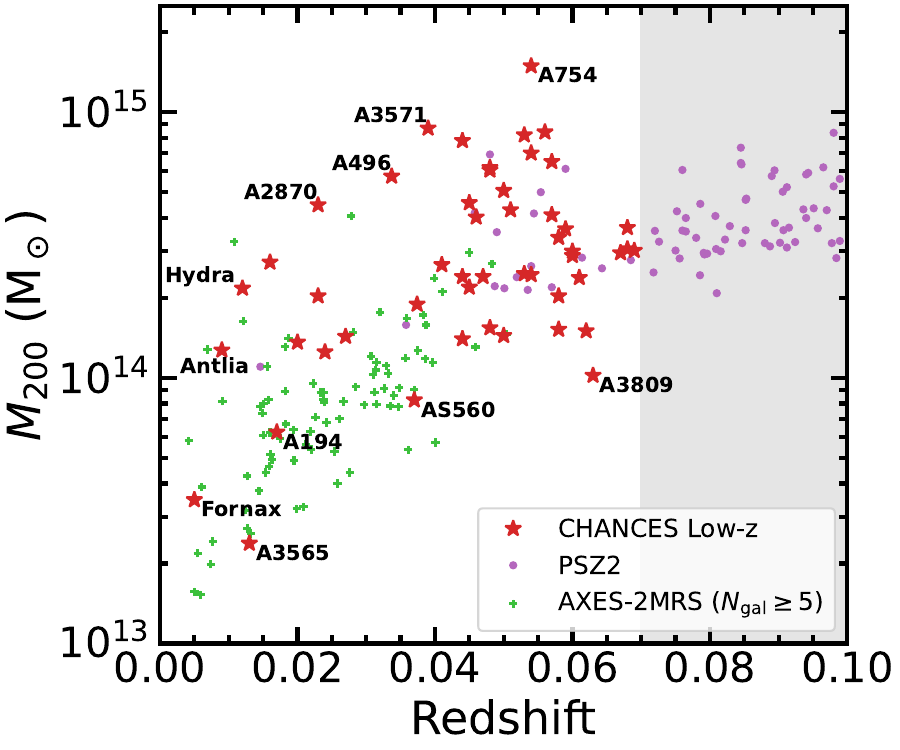}
    \includegraphics[width=0.48\linewidth]{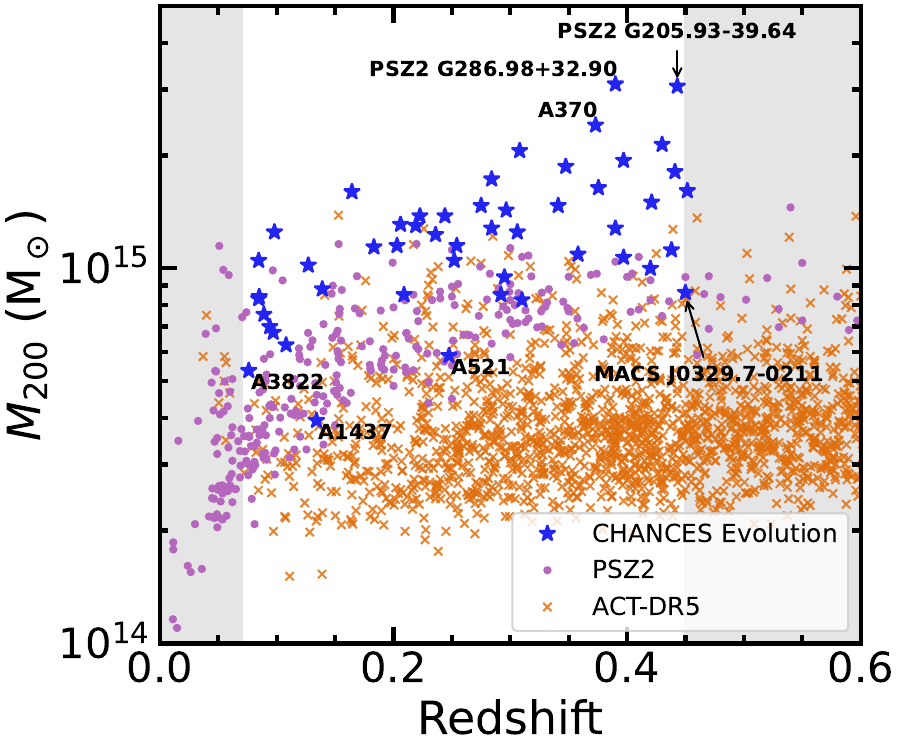}
    \caption{CHANCES cluster samples (stars) in context. Left: Low-z subsurvey, excluding clusters in the Shapley and Horologium-Reticulum superclusters, compared with southern (Dec $<+5^\circ$) clusters in AXES-2MRS (with at least five spectroscopic members) and PSZ2, also excluding clusters in the supercluster regions. Right: Evolution subsurvey, compared with southern clusters in PSZ2 and ACT-DR5. In both cases, we removed duplicate clusters by only keeping the higher-priority mass estimate (see Sect. \ref{s:masses}), and we removed clusters with Galactic latitudes $|b|<20^\circ$. We give the names of the least and most massive clusters in each sample, as well as the names of some well-known clusters.
    In the right panel, we also highlight MACS~J0329.7$-$0211, which is the only Evolution cluster that is not in PSZ2. The redshift range of each CHANCES subsurvey is shown with the white background. The CHANCES cluster masses correspond to those listed in \Cref{t:lowz,t:evolution}.}
    \label{f:sample}
\end{figure*}

As described in the previous section, we selected clusters homogeneously based on mass proxies from X-rays (Low-z) and the thermal SZ effect (evolution). However, we used masses from a variety of sources in an attempt to obtain the best $r_{200}$ estimate for each cluster while still maintaining some homogeneity in the mass estimates. To this end, we cross-matched our cluster samples with several catalogues that provide masses with different techniques, as summarised in \Cref{t:sources_lowz,t:sources_evol} (explained below).
Because we combined several cluster surveys, it was important to homogenise their mass estimates to provide uniformity. We emphasise here that the goal of this exercise was not to obtain accurate and precise mass estimates based on all the available data for each cluster, but to establish a preliminary mass scale from which to obtain an estimate of $r_{200}$, which was used to define the extent of our target galaxy catalogues.

We retrieved mass estimates from various surveys and applied a normalisation factor to each one to ensure statistical consistency. Because different assumptions were made to estimate the masses in each of the catalogues, this is preferable to combining masses from different catalogues. In order to estimate the normalisation factors, we matched clusters across catalogues with a $5'$ matching radius and a maximum redshift difference of 10\%. The former matches the PSZ2 beam size and is small compared to the extent of Low-z clusters, while the latter allows photometric redshift errors, and we tested that more stringent matching criteria do not change the outcome.
We show examples of cross-matched catalogues and the resulting mass normalisations in Fig. \ref{f:mass_comparison}, and we summarise the normalisations we applied to each external catalogue in \Cref{t:sources_lowz,t:sources_evol}.
We fit the normalisations, $a$, by fitting a linear relation using a maximum likelihood procedure including intrinsic scatter,
\begin{equation}
    \mathcal{L} = \prod_i\frac1{\sqrt{2\pi}w_i}
    \exp\left(-\frac{[\log M_{200}^{y,i}-(a+\log M_{200}^{x,i})]^2}{2w_i^2}\right).
\end{equation}
The weights $w_i$ account for the uncertainties in the two quantities $\delta x_i$ and $\delta y_i$ and for the intrinsic scatter, $\sigma_\mathrm{int}$,
\begin{equation}
    w_i^2 = (\delta x_i)^2 + (\delta y_i)^2 + \sigma_\mathrm{int}^2.
\end{equation}
The latter holds for a linear relation \citep[e.g.][]{Hoekstra2012}. The inclusion of intrinsic scatter as a free parameter alleviates the effect of unrealistically small reported uncertainties, such as those in the left panel of Fig. \ref{f:mass_comparison}, which only include the statistical uncertainties and not the uncertainties from the pressure profile fit or the scaling relations.

In order to assign a mass to each cluster, we preferred the weak lensing masses at all redshifts. When no weak lensing masses were available, we preferred AXES at low redshifts for the reasons given above, followed by SZ-based masses because they are more stable and introduce less scatter than traditional X-ray masses \citep[e.g.][]{Rozo2014,Kugel2025}. Within the remaining catalogues, we preferred those containing more CHANCES clusters for homogeneity while prioritising WINGS over MCXC at low redshift because its selection is similar to that of CHANCES. As shown in Fig. \ref{f:mass_comparison}, the exact choice introduces a 20--30\% scatter in mass, which translates into a $\lesssim$10\% scatter in $r_{200}$.

As a reference scale for the Low-z subsurvey masses, we therefore used the weak lensing mass estimates from MENeaCS \citep{Herbonnet2020}, while for the Evolution subsurvey, we also used the masses derived from ACT-DR5 SZ effect mass estimates, which were normalised to match the scale of weak lensing mass estimates as described by \cite{Hilton2021}\footnote{That is, we used the \texttt{M500cCal} column.}. While MENeaCS masses are given as $M_{200}$, ACT-DR5 masses are given as $M_{500}$. We converted these and other mass definitions as appropriate into $M_{200}$ using \texttt{colossus}\footnote{\url{https://bdiemer.bitbucket.io/colossus/}} \citep{Diemer2018}. For this, we assumed an NFW profile \citep{Navarro1996} with the \cite{Ishiyama2021} mass-concentration relation.

We therefore defined the CHANCES mass of each CHANCES cluster following the order in \Cref{t:sources_lowz,t:sources_evol} for the Low-z and Evolution samples, respectively. In each table, the last column lists the number of clusters that derived their CHANCES mass from that catalogue, while the previous column lists the total number of CHANCES clusters in each catalogue.
In the case of Low-z, this means, for instance, that all 5 clusters in MENeaCS are assigned the MENeaCS masses, and the 11 AXES-2MRS clusters with $z\leq0.04$ that are not in MENeaCS are assigned the AXES-2MRS mass multiplied by 0.86, and so on. We only considered AXES-2MRS and AXES-LEGACY for clusters at $z\leq0.04$, as discussed in Sect. \ref{s:lowz}. 

When there are not enough cross-matches between a particular catalogue and the reference catalogue, we applied successive normalisations. For example, the normalisation of 0.86 for AXES-2MRS (and AXES-LEGACY) comes from the combination of $M_{200}^\mathrm{PSZ2}=0.79M_{200}^\mathrm{AXES-2MRS}$ (shown in the left panel of Fig. \ref{f:mass_comparison}) and $M_{200}^\mathrm{ACT-DR5}=1.09M_{200}^\mathrm{PSZ2}$ (not shown).  The ACT-DR5 and the two SPT catalogues were shown to be consistent with each other and with weak lensing mass estimates \citep{Bleem2020,Hilton2021}, and we therefore set all these normalisations to one.

The two outliers in the right panel of Fig. \ref{f:mass_comparison} are Abell~536 at $z=0.040$ and Abell~S560 at $z=0.037$, with ACT-DR5 masses 11 and 13 times the MCXC masses, respectively. At these low redshifts, the ACT-DR5 catalogue is highly incomplete, and it is therefore expected that some clusters are up-scattered due to either statistical or intrinsic noise. This may raise concerns about using ACT-DR5 in the Low-z sample. Reassuringly, the mass estimates of the two Low-z clusters assigned ACT-DR5 masses (Abell~3667 at $z=0.053$ and Abell~4059 at $z=0.048$) from ACT-DR5, PSZ2, MCXC, AXES-LEGACY, and WINGS, with standard deviations of 16\% and 29\%, respectively. The catalogue choice therefore affects the assigned $r_{200}$ only little (namely, 5\% and 10\%, respectively).

There are two exceptions to the above scheme, and both pertain to the Low-z subsurvey. The first is Abell~3395, which is a well-known merging system that also lies in close proximity to Abell~3391 \citep[e.g.][]{Reiprich2021,Dietl2024,Veronica2024}. Abell~3395 has a mass estimate from the WINGS survey of $M_{200}=2.9\times10^{15}\,\Msun$ \citep{Moretti2017}, while all ICM-based estimates suggest $M_{200}\sim5\times10^{14}\,\Msun$ \citep{Piffaretti2011,PlanckCollaboration2016,Bulbul2024,Damsted2024}. Because mass estimates based on the velocity dispersion are known to be highly biased for merging clusters \citep[as was indeed pointed out by][]{Moretti2017}, we instead chose use the PSZ2 mass for Abell~3395, $M_{200}=5.1\times10^{14}\,\Msun$. The other exception is Abell~3490, for which we used the rescaled AXES-LEGACY mass even though it is at $z=0.069$, because it is not matched to any other catalogue. The resulting mass estimates for each cluster and the associated $r_{200}$ are listed in \Cref{t:lowz,t:evolution}.

\begin{table}
\caption{Mass sources for the Low-z subsurvey, ordered by priority.}
\label{t:sources_lowz}
\begin{tabular}{l|ccc}
\hline\hline
  & (1) & (2) & (3) \\
Survey & \multirow{2}{*}{Normalisation} & No.\ of & \multirow{2}{*}{Source for} \\
 &  & clusters & \\
 \hline
MENeaCS$^a$      & 1 & 5 & 5 \\
AXES-2MRS$^b$    & 0.86 & 21 & 11 \\
AXES-LEGACY$^c$  & 0.86 & 36 & 1 \\
ACT-DR5$^d$      & 1 & 5 & 2 \\
SPT-ECS$^e$      & 1 & 1 & 0 \\
SPT-SZ$^f$       & 1 & 2 & 1 \\
WINGS$^g$        & 0.76 & 18 & 13 \\
MCXC$^h$         & 1.37 & 43 & 14 \\
PSZ2$^i$         & 1.09 & 27 & 2 \\
CODEX$^j$        & 0.69 & 34 & 1 \\[0.5ex]
 \hline
\end{tabular}
\tablefoot{
The columns list (1) the correction factor applied to masses provided by the survey, according to the comparison described in Sect. \ref{s:masses}; (2) the total number of CHANCES clusters in each catalogue; and (3) the number of clusters with CHANCES mass estimates from each source.
\tablefoottext{a}{\cite{Herbonnet2020}}
\tablefoottext{b}{\cite{Khalil2024}. Used only as source for clusters at $z \leq 0.04$.}
\tablefoottext{c}{Finoguenov et al.\ (in prep). Used only for clusters at $z \leq 0.04$, except for Abell~3490 ($z=0.069$), which is not matched in any other catalogue.}
\tablefoottext{d}{\cite{Hilton2021}}
\tablefoottext{e}{\cite{Bleem2020}}
\tablefoottext{f}{\cite{Bleem2015}}
\tablefoottext{g}{\cite{Moretti2017}}
\tablefoottext{h}{\cite{Piffaretti2011}}
\tablefoottext{i}{\cite{PlanckCollaboration2016}}
\tablefoottext{j}{\cite{Damsted2023}.}
}
\end{table}

\begin{table}
\caption{Mass sources for the Evolution sub-survey, ordered by priority}
\label{t:sources_evol}
\centering
\begin{tabular}{l|ccc}
\hline\hline
\multirow{2}{*}{Survey} & \multirow{2}{*}{Normalisation} & No.\ of & \multirow{2}{*}{Source for} \\
 &  & clusters & \\
 \hline
LoCuSS$^a$   & 1 & 4 & 4 \\
MENeaCS      & 1 & 11 & 8 \\
CoMaLit$^b$  & 1 & 25 & 13 \\
ACT-DR5      & 1 & 34 & 14 \\
SPT-ECS      & 1 & 14 & 3 \\
SPT-SZ       & 1 & 12 & 2 \\
PSZ2         & 1.09 & 49 & 6 \\
CODEX        & 0.69 & 40 & 0 \\
MCXC         & 1.37 & 40 & 0 \\[0.5ex]
\hline
\end{tabular}
\tablefoot{
 See \Cref{t:sources_lowz} for details and additional references.
\tablefoottext{a}{\citet{Okabe2016}. Does not contain any clusters in the Low-z redshift range.}
\tablefoottext{b}{\cite{Sereno2015}.}
}
\end{table}

We show the CHANCES clusters in mass--redshift space in Fig. \ref{f:sample}. We place the sample of each subsurvey in context by comparing it to the mass and redshift distributions of clusters in PSZ2 and AXES-2MRS (Low-z, left panel) and PSZ2 and ACT-DR5 (Evolution, right panel) over the sky available to CHANCES (i.e.\ $\delta<+5^\circ$, $|b|>20^\circ$). We removed duplicates from Fig. \ref{f:sample} by matching clusters as described above (a matching radius of $5'$ and a redshift difference of $<10\%$), following the priority scheme already described. That is, in the left panel, we only show as green crosses AXES-2MRS clusters that are not in CHANCES and only show as purple circles PSZ2 clusters that are neither in AXES-2MRS nor CHANCES. This is done analogously for the Evolution sample in the right panel. 

CHANCES includes all of the massive nearby clusters with LSDR10 coverage in the southern hemisphere at $|b|>20^\circ$. The massive AXES-2MRS clusters that are not part of CHANCES  all lack LSDR10 coverage in a significant fraction of their $5r_{200}$ area. The most massive of these are Abell~3526 at $z=0.011$ and Abell~4038 at $z=0.029$.
CHANCES Low-z clusters cover the entire mass range probed by AXES and PSZ2 over $0<z<0.07$. With a few exceptions, in contrast, the Evolution sample specifically targets the most massive systems, although as mentioned, we enforced homogeneous coverage of the full Evolution redshift range.
This continuous sampling of the most massive clusters from $z=0$ to $z=0.45$ is not designed to track the evolution of clusters: It is evident that clusters in the Low-z sample are not the descendants of clusters in the Evolution sample. By targeting the entire volume around each cluster that is detached from the Hubble-Lema\^itre flow, we will instead determine the effect of the time of arrival on the infall patterns and its consequences on galaxy evolution. This is known as the Butcher-Oemler effect \citep{Butcher1984}. The superclusters then provide a natural extension by probing larger-scale environments including the full region that will become detached from the Hubble-Lema\^itre flow in the distant future \citep{Dunner2006,ArayaMelo2009}.

\begin{figure}
    \centering
    \includegraphics[width=\linewidth]{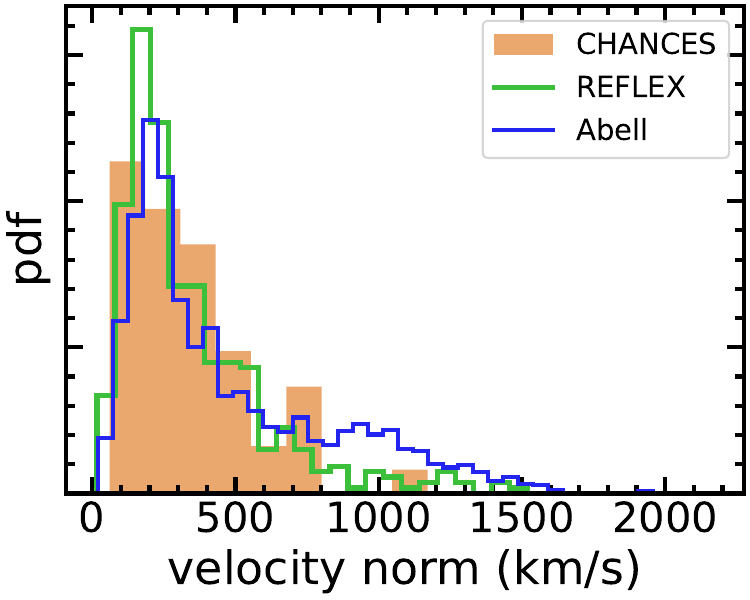}
    \caption{Three-dimensional peculiar velocity norm distribution of CHANCES clusters, compared to all Abell and REFLEX clusters with known redshifts. These have been reconstructed by \cite{Dupuy2023} from distances estimated by Cosmicflows-4 \citep{Courtois2023}. Bin sizes were chosen following \cite{Knuth2006}.}
    \label{f:vpec}
\end{figure}

We expand upon the previous point by looking at the reconstruction of the peculiar velocities of CHANCES clusters. These peculiar velocities were derived by \cite{Dupuy2023} by reconstructing the large-scale density using the distances estimated by the Cosmicflows-4 project \citep{Courtois2023}. Peculiar velocities trace the location in the large-scale structure of the Universe: Generally speaking, large peculiar-velocity structures fall towards small peculiar-velocity structures. Figure \ref{f:vpec} shows the distribution\footnote{We defined the bin widths using the rule described by \cite{Knuth2006}, which optimises the bin size of a piecewise-constant probability distribution in a Bayesian framework. We used the implementation in \texttt{astropy.visualization}.} of peculiar velocities of all CHANCES clusters (excluding superclusters), compared to all Abell clusters \citep{Abell1989} with known redshifts\footnote{As listed in the NASA/IPAC Extragalactic Database, \url{http://ned.ipac.caltech.edu/}.} as well as the ROSAT-ESO Flux Limited X-ray (REFLEX) cluster sample \citep{Boehringer2004}. The peculiar velocities of most CHANCES clusters are lower than $500\,\mathrm{km\,s^{-1}}$. While a few have somewhat higher peculiar velocities, the distribution is much more skewed to lower values than the other samples. This suggests CHANCES clusters are preferentially local attractors. These peculiar-velocity estimates offer yet another aspect whose impact on galaxy evolution will be uniquely assessed by CHANCES.

\section{Cluster environments: A sneak peek}
\label{s:environment}

\subsection{A preliminary census of infalling structures around CHANCES clusters}\label{s:infalling}

\begin{figure*}
    \centering
    \includegraphics[width=0.31\linewidth]{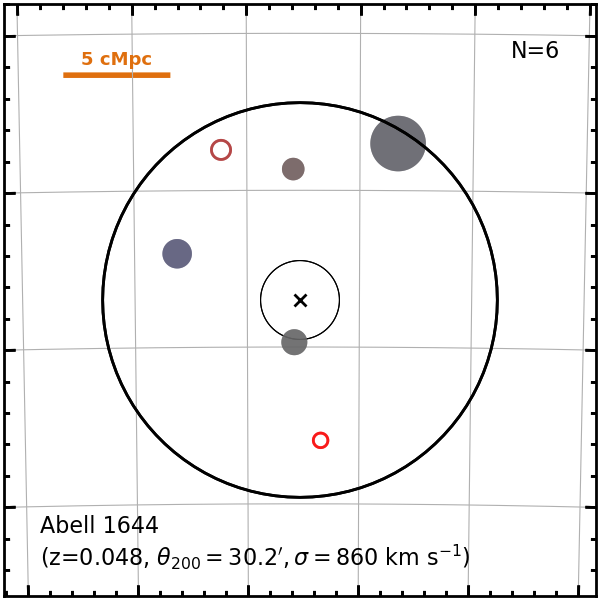}
    \includegraphics[width=0.31\linewidth]{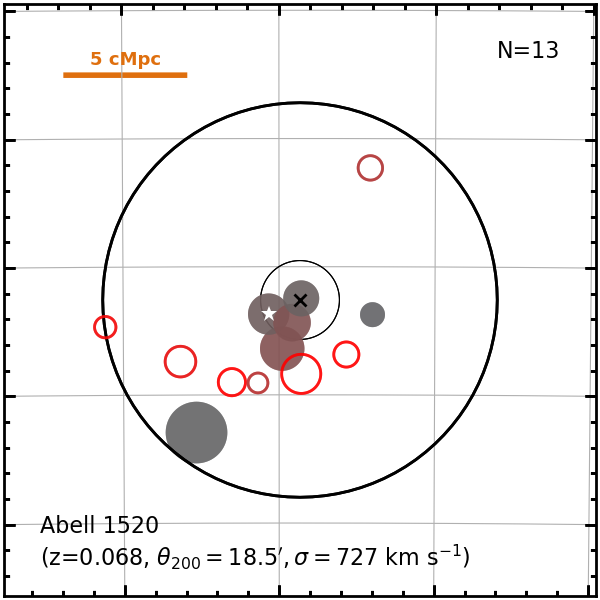}
    \includegraphics[width=0.31\linewidth]{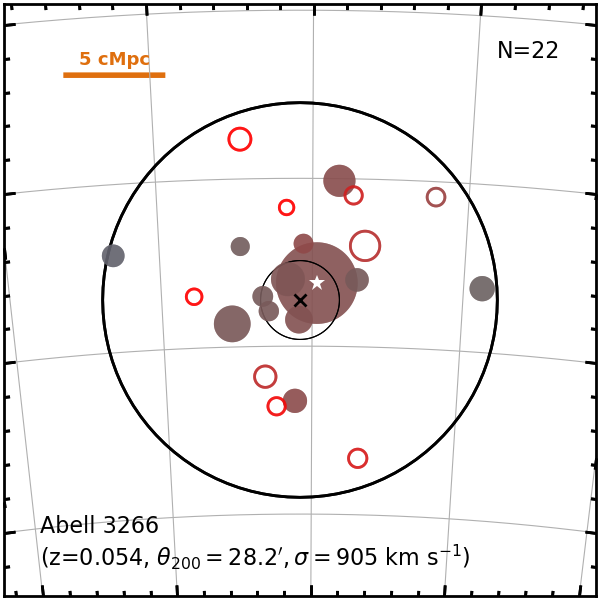}
    \includegraphics[width=0.31\linewidth]{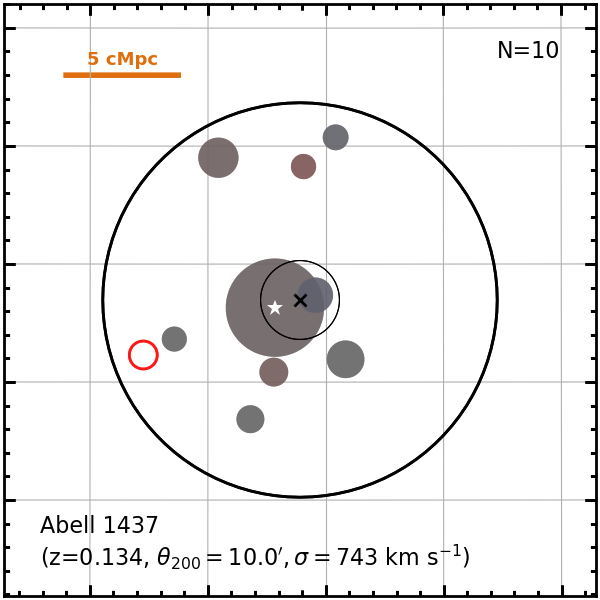}
    \includegraphics[width=0.31\linewidth]{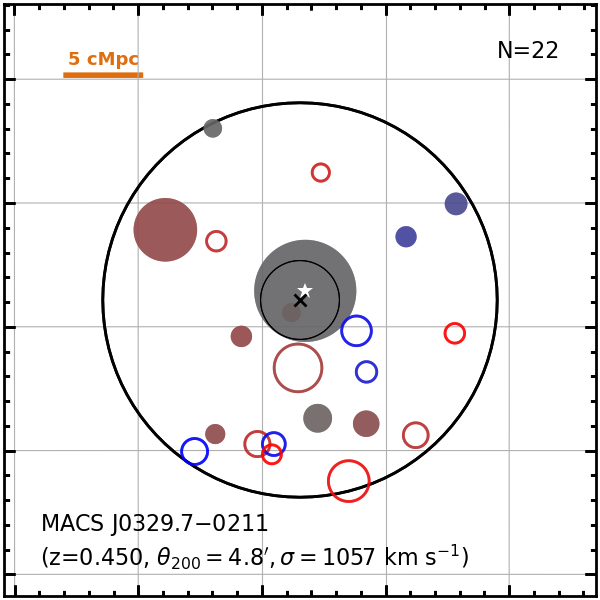}
    \includegraphics[width=0.31\linewidth]{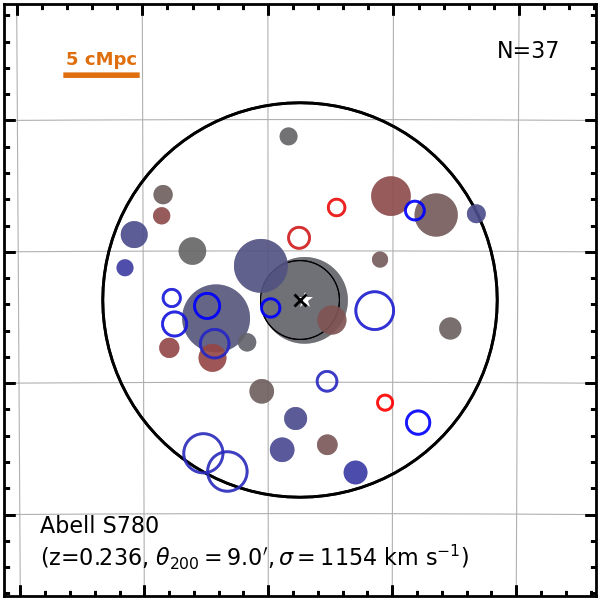}
    \includegraphics[width=0.99\linewidth]{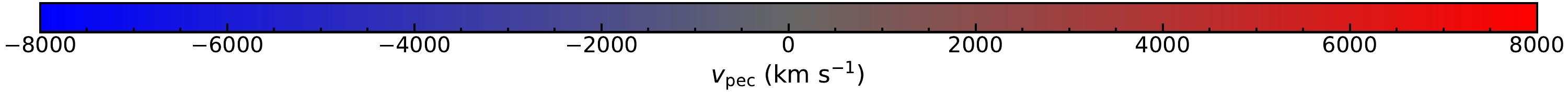}
    \caption{Infalling structures in example CHANCES Low-z (top) and Evolution (bottom) clusters as traced by associations of red galaxies identified with \redmapper\ within 8,000 km s$^{-1}$ of the nominal redshift of each CHANCES cluster. The filled circles correspond to \redmapper\ groups with velocities within three times the CHANCES velocity dispersion. For each subsurvey, we show a cluster with a small (left), typical (middle), and large (right) number of \redmapper\ groups around them. The concentric black cross and empty thin and thick circles mark the centre of each CHANCES cluster, $r_{200}$, and $5r_{200}$, respectively. The coloured circle sizes correspond to the $r_{200}$ of each group based on its richness. The colour scale shows the peculiar velocity with respect to the main cluster. When present, the \redmapper\ system associated with the main CHANCES cluster is shown with a white star. In the bottom of each panel, we list the redshift, angular size corresponding to $r_{200}$, and richness-derived velocity dispersion of each cluster. The numbers in the top right corners correspond to the number of \redmapper\ groups, and the orange bars in the top left corners indicate 5 comoving Mpc at the redshift of each cluster.
    }
    \label{f:lss_sky}
\end{figure*}

One of the primary goals of CHANCES is to study the preprocessing of galaxies prior to cluster infall. It is thus important to characterise not only the main clusters, but also the surrounding structures.
As a proof of concept, we explored the network of groups\footnote{Traditionally, the division between the terms cluster and group is set by a mass threshold, typically around $10^{14}\,\Msun$. In an attempt to reduce confusion and redundancy, we use both terms hierarchically in this section: CHANCES clusters are surrounded by \redmapper\ groups, regardless of mass.} around CHANCES clusters using the \redmapper\ \citep{Rykoff2014} group catalogue generated using the LSDR10. This catalogue was constructed by \cite{Kluge2024} to provide targets for the 4MOST eROSITA Galaxy Cluster Redshift Survey \citep{Finoguenov2019} following the strategy described by \cite{Clerc2020}.
\redmapper\ identifies associations of red galaxies and calculates the richness, $\lambda$, as the total membership probability across all galaxies within a richness-dependent cluster radius that is about 1 Mpc. The cluster centre is chosen as the location of the most likely central galaxy based on galaxy position, magnitude, and colour, as described by \cite{Rykoff2014}.
By construction, the \redmapper\ catalogue does not contain systems lacking a developed red sequence. This selection effect only impacts low-mass systems, and overcoming it requires the spectroscopic completeness that CHANCES will provide.
We describe all of the \redmapper\ group properties used in this demonstration (namely redshifts $z$ and their uncertainties $\delta z$, peculiar velocities $v_\mathrm{pec}$, masses $M_{200}$, and velocity dispersions $\sigma_v$) in Appendix \ref{s:redmapper}. For the purpose of this demonstration, we also calculated the one-dimensional velocity dispersion of each CHANCES cluster given $M_{200}$, using the relation by \cite{Munari2013}. We call this velocity dispersion $\sigma_\mathrm{main}$.

We considered all \redmapper\ groups within $5r_{200}$ of each CHANCES clusters and with 
$|v_\mathrm{pec}|<8000\,\mathrm{km\,s^{-1}}$. Figure \ref{f:lss_sky} shows the \redmapper\ groups thus selected around example Low-z and Evolution clusters. 
There are a total 1,406 \redmapper\ groups thus selected, which is an average 13.5 \redmapper\ groups within $5r_{200}$ of each CHANCES cluster.
We then identified the most massive \redmapper\ group within the nominal CHANCES $r_{200}$ (see\ \Cref{t:lowz,t:evolution}) as the CHANCES cluster itself. Automatic cluster finding is notoriously difficult at very low redshift, where the mean separation between galaxies is several arcminutes on the sky. The LSDR10 data in addition lack $u$-band data, which is critical for photometric redshift estimation at $z<0.1$, where it is needed to detect the 4000\AA\ break and to identify star-forming contaminants \citep{Rozo2016}. We assigned a \redmapper\ group with the main CHANCES cluster in 30 Low-z clusters, all at $z>0.037$, while the rest (all at $z<0.048$) have no \redmapper\ matches within $r_{200}$. One such example is Abell~1644, which we show in the top left panel of Fig. \ref{f:lss_sky}.
Some clusters, however, may be matched to the wrong main cluster, such as in the case of Abell~1520 in the top centre panel of Fig. \ref{f:lss_sky}. All Evolution clusters have a \redmapper\ match, except for RXJ~1347.5$-$1144, which is not within the LSDR10 \redmapper\ footprint because it is covered in $giz$, but not in $r$-band. 
As expected, some CHANCES clusters contain a wealth of infalling groups, although this is not apparent in others from the \redmapper\ catalogue. This may be due either to incompleteness in the \redmapper\ catalogue (e.g.\ because these low-mass systems might not contain a red sequence) or because some CHANCES clusters really reside in regions that are devoid of significant galaxy overdensities. This question can only be answered by the high spectroscopic completeness provided by CHANCES.

\begin{figure}
    \centering
    \includegraphics[width=\linewidth]{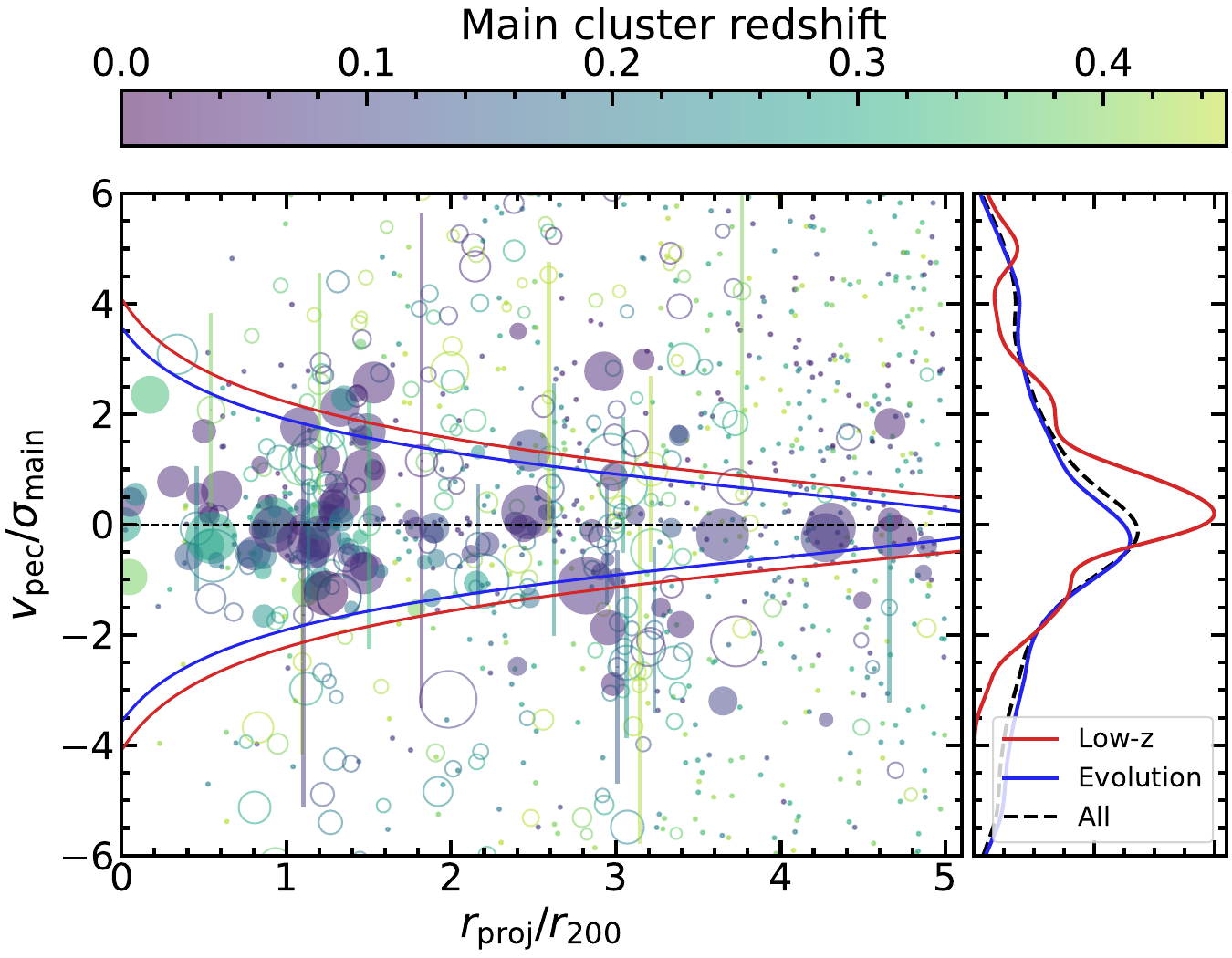}
    \caption{Line-of-sight velocity vs.\ projected distance of \redmapper\ groups around all CHANCES clusters, normalised by the velocity dispersion and $r_{200}$ of each CHANCES cluster, respectively. Each circle shows a \redmapper\ group with a spectroscopic redshift, has a radius proportional to the group richness-derived $r_{200}$ in units of each CHANCES cluster $r_{200}$, and is colour-coded by redshift. The filled and open circles show clusters with uncertainties $\delta v_\mathrm{pec}$ smaller and larger than $\sigma_\mathrm{main}$, respectively. We show uncertainties for a random 10\% of the latter for illustration; the uncertainties for filled circles are about the circle size and are not shown. The small points show groups with photometric redshifts. \redmapper\ groups corresponding to the main CHANCES cluster are excluded from this plot. The red and blue lines in the main panel show the escape velocities for typical Low-z and Evolution clusters; groups with higher peculiar velocities are expected to be unbound from the respective CHANCES cluster. The right panel shows kernel density estimates of the peculiar-velocity distributions of \redmapper\ groups with spectroscopic redshifts only (filled and empty circles in the main panel), using a Gaussian kernel with a width of 0.2 for the Low-z (red) and Evolution (blue) samples, and the combined distribution (dashed black).}
    \label{f:phase_space}
\end{figure}

We now describe the statistical properties of the \redmapper\ sample and its relation to the CHANCES clusters. In Fig. \ref{f:phase_space} we show the  line-of-sight velocity against the projected cluster-centric distance (this is generally referred to as a phase-space diagram) for the \redmapper\ groups that surround the CHANCES clusters.
Phase-space diagrams are commonly used to infer the infall history of galaxy populations within clusters \citep[e.g.][]{Oman2013,Muzzin2014,Haines2015,Jaffe2015}. As with galaxies, the phase-space diagram of infalling groups reveals details of their infall history \citep[e.g.][]{Jaffe2016,Einasto2018,Haines2018,PirainoCerda2024}.
Each data point in Fig. \ref{f:phase_space} is a \redmapper\ group near a CHANCES cluster. The large number of groups with either spectroscopic redshifts with large uncertainties or photometric redshifts highlight the rich structure potentially available for CHANCES to uncover.
The main CHANCES clusters were excluded from Fig. \ref{f:phase_space} following the associations described above. However, some \redmapper\ groups are located right at the CHANCES cluster centres. These are cases where the simple association above clearly failed, possibly due to the fragmentation of the system by \redmapper\ (e.g.\ Abell~1520 and Abell~1437 in Fig. \ref{f:lss_sky}). We treat these failures as nuisance in the context of the proof of concept presented here.

We also show in Fig. \ref{f:phase_space} the escape velocity profiles for typical Low-z and Evolution clusters (with $M_\mathrm{200}=\{3,10\}\times10^{14}\,\Msun$ and $z=\{0.05,0.25\}$, respectively), calculated using Eq. 7 in \cite{Miller2016}, appropriate for clusters in an expanding universe. This calculation shows that $5r_{200}$ is indeed similar to the turn-around radius,as discussed above.
The peculiar velocities of many \redmapper\ groups significantly exceed the escape velocity.
These objects are not falling into the main clusters, although some of them might be part of the same large-scale structure. 
CHANCES will enable us to assess the filamentary structure in detail, including embedded groups that are associated with the clusters in our sample. 
In-depth analyses of the filamentary structure around the CHANCES clusters from the available photometric data will be presented by Baier et al.\ (in prep.) and Piraino-Cerda et al.\ (in prep.). Combined with the optical properties of the member galaxies from ancillary imaging and the CHANCES spectra, and combined with X-ray and SZ measurements of the intragroup medium properties (in some cases individually and in some cases through stacking), CHANCES will allow us to closely link the infall history of galaxies that fall into CHANCES clusters as members of groups or as part of the filamentary network but not a group, and also with field galaxies outside this network.

\begin{figure}
    \centering
    \includegraphics[width=0.9\linewidth]{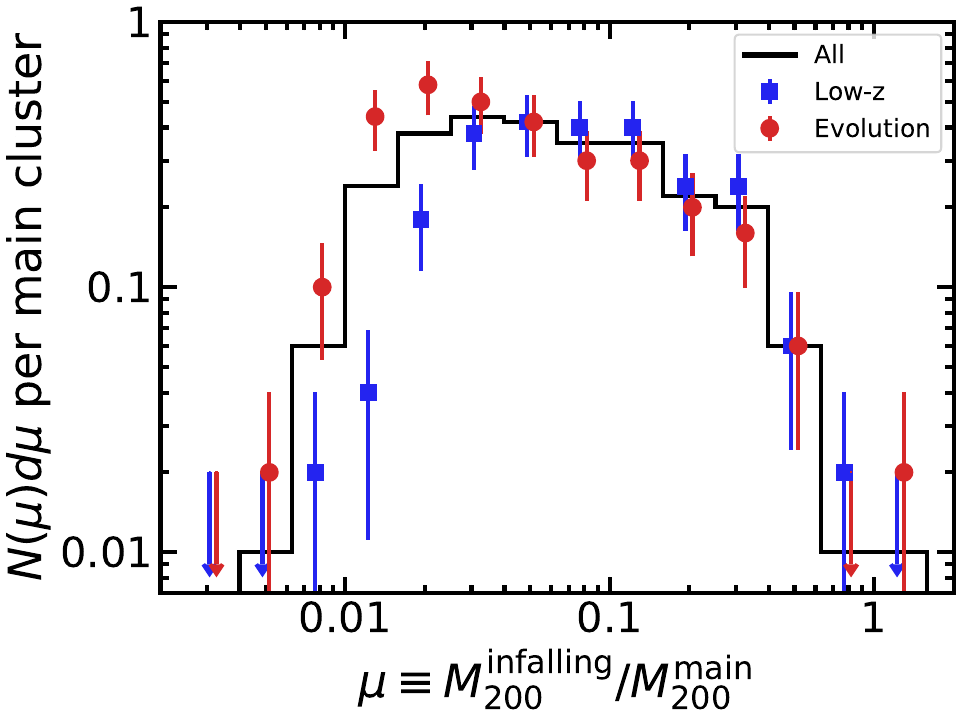}
    \caption{Mass functions of infalling groups for the Low-z (blue) and Evolution (red) samples and for the two combined (black), including only \redmapper\ groups with spectroscopic redshifts and $|v_\mathrm{pec}|<3\sigma_\mathrm{main}$. Differential counts are normalised to the number of CHANCES clusters in each sample and are not corrected for incompleteness. The error bars are Poisson uncertainties, and the arrows are $1\sigma$ upper limits. We show the full sample as a histogram and omit the error bars to avoid cluttering. The blue and red data points are slightly shifted horizontally for clarity. 
    }
    \label{f:imf}
\end{figure}

Finally, we show in Fig. \ref{f:imf} the mass function of \redmapper\ groups with $|v_\mathrm{pec}| < 3\sigma_\mathrm{main}$ (a common first selection of member galaxies within clusters), again taking the richness-derived masses and excluding main CHANCES clusters. We show the mass function in units of the main cluster mass as listed in \Cref{t:lowz,t:evolution}, whereas the masses of infalling groups are the richness-derived masses. As described in Appendix \ref{s:redmapper}, the two mass estimates have consistent medians, with a scatter about each other of 0.2--0.3 dex. Interestingly, the Low-z and Evolution populations both follow the same mass distribution for mass ratios $\mu\equiv M_{200}^\mathrm{infalling}/M_{200}^\mathrm{main}>0.03$, preliminarily suggesting a lack of evolution of the high-mass end of the infall mass function over the past 5 Gyr. The most natural interpretation of the abrupt drop at lower mass ratios is that the \redmapper\ catalogue is incomplete at low masses: The points at which both distributions turn over correspond approximately to $M_{200}^\mathrm{infalling}\sim10^{13}\,\Msun$, in which mass regime the red sequence is not expected to be as dominant as required by \redmapper.

\subsection{Multiple-cluster systems}
\label{s:overlapping}

\begin{table}[]
    \caption{Multiple-cluster systems that overlap on the sky and in redshift space.}
    \centering
    \begin{tabular}{cc}
    \hline\hline
        CHANCES clusters & Redshift \\
        \hline
        Antlia / Hydra & 0.012 / 0.012 \\
        A119 / A147 / A168 &  0.044 / 0.044 / 0.044 \\
         A754 /  A780 & 0.054 / 0.057 \\
        A1631 / A1644 & 0.047 / 0.048 \\
        A1650 / A1651 & 0.085 / 0.085 \\
        A2717 / A4059 & 0.050 / 0.048 \\
        A2870 / A2877 & 0.023 / 0.024 \\
        A3651 / A3667 & 0.060 / 0.053 \\
    \hline
    \end{tabular}
    \label{t:overlapping}
\end{table}

\begin{figure}
    \centering
    \includegraphics[width=0.9\linewidth]{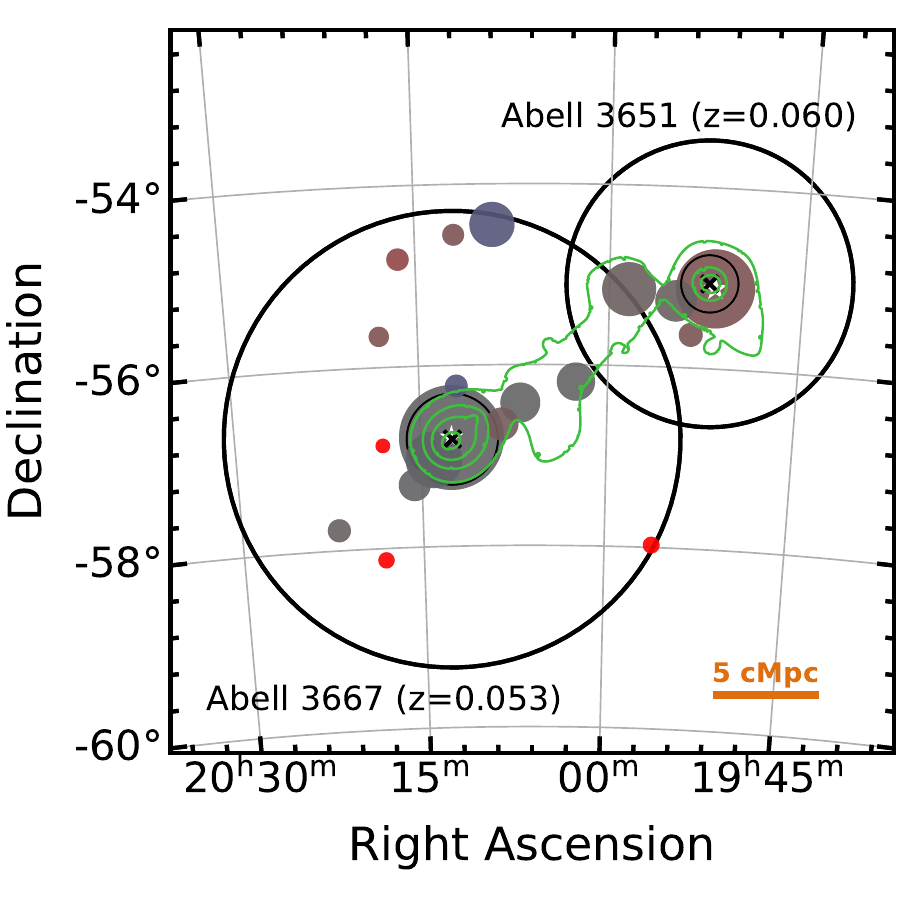}
    \caption{The Abell~3651/3667 system as an example of overlapping CHANCES clusters (see Sect. \ref{s:overlapping}). The symbols and colours are the same as in Fig. \ref{f:lss_sky}, but we do not highlight \redmapper\ systems according to the CHANCES velocity dispersions. The circle colours are with respect to a reference $z=0.056$, and the orange bar at the bottom right shows 5 comoving Mpc at the same redshift. The green contours show the eROSITA detection of hot gas within and between the clusters \citep{Dietl2024}.}
    \label{f:overlapping}
\end{figure}

In addition to the rich array of infalling groups, a few CHANCES clusters are associated with each other: They overlap with other CHANCES clusters on the sky and in redshift. We defined a multiple-cluster system of clusters as a system in which the $5r_{200}$ circles of multiple clusters overlap. These systems are listed in \Cref{t:overlapping}, and they will offer an even larger view of the cosmic web that surrounds massive clusters. They will also serve as an intermediate regime between single CHANCES clusters and superclusters. All of these overlapping systems are part of the Low-z subsurvey, except for Abell~1650/1651. 

We show the Abell~3651/3667 system in \Cref{f:overlapping} as an example. \cite{Arp2001} recognised a possible connection between these two clusters based on the distributions of galaxies in optical images and of X-ray point sources from ROSAT data. \cite{Dietl2024} detected a bridge of X-ray emission that connects these two clusters in the eROSITA X-ray images, which is nicely traced by the structure of \redmapper\ groups shown in Fig. \ref{f:overlapping}. Furthermore, like many other CHANCES clusters, Abell~3667 is a well-known merging system with a spectacular morphology at radio wavelengths that traces non-thermal phenomena in the ICM \citep{DeGasperin2022}. In combination with ICM data, CHANCES will reveal the full extent of the network that connects these two clusters and the others in \Cref{t:overlapping}.

\subsection{Superclusters}

\begin{figure*}
    \centering\includegraphics[width=0.45\linewidth]{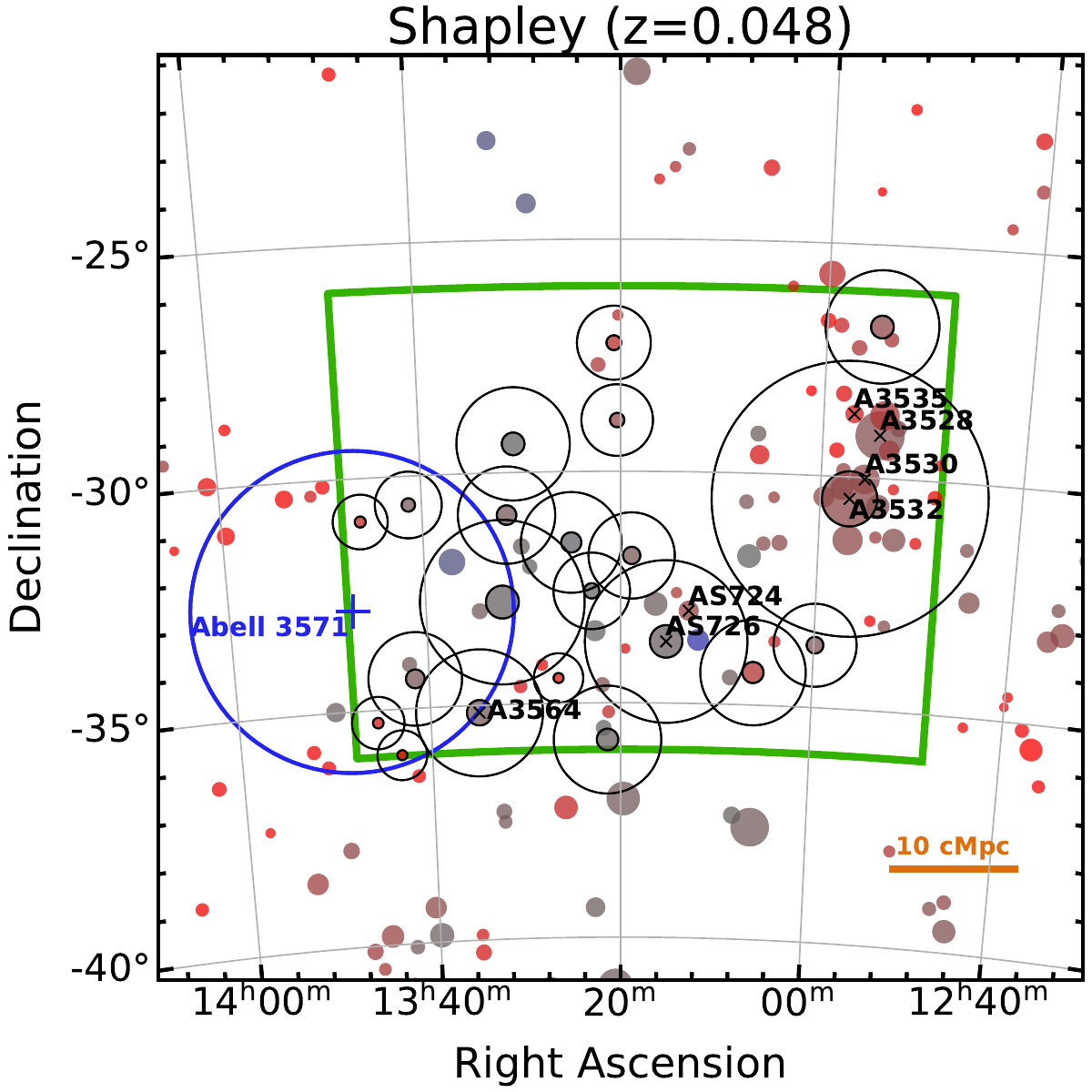}
    \includegraphics[width=0.45\linewidth]{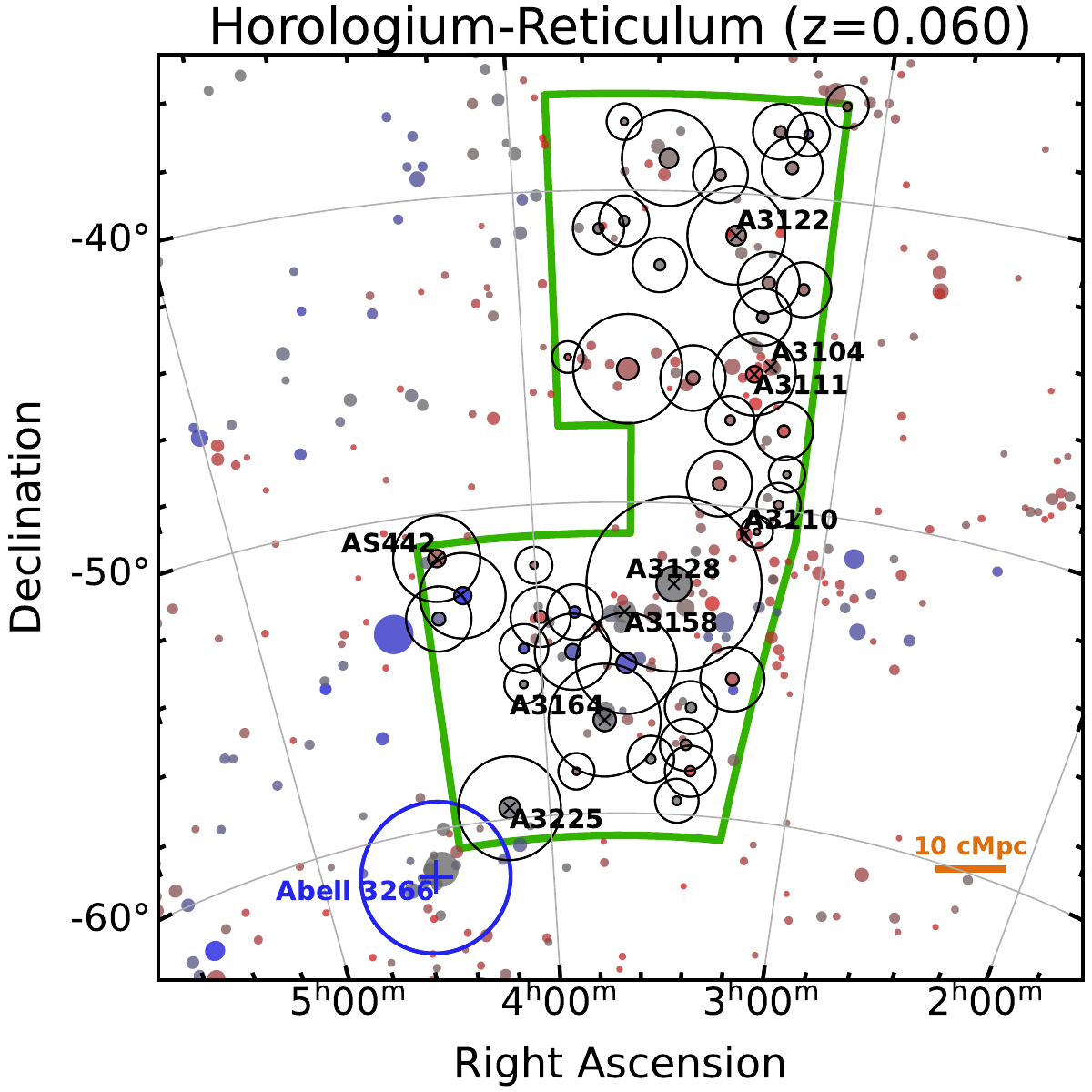}
    \includegraphics[width=0.9\linewidth]{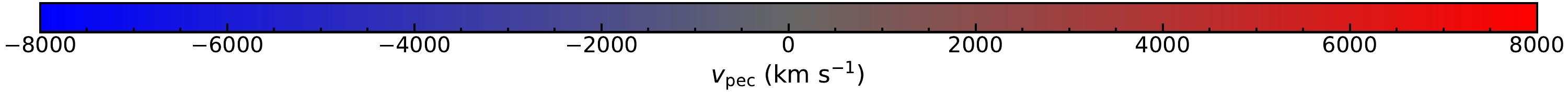}
        \caption{Optical groups and clusters identified with \redmapper within and around the Shapley (left) and Horologium-Reticulum (right) superclusters. The colours represent the velocity with respect to the mean supercluster redshift (titles). The sizes are equal to $r_{200}$ determined through the \redmapper\ richness. The coloured circles with black outlines are primary clusters (only shown within the supercluster regions), and circles without the black outline are secondary clusters or clusters outside the CHANCES supercluster survey areas, which are marked with green polygons. For primary clusters, we also show $5r_{200}$ with a black circle. The blue crosses and circles mark clusters Abell~3571 (left) and Abell~3266 (right), which are at the supercluster redshifts, but are targeted individually as CHANCES main clusters. We also mark the most massive \redmapper\ systems associated with Abell clusters with black crosses (not all of them are primaries).} 
    \label{f:superclusters}
\end{figure*}

We end this sneak peek by showing in Fig. \ref{f:superclusters} all \redmapper\ clusters within and around the two supercluster regions. We illustrate the hierarchy of clusters in each supercluster by defining primary clusters as those that are the most massive system within their $5r_{200}$ and do not lie within the $5r_{200}$ of any more massive cluster. All clusters that are not primaries are labelled as secondary. When we consider only the regions to be surveyed by CHANCES, there are 21 primary and 53 secondary clusters in Shapley and 44 primary and 92 secondary clusters in Horologium-Reticulum. 
Although we drew full circles around all primary clusters within the survey regions in Fig. \ref{f:superclusters}, only the area within the thick polygons will be observed. The exceptions are the two clusters shown with blue crosses in Fig. \ref{f:superclusters}: Abell~3571 in Shapley, and Abell~3266 in Horologium-Reticulum. In these cases, CHANCES will cover the entire $5r_{200}$ area and add the non-overlapping area to the supercluster coverage. The difficulty of characterising low-redshift structures with photometric redshifts is again highlighted in the lack of a match in the \redmapper\ catalogue for Abell~3571: This cluster is at $z=0.039$, but the only $z<0.2$ \redmapper\ cluster within $20'$ has $z_\mathrm{phot}=0.19$ and $\lambda=5$. Evidently, the absence of Abell~3571 from the \redmapper\ catalogue affects our census of primary and secondary clusters. This highlights once again the need for CHANCES spectroscopy.

As mentioned in Sect. \ref{s:design_superclusters}, Abell~3574 is located just north of Abell~3571, but at $z=0.016$, it is at a significantly lower redshift than the Shapley supercluster (and Abell~3571). This means that target overlap will be negligible in this case, even though both targets (i.e.\ Abell~3574 and the Shapley supercluster) overlap in the sky. A similar situation occurs in Horologium-Reticulum, which significantly overlaps the Fornax cluster (whose $5r_{200}$ reaches as far south as $-46^\circ$) on the sky, but not in redshift.

\section{Summary}
\label{s:conclusions}

We presented the cluster sample for CHANCES, the Chilean Cluster galaxy Evolution Survey. This is one of 18 4MOST public surveys that will use the 4MOST Spectroscopic Survey Facility on the VISTA 4m telescope over the next five years. CHANCES will obtain spectra for 500,000 galaxies out to $5r_{200}$ around galaxy clusters at $z<0.45$ (Fig. \ref{f:sky}) in three different regimes that we list below.
\begin{enumerate}[label=(\roman*)]
 \item $m^\star>10^{8.5}\,\Msun$ galaxies around 50 clusters at $z<0.07$, covering an order of magnitude in cluster masses (Fig. \ref{f:sample}, left panel).
 \item $m^\star>10^{8.5}\,\Msun$ galaxies in large contiguous regions over a total of 353 deg$^2$, covering the Shapley ($z=0.048$) and Horologium-Reticulum ($z=0.060$) superclusters (Fig. \ref{f:superclusters}).
 \item $m^\star>10^{10}\,\Msun$ galaxies around 50 of the most massive clusters at $0.07<z<0.45$ (Fig. \ref{f:sample}, right panel).
 \end{enumerate}

Points (i) and (ii) comprise the Low-z subsurvey (Sect. \ref{s:lowz}), and point (iii) refers to the Evolution subsurvey (Sect. \ref{s:evolution}).
Clusters in the Low-z subsurvey were selected by combining the AXES-2MRS catalogue \citep{Khalil2024} at $z<0.04$ with the WINGS survey at $0.04<z<0.07$ \citep{Fasano2006}. The clusters in the Evolution subsurvey were selected as the most massive southern clusters from the second Planck SZ-selected cluster catalogue \citep{PlanckCollaboration2016} in five bins across $0.07<z<0.45$ to ensure uniform redshift coverage. With the exception of the Antlia cluster at $z=0.0087$, we require multi-band optical imaging from DECaLS DR10 for a uniform photometry and astrometry to select clusters in the two subsurveys.

We produced consistent mass estimates for all clusters by rescaling masses from a number of literature sources (Fig. \ref{f:mass_comparison}) in order to obtain mass estimates that are consistent with weak lensing measurements from MENeaCS \citep{Herbonnet2020} and LoCuSS \citep{Okabe2016}, which in turn are consistent with a number of other weak lensing mass estimates, as well as calibrated SZ-based mass estimates from ACT and SPT.

As a proof of concept for some of the main CHANCES goals, we used the \redmapper\ red-sequence catalogue, which we ran on the LSDR10 imaging data to present a preliminary census of groups surrounding the CHANCES clusters (Figs. \ref{f:lss_sky} and \ref{f:overlapping}) as well as the cluster and group distribution composing the CHANCES superclusters (Fig. \ref{f:superclusters}). The reconstructed peculiar velocities suggest that the CHANCES clusters tend to reside in local attractors, although they cover a wide range of large-scale environments. We also presented preliminary measurements of the stacked phase-space (Fig. \ref{f:phase_space}) and mass function (Fig. \ref{f:imf}) of groups associated with the CHANCES clusters. The unique wide coverage and high completeness of CHANCES will provide an unprecedented view of the evolution of galaxies and the growth of structure that is facilitated by the cosmic web. Beyond our census of infalling groups, CHANCES will reveal the full extent of the multi-phase cosmic web that surrounds massive clusters, including the sheets and filaments that funnel these groups into their more massive neighbours.

In a series of forthcoming papers, we will present the target selection procedure in detail. It involves assessing cluster membership from photometric data as well as thorough analyses of the substructure and large-scale structure based on the resulting target catalogues. 

\section*{Data availability}

The code used for the analysis and figures in this paper is publicly available at \url{https://github.com/4MOST-CHANCES/cluster-catalogues}.

\begin{acknowledgements}
We thank the anonymous referee for their thorough, constructive review, which helped improve the clarity of the paper, and Jakob Dietl for sharing the eROSITA contours for the Abell 3651/3667 system.
CS, YJ, RD, AM, and HMH acknowledge support from the Agencia Nacional de Investigaci\'on y Desarrollo (ANID) through Basal project FB210003. CS acknowledges support from FONDECYT Iniciación grant no.\ 11191125.
AF thanks FINCA, USM, UdA for the travel support.
CPH, YJ, HMH, RBS, DP, FP-C, and RS gratefully acknowledge financial support from ANID - MILENIO - NCN2024\_112.
CPH acknowledges support from ANID through Fondecyt Regular 2021 project no. 1211909.
YJ acknowledges support from FONDECYT Regular grants no.\ 1241426 and 1230441.
BMA acknowledges the Universidad de Atacama PhD Scholarship and the Fondecyt Regular 2021 Grant of Dr. Christopher P. Haines, PhD Supervisor, for the expenses to successfully conduct this research.
RD gratefully acknowledges support by the ANID BASAL project FB210003.
ERVL acknowledges the financial support given by Coordenação de Aperfeiçoamento de Pessoal de Nível Superior (CAPES, grant 88887.470064/2019-00) and Conselho Nacional de Desenvolvimento Científico e Tecnológico (CNPq, grant 169181/2017-0).
CL-D acknowledges a grant from the ESO Comite Mixto 2022.
HMH acknowledges support from National Fund for Scientific and Technological Research of Chile (FONDECYT) through grant no. 3230176.
AM acknowledges support by the FONDECYT Regular grant 1212046 and funding from the Max Planck Society through a ``PartnerGroup'' grant and the HORIZON-MSCA-2021-SE-01 Research and Innovation Programme under the Marie Sklodowska-Curie grant agreement number 101086388.
FA-F acknowledges funding for this work from FAPESP grant 2018/20977-2.
PA-A thanks the Coordenaç\~o de Aperfeiçoamento de Pessoal de Nível Superior – Brasil (CAPES), for supporting his PhD scholarship (project 88882.332909/2020-01).
MA-F acknowledges support from ANID FONDECYT iniciaci\'on project 11200107 and the Emergia program (EMERGIA20\_38888) from Consejer\'ia de Universidad, Investigaci\'on e Innovaci\'on de la Junta de Andaluc\'ia.
ANID / Subdirección de Capital Humano / Doctorado Nacional / 2023 - 21231017.
CRB acknowledges the financial support from CNPq (316072/2021-4) and from FAPERJ (grants 201.456/2022 and 210.330/2022) and the FINEP contract 01.22.0505.00 (ref. 1891/22).
HMC acknowledges support from the Institut Universitaire de France and from Centre National d’Etudes Spatiales (CNES), France.
GD acknowledges support by UKRI-STFC grants: ST/T003081/1 and ST/X001857/1.
AD is supported by a KIAS Individual Grant PG 087201 at the Korea Institute for Advanced Studies.
RFH acknowledge financial support from Consejo Nacional de Investigaciones Científicas y Tecnicas (CONICET), Agencia I+D+i (PICT 2019–03299), and Universidad Nacional de La Plata (Argentina).
FRH acknowledges support from FAPESP grants 2018/21661-9 and 2021/11345-5.
EI acknowledge funding by ANID FONDECYT Regular 1221846.
UK acknowledges financial support from the UK Science and Technology Facilities Council (STFC; grant ref: ST/T000171/1).
ARL acknowledges financial support from CONICET, Agencia I+D+i (PICT  2019-03299) and Universidad Nacional de La Plata (Argentina).
SL acknowledges support by FONDECYT grant 1231187.
CMdO acknowledges funding of the S-PLUS project through FAPESP grant 2019/26492-3.
LM  acknowledge  the  support  from  PROYECTOS  FONDO  de  ASTRONOMIA  ANID–ALMA  2021  Code:  ASTRO21-0007.
DP acknowledges financial support from ANID through FONDECYT Postdoctrorado Project 3230379.
FPC acknowledges financial support from Dirección de Postgrado (Universidad Técnica Federico Santa María, Chile) through Becas Internas para Doctorado y Magíster Científico-Tecnológicos.
AVSC acknowledges financial support from CONICET, Agencia I+D+i (PICT 2019-03299) and Universidad Nacional de La Plata (Argentina).
RS acknowledges financial support from FONDECYT Regular 2023 project No. 1230441
LSJ acknowledges the support from CNPq (308994/2021-3) and FAPESP (2011/51680-6).
ET was supported by the Estonian Ministry of Education and Research (grant TK202), Estonian Research Council grant (PRG1006) and the European Union's Horizon Europe research and innovation programme (EXCOSM, grant No. 101159513).
This work made use of Astropy: a community-developed core Python package and an ecosystem of tools and resources for astronomy \citep[\url{http://www.astropy.org}]{AstropyCollaboration2013, AstropyCollaboration2018, AstropyCollaboration2022}. This work has benefited from open-source software including Matplotlib \citep[\url{https://matplotlib.org/}]{Hunter2007}, NumPy \citep[\url{https://numpy.org/}]{Harris2020}, SciPy \citep[\url{https://scipy.org/}]{Virtanen2020}, and Pandas \citep[\url{https://pandas.pydata.org/}]{Team2024}. This work has been undertaken in the framework of the 4MOST collaboration (\url{https://www.4most.eu/cms/home/}).

\end{acknowledgements}

\bibliographystyle{aa}
\bibliography{chances_clusters}

\begin{thebibliography}{135}
\expandafter\ifx\csname natexlab\endcsname\relax\def\natexlab#1{#1}\fi

\bibitem[{{Abell} {et~al.}(1989){Abell}, {Corwin}, \& {Olowin}}]{Abell1989}
{Abell}, G.~O., {Corwin}, Jr., H.~G., \& {Olowin}, R.~P. 1989, \apjs, 70, 1

\bibitem[{{Ade} {et~al.}(2019){Ade}, {Aguirre}, {Ahmed}, {Aiola}, {Ali},
  {Alonso}, {Alvarez}, {Arnold}, {Ashton}, {Austermann}, {Awan}, {Baccigalupi},
  {Baildon}, {Barron}, {Battaglia}, {Battye}, {Baxter}, {Bazarko}, {Beall},
  {Bean}, {Beck}, {Beckman}, {Beringue}, {Bianchini}, {Boada}, {Boettger},
  {Bond}, {Borrill}, {Brown}, {Bruno}, {Bryan}, {Calabrese}, {Calafut},
  {Calisse}, {Carron}, {Challinor}, {Chesmore}, {Chinone}, {Chluba}, {Cho},
  {Choi}, {Coppi}, {Cothard}, {Coughlin}, {Crichton}, {Crowley}, {Crowley},
  {Cukierman}, {D'Ewart}, {D{\"u}nner}, {de Haan}, {Devlin}, {Dicker},
  {Didier}, {Dobbs}, {Dober}, {Duell}, {Duff}, {Duivenvoorden}, {Dunkley},
  {Dusatko}, {Errard}, {Fabbian}, {Feeney}, {Ferraro}, {Flux{\`a}}, {Freese},
  {Frisch}, {Frolov}, {Fuller}, {Fuzia}, {Galitzki}, {Gallardo}, {Tomas Galvez
  Ghersi}, {Gao}, {Gawiser}, {Gerbino}, {Gluscevic}, {Goeckner-Wald}, {Golec},
  {Gordon}, {Gralla}, {Green}, {Grigorian}, {Groh}, {Groppi}, {Guan},
  {Gudmundsson}, {Han}, {Hargrave}, {Hasegawa}, {Hasselfield}, {Hattori},
  {Haynes}, {Hazumi}, {He}, {Healy}, {Henderson}, {Hervias-Caimapo}, {Hill},
  {Hill}, {Hilton}, {Hilton}, {Hincks}, {Hinshaw}, {Hlo{\v{z}}ek}, {Ho}, {Ho},
  {Howe}, {Huang}, {Hubmayr}, {Huffenberger}, {Hughes}, {Ijjas}, {Ikape},
  {Irwin}, {Jaffe}, {Jain}, {Jeong}, {Kaneko}, {Karpel}, {Katayama}, {Keating},
  {Kernasovskiy}, {Keskitalo}, {Kisner}, {Kiuchi}, {Klein}, {Knowles},
  {Koopman}, {Kosowsky}, {Krachmalnicoff}, {Kuenstner}, {Kuo}, {Kusaka},
  {Lashner}, {Lee}, {Lee}, {Leon}, {Leung}, {Lewis}, {Li}, {Li}, {Limon},
  {Linder}, {Lopez-Caraballo}, {Louis}, {Lowry}, {Lungu}, {Madhavacheril},
  {Mak}, {Maldonado}, {Mani}, {Mates}, {Matsuda}, {Maurin}, {Mauskopf}, {May},
  {McCallum}, {McKenney}, {McMahon}, {Meerburg}, {Meyers}, {Miller},
  {Mirmelstein}, {Moodley}, {Munchmeyer}, {Munson}, {Naess}, {Nati},
  {Navaroli}, {Newburgh}, {Nguyen}, {Niemack}, {Nishino}, {Orlowski-Scherer},
  {Page}, {Partridge}, {Peloton}, {Perrotta}, {Piccirillo}, {Pisano},
  {Poletti}, {Puddu}, {Puglisi}, {Raum}, {Reichardt}, {Remazeilles},
  {Rephaeli}, {Riechers}, {Rojas}, {Roy}, {Sadeh}, {Sakurai}, {Salatino},
  {Sathyanarayana Rao}, {Schaan}, {Schmittfull}, {Sehgal}, {Seibert}, {Seljak},
  {Sherwin}, {Shimon}, {Sierra}, {Sievers}, {Sikhosana}, {Silva-Feaver},
  {Simon}, {Sinclair}, {Siritanasak}, {Smith}, {Smith}, {Spergel}, {Staggs},
  {Stein}, {Stevens}, {Stompor}, {Suzuki}, {Tajima}, {Takakura}, {Teply},
  {Thomas}, {Thorne}, {Thornton}, {Trac}, {Tsai}, {Tucker}, {Ullom},
  {Vagnozzi}, {van Engelen}, {Van Lanen}, {Van Winkle}, {Vavagiakis},
  {Verg{\`e}s}, {Vissers}, {Wagoner}, {Walker}, {Ward}, {Westbrook},
  {Whitehorn}, {Williams}, {Williams}, {Wollack}, {Xu}, {Yu}, {Yu}, {Zago},
  {Zhang}, {Zhu}, \& {Simons Observatory Collaboration}}]{Ade2019}
{Ade}, P., {Aguirre}, J., {Ahmed}, Z., {et~al.} 2019, \jcap, 2019, 056

\bibitem[{{Alberts} \& {Noble}(2022)}]{Alberts2022}
{Alberts}, S. \& {Noble}, A. 2022, Universe, 8, 554

\bibitem[{{Anand} {et~al.}(2021){Anand}, {Nelson}, \& {Kauffmann}}]{Anand2021}
{Anand}, A., {Nelson}, D., \& {Kauffmann}, G. 2021, \mnras, 504, 65

\bibitem[{{Andrade-Santos} {et~al.}(2017){Andrade-Santos}, {Jones}, {Forman},
  {Lovisari}, {Vikhlinin}, {van Weeren}, {Murray}, {Arnaud}, {Pratt},
  {D{\'e}mocl{\`e}s}, {Kraft}, {Mazzotta}, {B{\"o}hringer}, {Chon},
  {Giacintucci}, {Clarke}, {Borgani}, {David}, {Douspis}, {Pointecouteau},
  {Dahle}, {Brown}, {Aghanim}, \& {Rasia}}]{AndradeSantos2017}
{Andrade-Santos}, F., {Jones}, C., {Forman}, W.~R., {et~al.} 2017, \apj, 843,
  76

\bibitem[{{Araya-Melo} {et~al.}(2009){Araya-Melo}, {Reisenegger}, {Meza}, {van
  de Weygaert}, {D{\"u}nner}, \& {Quintana}}]{ArayaMelo2009}
{Araya-Melo}, P.~A., {Reisenegger}, A., {Meza}, A., {et~al.} 2009, \mnras, 399,
  97

\bibitem[{{Arp} \& {Russell}(2001)}]{Arp2001}
{Arp}, H. \& {Russell}, D. 2001, \apj, 549, 802

\bibitem[{{Astropy Collaboration} {et~al.}(2022){Astropy Collaboration},
  {Price-Whelan}, {Lim}, {Earl}, {Starkman}, {Bradley}, {Shupe}, {Patil},
  {Corrales}, {Brasseur}, {N{\"o}the}, {Donath}, {Tollerud}, {Morris},
  {Ginsburg}, {Vaher}, {Weaver}, {Tocknell}, {Jamieson}, {van Kerkwijk},
  {Robitaille}, {Merry}, {Bachetti}, {G{\"u}nther}, {Aldcroft},
  {Alvarado-Montes}, {Archibald}, {B{\'o}di}, {Bapat}, {Barentsen},
  {Baz{\'a}n}, {Biswas}, {Boquien}, {Burke}, {Cara}, {Cara}, {Conroy},
  {Conseil}, {Craig}, {Cross}, {Cruz}, {D'Eugenio}, {Dencheva}, {Devillepoix},
  {Dietrich}, {Eigenbrot}, {Erben}, {Ferreira}, {Foreman-Mackey}, {Fox},
  {Freij}, {Garg}, {Geda}, {Glattly}, {Gondhalekar}, {Gordon}, {Grant},
  {Greenfield}, {Groener}, {Guest}, {Gurovich}, {Handberg}, {Hart},
  {Hatfield-Dodds}, {Homeier}, {Hosseinzadeh}, {Jenness}, {Jones}, {Joseph},
  {Kalmbach}, {Karamehmetoglu}, {Ka{\l}uszy{\'n}ski}, {Kelley}, {Kern},
  {Kerzendorf}, {Koch}, {Kulumani}, {Lee}, {Ly}, {Ma}, {MacBride}, {Maljaars},
  {Muna}, {Murphy}, {Norman}, {O'Steen}, {Oman}, {Pacifici}, {Pascual},
  {Pascual-Granado}, {Patil}, {Perren}, {Pickering}, {Rastogi}, {Roulston},
  {Ryan}, {Rykoff}, {Sabater}, {Sakurikar}, {Salgado}, {Sanghi}, {Saunders},
  {Savchenko}, {Schwardt}, {Seifert-Eckert}, {Shih}, {Jain}, {Shukla}, {Sick},
  {Simpson}, {Singanamalla}, {Singer}, {Singhal}, {Sinha}, {Sip{\H{o}}cz},
  {Spitler}, {Stansby}, {Streicher}, {{\v{S}}umak}, {Swinbank}, {Taranu},
  {Tewary}, {Tremblay}, {de Val-Borro}, {Van Kooten}, {Vasovi{\'c}}, {Verma},
  {de Miranda Cardoso}, {Williams}, {Wilson}, {Winkel}, {Wood-Vasey}, {Xue},
  {Yoachim}, {Zhang}, {Zonca}, \& {Astropy Project
  Contributors}}]{AstropyCollaboration2022}
{Astropy Collaboration}, {Price-Whelan}, A.~M., {Lim}, P.~L., {et~al.} 2022,
  \apj, 935, 167

\bibitem[{{Astropy Collaboration} {et~al.}(2018){Astropy Collaboration},
  {Price-Whelan}, {Sip{\H{o}}cz}, {G{\"u}nther}, {Lim}, {Crawford}, {Conseil},
  {Shupe}, {Craig}, {Dencheva}, {Ginsburg}, {VanderPlas}, {Bradley},
  {P{\'e}rez-Su{\'a}rez}, {de Val-Borro}, {Aldcroft}, {Cruz}, {Robitaille},
  {Tollerud}, {Ardelean}, {Babej}, {Bach}, {Bachetti}, {Bakanov}, {Bamford},
  {Barentsen}, {Barmby}, {Baumbach}, {Berry}, {Biscani}, {Boquien}, {Bostroem},
  {Bouma}, {Brammer}, {Bray}, {Breytenbach}, {Buddelmeijer}, {Burke},
  {Calderone}, {Cano Rodr{\'\i}guez}, {Cara}, {Cardoso}, {Cheedella}, {Copin},
  {Corrales}, {Crichton}, {D'Avella}, {Deil}, {Depagne}, {Dietrich}, {Donath},
  {Droettboom}, {Earl}, {Erben}, {Fabbro}, {Ferreira}, {Finethy}, {Fox},
  {Garrison}, {Gibbons}, {Goldstein}, {Gommers}, {Greco}, {Greenfield},
  {Groener}, {Grollier}, {Hagen}, {Hirst}, {Homeier}, {Horton}, {Hosseinzadeh},
  {Hu}, {Hunkeler}, {Ivezi{\'c}}, {Jain}, {Jenness}, {Kanarek}, {Kendrew},
  {Kern}, {Kerzendorf}, {Khvalko}, {King}, {Kirkby}, {Kulkarni}, {Kumar},
  {Lee}, {Lenz}, {Littlefair}, {Ma}, {Macleod}, {Mastropietro}, {McCully},
  {Montagnac}, {Morris}, {Mueller}, {Mumford}, {Muna}, {Murphy}, {Nelson},
  {Nguyen}, {Ninan}, {N{\"o}the}, {Ogaz}, {Oh}, {Parejko}, {Parley}, {Pascual},
  {Patil}, {Patil}, {Plunkett}, {Prochaska}, {Rastogi}, {Reddy Janga},
  {Sabater}, {Sakurikar}, {Seifert}, {Sherbert}, {Sherwood-Taylor}, {Shih},
  {Sick}, {Silbiger}, {Singanamalla}, {Singer}, {Sladen}, {Sooley},
  {Sornarajah}, {Streicher}, {Teuben}, {Thomas}, {Tremblay}, {Turner},
  {Terr{\'o}n}, {van Kerkwijk}, {de la Vega}, {Watkins}, {Weaver}, {Whitmore},
  {Woillez}, {Zabalza}, \& {Astropy Contributors}}]{AstropyCollaboration2018}
{Astropy Collaboration}, {Price-Whelan}, A.~M., {Sip{\H{o}}cz}, B.~M., {et~al.}
  2018, \aj, 156, 123

\bibitem[{{Astropy Collaboration} {et~al.}(2013){Astropy Collaboration},
  {Robitaille}, {Tollerud}, {Greenfield}, {Droettboom}, {Bray}, {Aldcroft},
  {Davis}, {Ginsburg}, {Price-Whelan}, {Kerzendorf}, {Conley}, {Crighton},
  {Barbary}, {Muna}, {Ferguson}, {Grollier}, {Parikh}, {Nair}, {Unther},
  {Deil}, {Woillez}, {Conseil}, {Kramer}, {Turner}, {Singer}, {Fox}, {Weaver},
  {Zabalza}, {Edwards}, {Azalee Bostroem}, {Burke}, {Casey}, {Crawford},
  {Dencheva}, {Ely}, {Jenness}, {Labrie}, {Lim}, {Pierfederici}, {Pontzen},
  {Ptak}, {Refsdal}, {Servillat}, \& {Streicher}}]{AstropyCollaboration2013}
{Astropy Collaboration}, {Robitaille}, T.~P., {Tollerud}, E.~J., {et~al.} 2013,
  \aap, 558, A33

\bibitem[{{Bah{\'e}} {et~al.}(2013){Bah{\'e}}, {McCarthy}, {Balogh}, \&
  {Font}}]{Bahe2013}
{Bah{\'e}}, Y.~M., {McCarthy}, I.~G., {Balogh}, M.~L., \& {Font}, A.~S. 2013,
  \mnras, 430, 3017

\bibitem[{{Bah{\'e}} {et~al.}(2019){Bah{\'e}}, {Schaye}, {Barnes}, {Dalla
  Vecchia}, {Kay}, {Bower}, {Hoekstra}, {McGee}, \& {Theuns}}]{Bahe2019}
{Bah{\'e}}, Y.~M., {Schaye}, J., {Barnes}, D.~J., {et~al.} 2019, \mnras, 485,
  2287

\bibitem[{{Bertschinger}(1985)}]{Bertschinger1985}
{Bertschinger}, E. 1985, \apjs, 58, 39

\bibitem[{{Bianconi} {et~al.}(2018){Bianconi}, {Smith}, {Haines}, {McGee},
  {Finoguenov}, \& {Egami}}]{Bianconi2018}
{Bianconi}, M., {Smith}, G.~P., {Haines}, C.~P., {et~al.} 2018, \mnras, 473,
  L79

\bibitem[{{Bleem} {et~al.}(2020){Bleem}, {Bocquet}, {Stalder}, {Gladders},
  {Ade}, {Allen}, {Anderson}, {Annis}, {Ashby}, {Austermann}, {Avila}, {Avva},
  {Bayliss}, {Beall}, {Bechtol}, {Bender}, {Benson}, {Bertin}, {Bianchini},
  {Blake}, {Brodwin}, {Brooks}, {Buckley-Geer}, {Burke}, {Carlstrom}, {Rosell},
  {Carrasco Kind}, {Carretero}, {Chang}, {Chiang}, {Citron}, {Moran},
  {Costanzi}, {Crawford}, {Crites}, {da Costa}, {de Haan}, {De Vicente},
  {Desai}, {Diehl}, {Dietrich}, {Dobbs}, {Eifler}, {Everett}, {Flaugher},
  {Floyd}, {Frieman}, {Gallicchio}, {Garc{\'\i}a-Bellido}, {George}, {Gerdes},
  {Gilbert}, {Gruen}, {Gruendl}, {Gschwend}, {Gupta}, {Gutierrez}, {Halverson},
  {Harrington}, {Henning}, {Heymans}, {Holder}, {Hollowood}, {Holzapfel},
  {Honscheid}, {Hrubes}, {Huang}, {Hubmayr}, {Irwin}, {James}, {Jeltema},
  {Joudaki}, {Khullar}, {Klein}, {Knox}, {Kuropatkin}, {Lee}, {Li}, {Lidman},
  {Lowitz}, {MacCrann}, {Mahler}, {Maia}, {Marshall}, {McDonald}, {McMahon},
  {Melchior}, {Menanteau}, {Meyer}, {Miquel}, {Mocanu}, {Mohr}, {Montgomery},
  {Nadolski}, {Natoli}, {Nibarger}, {Noble}, {Novosad}, {Padin}, {Palmese},
  {Parkinson}, {Patil}, {Paz-Chinch{\'o}n}, {Plazas}, {Pryke}, {Ramachandra},
  {Reichardt}, {Remolina Gonz{\'a}lez}, {Romer}, {Roodman}, {Ruhl}, {Rykoff},
  {Saliwanchik}, {Sanchez}, {Saro}, {Sayre}, {Schaffer}, {Schrabback},
  {Serrano}, {Sharon}, {Sievers}, {Smecher}, {Smith}, {Soares-Santos}, {Stark},
  {Story}, {Suchyta}, {Tarle}, {Tucker}, {Vanderlinde}, {Veach}, {Vieira},
  {Wang}, {Weller}, {Whitehorn}, {Wu}, {Yefremenko}, \& {Zhang}}]{Bleem2020}
{Bleem}, L.~E., {Bocquet}, S., {Stalder}, B., {et~al.} 2020, \apjs, 247, 25

\bibitem[{{Bleem} {et~al.}(2015){Bleem}, {Stalder}, {de Haan}, {Aird}, {Allen},
  {Applegate}, {Ashby}, {Bautz}, {Bayliss}, {Benson}, {Bocquet}, {Brodwin},
  {Carlstrom}, {Chang}, {Chiu}, {Cho}, {Clocchiatti}, {Crawford}, {Crites},
  {Desai}, {Dietrich}, {Dobbs}, {Foley}, {Forman}, {George}, {Gladders},
  {Gonzalez}, {Halverson}, {Hennig}, {Hoekstra}, {Holder}, {Holzapfel},
  {Hrubes}, {Jones}, {Keisler}, {Knox}, {Lee}, {Leitch}, {Liu}, {Lueker},
  {Luong-Van}, {Mantz}, {Marrone}, {McDonald}, {McMahon}, {Meyer}, {Mocanu},
  {Mohr}, {Murray}, {Padin}, {Pryke}, {Reichardt}, {Rest}, {Ruel}, {Ruhl},
  {Saliwanchik}, {Saro}, {Sayre}, {Schaffer}, {Schrabback}, {Shirokoff},
  {Song}, {Spieler}, {Stanford}, {Staniszewski}, {Stark}, {Story}, {Stubbs},
  {Vanderlinde}, {Vieira}, {Vikhlinin}, {Williamson}, {Zahn}, \&
  {Zenteno}}]{Bleem2015}
{Bleem}, L.~E., {Stalder}, B., {de Haan}, T., {et~al.} 2015, \apjs, 216, 27

\bibitem[{{B{\"o}hringer} {et~al.}(2004){B{\"o}hringer}, {Schuecker}, {Guzzo},
  {Collins}, {Voges}, {Cruddace}, {Ortiz-Gil}, {Chincarini}, {De Grandi},
  {Edge}, {MacGillivray}, {Neumann}, {Schindler}, \& {Shaver}}]{Boehringer2004}
{B{\"o}hringer}, H., {Schuecker}, P., {Guzzo}, L., {et~al.} 2004, \aap, 425,
  367

\bibitem[{{Boselli} {et~al.}(2022){Boselli}, {Fossati}, \& {Sun}}]{Boselli2022}
{Boselli}, A., {Fossati}, M., \& {Sun}, M. 2022, \aapr, 30, 3

\bibitem[{{Brown} {et~al.}(2017){Brown}, {Catinella}, {Cortese}, {Lagos},
  {Dav{\'e}}, {Kilborn}, {Haynes}, {Giovanelli}, \&
  {Rafieferantsoa}}]{Brown2017}
{Brown}, T., {Catinella}, B., {Cortese}, L., {et~al.} 2017, \mnras, 466, 1275

\bibitem[{{Bruzual} \& {Charlot}(2003)}]{Bruzual2003}
{Bruzual}, G. \& {Charlot}, S. 2003, \mnras, 344, 1000

\bibitem[{{Bulbul} {et~al.}(2024){Bulbul}, {Liu}, {Kluge}, {Zhang}, {Sanders},
  {Bahar}, {Ghirardini}, {Artis}, {Seppi}, {Garrel}, {Ramos-Ceja}, {Comparat},
  {Balzer}, {B{\"o}ckmann}, {Br{\"u}ggen}, {Clerc}, {Dennerl}, {Dolag},
  {Freyberg}, {Grandis}, {Gruen}, {Kleinebreil}, {Krippendorf}, {Lamer},
  {Merloni}, {Migkas}, {Nandra}, {Pacaud}, {Predehl}, {Reiprich}, {Schrabback},
  {Veronica}, {Weller}, \& {Zelmer}}]{Bulbul2024}
{Bulbul}, E., {Liu}, A., {Kluge}, M., {et~al.} 2024, \aap, 685, A106

\bibitem[{{Butcher} \& {Oemler}(1984)}]{Butcher1984}
{Butcher}, H. \& {Oemler}, A., J. 1984, \apj, 285, 426

\bibitem[{{Cautun} {et~al.}(2014){Cautun}, {van de Weygaert}, {Jones}, \&
  {Frenk}}]{Cautun2014}
{Cautun}, M., {van de Weygaert}, R., {Jones}, B. J.~T., \& {Frenk}, C.~S. 2014,
  \mnras, 441, 2923

\bibitem[{{Cava} {et~al.}(2009){Cava}, {Bettoni}, {Poggianti}, {Couch},
  {Moles}, {Varela}, {Biviano}, {D'Onofrio}, {Dressler}, {Fasano}, {Fritz},
  {Kj{\ae}rgaard}, {Ramella}, \& {Valentinuzzi}}]{Cava2009}
{Cava}, A., {Bettoni}, D., {Poggianti}, B.~M., {et~al.} 2009, \aap, 495, 707

\bibitem[{{CCAT-Prime Collaboration} {et~al.}(2023){CCAT-Prime Collaboration},
  {Aravena}, {Austermann}, {Basu}, {Battaglia}, {Beringue}, {Bertoldi},
  {Bigiel}, {Bond}, {Breysse}, {Broughton}, {Bustos}, {Chapman}, {Charmetant},
  {Choi}, {Chung}, {Clark}, {Cothard}, {Crites}, {Dev}, {Douglas}, {Duell},
  {D{\"u}nner}, {Ebina}, {Erler}, {Fich}, {Fissel}, {Foreman}, {Freundt},
  {Gallardo}, {Gao}, {Garc{\'\i}a}, {Giovanelli}, {Golec}, {Groppi}, {Haynes},
  {Henke}, {Hensley}, {Herter}, {Higgins}, {Hlo{\v{z}}ek}, {Huber}, {Huber},
  {Hubmayr}, {Jackson}, {Johnstone}, {Karoumpis}, {Keating}, {Komatsu}, {Li},
  {Magnelli}, {Matthews}, {Mauskopf}, {McMahon}, {Meerburg}, {Meyers},
  {Muralidhara}, {Murray}, {Niemack}, {Nikola}, {Okada}, {Puddu}, {Riechers},
  {Rosolowsky}, {Rossi}, {Rotermund}, {Roy}, {Sadavoy}, {Schaaf}, {Schilke},
  {Scott}, {Simon}, {Sinclair}, {Sivakoff}, {Stacey}, {Stutz}, {Stutzki},
  {Tahani}, {Thanjavur}, {Timmermann}, {Ullom}, {van Engelen}, {Vavagiakis},
  {Vissers}, {Wheeler}, {White}, {Zhu}, \& {Zou}}]{CPC2023}
{CCAT-Prime Collaboration}, {Aravena}, M., {Austermann}, J.~E., {et~al.} 2023,
  \apjs, 264, 7

\bibitem[{{Chabrier}(2003)}]{Chabrier2003}
{Chabrier}, G. 2003, \pasp, 115, 763

\bibitem[{{CHEX-MATE Collaboration} {et~al.}(2021){CHEX-MATE Collaboration},
  {Arnaud}, {Ettori}, {Pratt}, {Rossetti}, {Eckert}, {Gastaldello}, {Gavazzi},
  {Kay}, {Lovisari}, {Maughan}, {Pointecouteau}, {Sereno}, {Bartalucci},
  {Bonafede}, {Bourdin}, {Cassano}, {Duffy}, {Iqbal}, {Maurogordato}, {Rasia},
  {Sayers}, {Andrade-Santos}, {Aussel}, {Barnes}, {Barrena}, {Borgani},
  {Burkutean}, {Clerc}, {Corasaniti}, {Cuillandre}, {De Grandi}, {De Petris},
  {Dolag}, {Donahue}, {Ferragamo}, {Gaspari}, {Ghizzardi}, {Gitti}, {Haines},
  {Jauzac}, {Johnston-Hollitt}, {Jones}, {K{\'e}ruzor{\'e}}, {Le Brun},
  {Mayet}, {Mazzotta}, {Melin}, {Molendi}, {Nonino}, {Okabe}, {Paltani},
  {Perotto}, {Pires}, {Radovich}, {Rubino-Martin}, {Salvati}, {Saro},
  {Sartoris}, {Schellenberger}, {Streblyanska}, {Tarr{\'\i}o}, {Tozzi},
  {Umetsu}, {van der Burg}, {Vazza}, {Venturi}, {Yepes}, \&
  {Zarattini}}]{CMC2021}
{CHEX-MATE Collaboration}, {Arnaud}, M., {Ettori}, S., {et~al.} 2021, \aap,
  650, A104

\bibitem[{{Clerc} {et~al.}(2020){Clerc}, {Kirkpatrick}, {Finoguenov},
  {Capasso}, {Comparat}, {Damsted}, {Furnell}, {Kukkola}, {Ider Chitham},
  {Merloni}, {Salvato}, {Gueguen}, {Dwelly}, {Collins}, {Saro}, {Erfanianfar},
  {Schneider}, {Brownstein}, {Mamon}, {Padilla}, {Jullo}, \&
  {Bizyaev}}]{Clerc2020}
{Clerc}, N., {Kirkpatrick}, C.~C., {Finoguenov}, A., {et~al.} 2020, \mnras,
  497, 3976

\bibitem[{{Courtois} {et~al.}(2023){Courtois}, {Dupuy}, {Guinet}, {Baulieu},
  {Ruppin}, \& {Brenas}}]{Courtois2023}
{Courtois}, H.~M., {Dupuy}, A., {Guinet}, D., {et~al.} 2023, \aap, 670, L15

\bibitem[{{Damsted} {et~al.}(2023){Damsted}, {Finoguenov}, {Clerc},
  {Davalgait{\.{e}}}, {Kirkpatrick}, {Mamon}, {Ider Chitham}, {Kiiveri},
  {Comparat}, \& {Collins}}]{Damsted2023}
{Damsted}, S., {Finoguenov}, A., {Clerc}, N., {et~al.} 2023, \aap, 676, A127

\bibitem[{{Damsted} {et~al.}(2024){Damsted}, {Finoguenov}, {Lietzen}, {Mamon},
  {Comparat}, {Tempel}, {Dmitrieva}, {Clerc}, {Collins}, {Gozaliasl}, \&
  {Eckert}}]{Damsted2024}
{Damsted}, S., {Finoguenov}, A., {Lietzen}, H., {et~al.} 2024, \aap, 690, A52

\bibitem[{{de Gasperin} {et~al.}(2022){de Gasperin}, {Rudnick}, {Finoguenov},
  {Wittor}, {Akamatsu}, {Br{\"u}ggen}, {Chibueze}, {Clarke}, {Cotton},
  {Cuciti}, {Dom{\'\i}nguez-Fern{\'a}ndez}, {Knowles}, {O'Sullivan}, \&
  {Sebokolodi}}]{DeGasperin2022}
{de Gasperin}, F., {Rudnick}, L., {Finoguenov}, A., {et~al.} 2022, \aap, 659,
  A146

\bibitem[{{de Jong} {et~al.}(2019){de Jong}, {Agertz}, {Berbel}, {Aird},
  {Alexander}, {Amarsi}, {Anders}, {Andrae}, {Ansarinejad}, {Ansorge},
  {Antilogus}, {Anwand-Heerwart}, {Arentsen}, {Arnadottir}, {Asplund}, {Auger},
  {Azais}, {Baade}, {Baker}, {Baker}, {Balbinot}, {Baldry}, {Banerji},
  {Barden}, {Barklem}, {Barth{\'e}l{\'e}my-Mazot}, {Battistini}, {Bauer},
  {Bell}, {Bellido-Tirado}, {Bellstedt}, {Belokurov}, {Bensby}, {Bergemann},
  {Bestenlehner}, {Bielby}, {Bilicki}, {Blake}, {Bland-Hawthorn}, {Boeche},
  {Boland}, {Boller}, {Bongard}, {Bongiorno}, {Bonifacio}, {Boudon}, {Brooks},
  {Brown}, {Brown}, {Br{\"u}ggen}, {Brynnel}, {Brzeski}, {Buchert},
  {Buschkamp}, {Caffau}, {Caillier}, {Carrick}, {Casagrande}, {Case}, {Casey},
  {Cesarini}, {Cescutti}, {Chapuis}, {Chiappini}, {Childress}, {Christlieb},
  {Church}, {Cioni}, {Cluver}, {Colless}, {Collett}, {Comparat}, {Cooper},
  {Couch}, {Courbin}, {Croom}, {Croton}, {Daguis{\'e}}, {Dalton}, {Davies},
  {Davis}, {de Laverny}, {Deason}, {Dionies}, {Disseau}, {Doel}, {D{\"o}scher},
  {Driver}, {Dwelly}, {Eckert}, {Edge}, {Edvardsson}, {Youssoufi}, {Elhaddad},
  {Enke}, {Erfanianfar}, {Farrell}, {Fechner}, {Feiz}, {Feltzing}, {Ferreras},
  {Feuerstein}, {Feuillet}, {Finoguenov}, {Ford}, {Fotopoulou}, {Fouesneau},
  {Frenk}, {Frey}, {Gaessler}, {Geier}, {Gentile Fusillo}, {Gerhard},
  {Giannantonio}, {Giannone}, {Gibson}, {Gillingham},
  {Gonz{\'a}lez-Fern{\'a}ndez}, {Gonzalez-Solares}, {Gottloeber}, {Gould},
  {Grebel}, {Gueguen}, {Guiglion}, {Haehnelt}, {Hahn}, {Hansen}, {Hartman},
  {Hauptner}, {Hawkins}, {Haynes}, {Haynes}, {Heiter}, {Helmi}, {Aguayo},
  {Hewett}, {Hinton}, {Hobbs}, {Hoenig}, {Hofman}, {Hook}, {Hopgood},
  {Hopkins}, {Hourihane}, {Howes}, {Howlett}, {Huet}, {Irwin}, {Iwert},
  {Jablonka}, {Jahn}, {Jahnke}, {Jarno}, {Jin}, {Jofre}, {Johl}, {Jones},
  {J{\"o}nsson}, {Jordan}, {Karovicova}, {Khalatyan}, {Kelz}, {Kennicutt},
  {King}, {Kitaura}, {Klar}, {Klauser}, {Kneib}, {Koch}, {Koposov},
  {Kordopatis}, {Korn}, {Kosmalski}, {Kotak}, {Kovalev}, {Kreckel}, {Kripak},
  {Krumpe}, {Kuijken}, {Kunder}, {Kushniruk}, {Lam}, {Lamer}, {Laurent},
  {Lawrence}, {Lehmitz}, {Lemasle}, {Lewis}, {Li}, {Lidman}, {Lind}, {Liske},
  {Lizon}, {Loveday}, {Ludwig}, {McDermid}, {Maguire}, {Mainieri}, {Mali},
  {Mandel}, {Mandel}, {Mannering}, {Martell}, {Martinez Delgado}, {Matijevic},
  {McGregor}, {McMahon}, {McMillan}, {Mena}, {Merloni}, {Meyer}, {Michel},
  {Micheva}, {Migniau}, {Minchev}, {Monari}, {Muller}, {Murphy},
  {Muthukrishna}, {Nandra}, {Navarro}, {Ness}, {Nichani}, {Nichol}, {Nicklas},
  {Niederhofer}, {Norberg}, {Obreschkow}, {Oliver}, {Owers}, {Pai},
  {Pankratow}, {Parkinson}, {Paschke}, {Paterson}, {Pecontal}, {Parry},
  {Phillips}, {Pillepich}, {Pinard}, {Pirard}, {Piskunov}, {Plank},
  {Pl{\"u}schke}, {Pons}, {Popesso}, {Power}, {Pragt}, {Pramskiy}, {Pryer},
  {Quattri}, {Queiroz}, {Quirrenbach}, {Rahurkar}, {Raichoor}, {Ramstedt},
  {Rau}, {Recio-Blanco}, {Reiss}, {Renaud}, {Revaz}, {Rhode}, {Richard},
  {Richter}, {Rix}, {Robotham}, {Roelfsema}, {Romaniello}, {Rosario},
  {Rothmaier}, {Roukema}, {Ruchti}, {Rupprecht}, {Rybizki}, {Ryde}, {Saar},
  {Sadler}, {Sahl{\'e}n}, {Salvato}, {Sassolas}, {Saunders}, {Saviauk},
  {Sbordone}, {Schmidt}, {Schnurr}, {Scholz}, {Schwope}, {Seifert}, {Shanks},
  {Sheinis}, {Sivov}, {Sk{\'u}lad{\'o}ttir}, {Smartt}, {Smedley}, {Smith},
  {Smith}, {Sorce}, {Spitler}, {Starkenburg}, {Steinmetz}, {Stilz}, {Storm},
  {Sullivan}, {Sutherland}, {Swann}, {Tamone}, {Taylor}, {Teillon}, {Tempel},
  {ter Horst}, {Thi}, {Tolstoy}, {Trager}, {Traven}, {Tremblay}, {Tresse},
  {Valentini}, {van de Weygaert}, {van den Ancker}, {Veljanoski}, {Venkatesan},
  {Wagner}, {Wagner}, {Walcher}, {Waller}, {Walton}, {Wang}, {Winkler},
  {Wisotzki}, {Worley}, {Worseck}, {Xiang}, {Xu}, {Yong}, {Zhao}, {Zheng},
  {Zscheyge}, \& {Zucker}}]{deJong2019}
{de Jong}, R.~S., {Agertz}, O., {Berbel}, A.~A., {et~al.} 2019, The Messenger,
  175, 3

\bibitem[{{de Jong} {et~al.}(2016){de Jong}, {Barden}, {Bellido-Tirado},
  {Brynnel}, {Frey}, {Giannone}, {Haynes}, {Johl}, {Phillips}, {Schnurr},
  {Walcher}, {Winkler}, {Ansorge}, {Feltzing}, {McMahon}, {Baker}, {Caillier},
  {Dwelly}, {Gaessler}, {Iwert}, {Mandel}, {Piskunov}, {Pragt}, {Walton},
  {Bensby}, {Bergemann}, {Chiappini}, {Christlieb}, {Cioni}, {Driver},
  {Finoguenov}, {Helmi}, {Irwin}, {Kitaura}, {Kneib}, {Liske}, {Merloni},
  {Minchev}, {Richard}, \& {Starkenburg}}]{deJong2016}
{de Jong}, R.~S., {Barden}, S.~C., {Bellido-Tirado}, O., {et~al.} 2016, in
  Society of Photo-Optical Instrumentation Engineers (SPIE) Conference Series,
  Vol. 9908, Ground-based and Airborne Instrumentation for Astronomy VI, ed.
  C.~J. {Evans}, L.~{Simard}, \& H.~{Takami}, 99081O

\bibitem[{{de Jong} {et~al.}(2022){de Jong}, {Bellido-Tirado}, {Brynnel},
  {Ezzati Amini}, {Frey}, {F{\"u}{\ss}lein}, {G{\"a}bler}, {Giannone}, {Johl},
  {Kuba}, {Lemke}, {Micheva}, {Saviauk}, {Steinmetz}, {Walcher}, {Winkler},
  {Lind}, {Loveday}, {Feltzing}, {McMahon}, {Mainieri}, {Pirard}, {Bensby},
  {Bergemann}, {Chiappini}, {Christlieb}, {Cioni}, {Comparat}, {Driver},
  {Hook}, {Irwin}, {Kneib}, {Liske}, {Merloni}, {Minchev}, {Richard},
  {Starkenburg}, {Sullivan}, {Worley}, {Gaessler}, {Laurent}, {Pragt},
  {Remillieux}, {Rothmaier}, {Smedley}, {Stilz}, {Walton}, {Alexander},
  {Church}, {Croom}, {Davies}, {Heneka}, {Kacharov}, {Knoche}, {Kordopatis},
  {Krumpe}, {Martell}, {Norberg}, {Pelisoli}, {Sharma}, {Storm}, \&
  {Tempel}}]{deJong2022}
{de Jong}, R.~S., {Bellido-Tirado}, O., {Brynnel}, J.~G., {et~al.} 2022, in
  Society of Photo-Optical Instrumentation Engineers (SPIE) Conference Series,
  Vol. 12184, Ground-based and Airborne Instrumentation for Astronomy IX, ed.
  C.~J. {Evans}, J.~J. {Bryant}, \& K.~{Motohara}, 1218414

\bibitem[{{de Vos} {et~al.}(2024){de Vos}, {Merrifield}, \&
  {Hatch}}]{deVos2024}
{de Vos}, K., {Merrifield}, M.~R., \& {Hatch}, N.~A. 2024, \mnras, 531, 4383

\bibitem[{{Dey} {et~al.}(2019){Dey}, {Schlegel}, {Lang}, {Blum}, {Burleigh},
  {Fan}, {Findlay}, {Finkbeiner}, {Herrera}, {Juneau}, {Landriau}, {Levi},
  {McGreer}, {Meisner}, {Myers}, {Moustakas}, {Nugent}, {Patej}, {Schlafly},
  {Walker}, {Valdes}, {Weaver}, {Y{\`e}che}, {Zou}, {Zhou}, {Abareshi},
  {Abbott}, {Abolfathi}, {Aguilera}, {Alam}, {Allen}, {Alvarez}, {Annis},
  {Ansarinejad}, {Aubert}, {Beechert}, {Bell}, {BenZvi}, {Beutler}, {Bielby},
  {Bolton}, {Brice{\~n}o}, {Buckley-Geer}, {Butler}, {Calamida}, {Carlberg},
  {Carter}, {Casas}, {Castander}, {Choi}, {Comparat}, {Cukanovaite}, {Delubac},
  {DeVries}, {Dey}, {Dhungana}, {Dickinson}, {Ding}, {Donaldson}, {Duan},
  {Duckworth}, {Eftekharzadeh}, {Eisenstein}, {Etourneau}, {Fagrelius},
  {Farihi}, {Fitzpatrick}, {Font-Ribera}, {Fulmer}, {G{\"a}nsicke},
  {Gaztanaga}, {George}, {Gerdes}, {Gontcho}, {Gorgoni}, {Green}, {Guy},
  {Harmer}, {Hernandez}, {Honscheid}, {Huang}, {James}, {Jannuzi}, {Jiang},
  {Joyce}, {Karcher}, {Karkar}, {Kehoe}, {Kneib}, {Kueter-Young}, {Lan},
  {Lauer}, {Le Guillou}, {Le Van Suu}, {Lee}, {Lesser}, {Perreault Levasseur},
  {Li}, {Mann}, {Marshall}, {Mart{\'\i}nez-V{\'a}zquez}, {Martini}, {du Mas des
  Bourboux}, {McManus}, {Meier}, {M{\'e}nard}, {Metcalfe},
  {Mu{\~n}oz-Guti{\'e}rrez}, {Najita}, {Napier}, {Narayan}, {Newman}, {Nie},
  {Nord}, {Norman}, {Olsen}, {Paat}, {Palanque-Delabrouille}, {Peng},
  {Poppett}, {Poremba}, {Prakash}, {Rabinowitz}, {Raichoor}, {Rezaie},
  {Robertson}, {Roe}, {Ross}, {Ross}, {Rudnick}, {Safonova}, {Saha},
  {S{\'a}nchez}, {Savary}, {Schweiker}, {Scott}, {Seo}, {Shan}, {Silva},
  {Slepian}, {Soto}, {Sprayberry}, {Staten}, {Stillman}, {Stupak}, {Summers},
  {Sien Tie}, {Tirado}, {Vargas-Maga{\~n}a}, {Vivas}, {Wechsler}, {Williams},
  {Yang}, {Yang}, {Yapici}, {Zaritsky}, {Zenteno}, {Zhang}, {Zhang}, {Zhou}, \&
  {Zhou}}]{Dey2019}
{Dey}, A., {Schlegel}, D.~J., {Lang}, D., {et~al.} 2019, \aj, 157, 168

\bibitem[{{Diemer}(2018)}]{Diemer2018}
{Diemer}, B. 2018, \apjs, 239, 35

\bibitem[{{Dietl} {et~al.}(2024){Dietl}, {Pacaud}, {Reiprich}, {Veronica},
  {Migkas}, {Spinelli}, {Dolag}, \& {Seidel}}]{Dietl2024}
{Dietl}, J., {Pacaud}, F., {Reiprich}, T.~H., {et~al.} 2024, \aap, 691, A286

\bibitem[{{Donnan} {et~al.}(2022){Donnan}, {Tojeiro}, \&
  {Kraljic}}]{Donnan2022}
{Donnan}, C.~T., {Tojeiro}, R., \& {Kraljic}, K. 2022, Nature Astronomy, 6, 599

\bibitem[{{D{\"u}nner} {et~al.}(2006){D{\"u}nner}, {Araya}, {Meza}, \&
  {Reisenegger}}]{Dunner2006}
{D{\"u}nner}, R., {Araya}, P.~A., {Meza}, A., \& {Reisenegger}, A. 2006,
  \mnras, 366, 803

\bibitem[{{Dupuy} \& {Courtois}(2023)}]{Dupuy2023}
{Dupuy}, A. \& {Courtois}, H.~M. 2023, \aap, 678, A176

\bibitem[{{Ebeling} {et~al.}(2014){Ebeling}, {Stephenson}, \&
  {Edge}}]{Ebeling2014}
{Ebeling}, H., {Stephenson}, L.~N., \& {Edge}, A.~C. 2014, \apjl, 781, L40

\bibitem[{{Einasto} {et~al.}(2018){Einasto}, {Gramann}, {Park}, {Kim},
  {Deshev}, {Tempel}, {Hein{\"a}m{\"a}ki}, {Lietzen}, {L{\"a}hteenm{\"a}ki},
  {Einasto}, \& {Saar}}]{Einasto2018}
{Einasto}, M., {Gramann}, M., {Park}, C., {et~al.} 2018, \aap, 620, A149

\bibitem[{{Fasano} {et~al.}(2006){Fasano}, {Marmo}, {Varela}, {D'Onofrio},
  {Poggianti}, {Moles}, {Pignatelli}, {Bettoni}, {Kj{\ae}rgaard}, {Rizzi},
  {Couch}, \& {Dressler}}]{Fasano2006}
{Fasano}, G., {Marmo}, C., {Varela}, J., {et~al.} 2006, \aap, 445, 805

\bibitem[{{Finoguenov} {et~al.}(2019){Finoguenov}, {Merloni}, {Comparat},
  {Nandra}, {Salvato}, {Tempel}, {Raichoor}, {Richard}, {Kneib}, {Pillepich},
  {Sahl{\'e}n}, {Popesso}, {Norberg}, {McMahon}, \& {4MOST
  Collaboration}}]{Finoguenov2019}
{Finoguenov}, A., {Merloni}, A., {Comparat}, J., {et~al.} 2019, The Messenger,
  175, 39

\bibitem[{{Fleenor} {et~al.}(2005){Fleenor}, {Rose}, {Christiansen},
  {Hunstead}, {Johnston-Hollitt}, {Drinkwater}, \& {Saunders}}]{Fleenor2005}
{Fleenor}, M.~C., {Rose}, J.~A., {Christiansen}, W.~A., {et~al.} 2005, \aj,
  130, 957

\bibitem[{{Fleenor} {et~al.}(2006){Fleenor}, {Rose}, {Christiansen},
  {Johnston-Hollitt}, {Hunstead}, {Drinkwater}, \& {Saunders}}]{Fleenor2006}
{Fleenor}, M.~C., {Rose}, J.~A., {Christiansen}, W.~A., {et~al.} 2006, \aj,
  131, 1280

\bibitem[{{Fujita}(2004)}]{Fujita2004}
{Fujita}, Y. 2004, \pasj, 56, 29

\bibitem[{{Girardi} {et~al.}(2024){Girardi}, {Boschin}, {Mercurio}, {Nocerino},
  {Nonino}, {Rosati}, {Biviano}, {Demarco}, {Grillo}, {Sartoris}, {Tozzi}, \&
  {Vanzella}}]{Girardi2024}
{Girardi}, M., {Boschin}, W., {Mercurio}, A., {et~al.} 2024, \aap, 692, A175

\bibitem[{{Gnedin}(2003)}]{Gnedin2003}
{Gnedin}, O.~Y. 2003, \apj, 582, 141

\bibitem[{{Gullieuszik} {et~al.}(2015){Gullieuszik}, {Poggianti}, {Fasano},
  {Zaggia}, {Paccagnella}, {Moretti}, {Bettoni}, {D'Onofrio}, {Couch},
  {Vulcani}, {Fritz}, {Omizzolo}, {Baruffolo}, {Schipani}, {Capaccioli}, \&
  {Varela}}]{Gullieuszik2015}
{Gullieuszik}, M., {Poggianti}, B., {Fasano}, G., {et~al.} 2015, \aap, 581, A41

\bibitem[{{Gunn} \& {Gott}(1972)}]{Gunn1972}
{Gunn}, J.~E. \& {Gott}, J.~Richard, I. 1972, \apj, 176, 1

\bibitem[{{Haines} {et~al.}(2023){Haines}, {Jaff{\'e}}, {Tejos}, {Monachesi},
  {Pompei}, {Finoguenov}, {Sif{\'o}n}, {Lopez}, {Manjunatha}, {Bilton},
  {Comparat}, {Cuellar}, {D'Ago}, {Demarco}, {Lima-Dias}, {L{\"o}sch},
  {Merluzzi}, {Smith Castelli}, {Sodre}, {Vinicius}, \& {CHANCES
  Team}}]{Haines2023}
{Haines}, C., {Jaff{\'e}}, Y., {Tejos}, N., {et~al.} 2023, The Messenger, 190,
  31

\bibitem[{{Haines} {et~al.}(2018){Haines}, {Busarello}, {Merluzzi}, {Pimbblet},
  {Vogt}, {Dopita}, {Mercurio}, {Grado}, \& {Limatola}}]{Haines2018}
{Haines}, C.~P., {Busarello}, G., {Merluzzi}, P., {et~al.} 2018, \mnras, 481,
  1055

\bibitem[{{Haines} {et~al.}(2015){Haines}, {Pereira}, {Smith}, {Egami},
  {Babul}, {Finoguenov}, {Ziparo}, {McGee}, {Rawle}, {Okabe}, \&
  {Moran}}]{Haines2015}
{Haines}, C.~P., {Pereira}, M.~J., {Smith}, G.~P., {et~al.} 2015, \apj, 806,
  101

\bibitem[{{Haines} {et~al.}(2013){Haines}, {Pereira}, {Smith}, {Egami},
  {Sanderson}, {Babul}, {Finoguenov}, {Merluzzi}, {Busarello}, {Rawle}, \&
  {Okabe}}]{Haines2013}
{Haines}, C.~P., {Pereira}, M.~J., {Smith}, G.~P., {et~al.} 2013, \apj, 775,
  126

\bibitem[{{Harris} {et~al.}(2020){Harris}, {Millman}, {van der Walt},
  {Gommers}, {Virtanen}, {Cournapeau}, {Wieser}, {Taylor}, {Berg}, {Smith},
  {Kern}, {Picus}, {Hoyer}, {van Kerkwijk}, {Brett}, {Haldane}, {del R{\'\i}o},
  {Wiebe}, {Peterson}, {G{\'e}rard-Marchant}, {Sheppard}, {Reddy}, {Weckesser},
  {Abbasi}, {Gohlke}, \& {Oliphant}}]{Harris2020}
{Harris}, C.~R., {Millman}, K.~J., {van der Walt}, S.~J., {et~al.} 2020, \nat,
  585, 357

\bibitem[{{Hasan} {et~al.}(2023){Hasan}, {Burchett}, {Abeyta}, {Hellinger},
  {Mandelker}, {Primack}, {Faber}, {Koo}, {Elek}, \& {Nagai}}]{Hasan2023}
{Hasan}, F., {Burchett}, J.~N., {Abeyta}, A., {et~al.} 2023, \apj, 950, 114

\bibitem[{{Herbonnet} {et~al.}(2020){Herbonnet}, {Sif{\'o}n}, {Hoekstra},
  {Bah{\'e}}, {van der Burg}, {Melin}, {von der Linden}, {Sand}, {Kay}, \&
  {Barnes}}]{Herbonnet2020}
{Herbonnet}, R., {Sif{\'o}n}, C., {Hoekstra}, H., {et~al.} 2020, \mnras, 497,
  4684

\bibitem[{{Hilton} {et~al.}(2021){Hilton}, {Sif{\'o}n}, {Naess},
  {Madhavacheril}, {Oguri}, {Rozo}, {Rykoff}, {Abbott}, {Adhikari}, {Aguena},
  {Aiola}, {Allam}, {Amodeo}, {Amon}, {Annis}, {Ansarinejad}, {Aros-Bunster},
  {Austermann}, {Avila}, {Bacon}, {Battaglia}, {Beall}, {Becker}, {Bernstein},
  {Bertin}, {Bhandarkar}, {Bhargava}, {Bond}, {Brooks}, {Burke}, {Calabrese},
  {Carrasco Kind}, {Carretero}, {Choi}, {Choi}, {Conselice}, {da Costa},
  {Costanzi}, {Crichton}, {Crowley}, {D{\"u}nner}, {Denison}, {Devlin},
  {Dicker}, {Diehl}, {Dietrich}, {Doel}, {Duff}, {Duivenvoorden}, {Dunkley},
  {Everett}, {Ferraro}, {Ferrero}, {Fert{\'e}}, {Flaugher}, {Frieman},
  {Gallardo}, {Garc{\'\i}a-Bellido}, {Gaztanaga}, {Gerdes}, {Giles}, {Golec},
  {Gralla}, {Grandis}, {Gruen}, {Gruendl}, {Gschwend}, {Gutierrez}, {Han},
  {Hartley}, {Hasselfield}, {Hill}, {Hilton}, {Hincks}, {Hinton}, {Ho},
  {Honscheid}, {Hoyle}, {Hubmayr}, {Huffenberger}, {Hughes}, {Jaelani}, {Jain},
  {James}, {Jeltema}, {Kent}, {Knowles}, {Koopman}, {Kuehn}, {Lahav}, {Lima},
  {Lin}, {Lokken}, {Loubser}, {MacCrann}, {Maia}, {Marriage}, {Martin},
  {McMahon}, {Melchior}, {Menanteau}, {Miquel}, {Miyatake}, {Moodley},
  {Morgan}, {Mroczkowski}, {Nati}, {Newburgh}, {Niemack}, {Nishizawa},
  {Ogando}, {Orlowski-Scherer}, {Page}, {Palmese}, {Partridge},
  {Paz-Chinch{\'o}n}, {Phakathi}, {Plazas}, {Robertson}, {Romer}, {Carnero
  Rosell}, {Salatino}, {Sanchez}, {Schaan}, {Schillaci}, {Sehgal}, {Serrano},
  {Shin}, {Simon}, {Smith}, {Soares-Santos}, {Spergel}, {Staggs}, {Storer},
  {Suchyta}, {Swanson}, {Tarle}, {Thomas}, {To}, {Trac}, {Ullom}, {Vale}, {Van
  Lanen}, {Vavagiakis}, {De Vicente}, {Wilkinson}, {Wollack}, {Xu}, \&
  {Zhang}}]{Hilton2021}
{Hilton}, M., {Sif{\'o}n}, C., {Naess}, S., {et~al.} 2021, \apjs, 253, 3

\bibitem[{{Hoekstra} {et~al.}(2012){Hoekstra}, {Mahdavi}, {Babul}, \&
  {Bildfell}}]{Hoekstra2012}
{Hoekstra}, H., {Mahdavi}, A., {Babul}, A., \& {Bildfell}, C. 2012, \mnras,
  427, 1298

\bibitem[{{Hou} {et~al.}(2014){Hou}, {Parker}, \& {Harris}}]{Hou2014}
{Hou}, A., {Parker}, L.~C., \& {Harris}, W.~E. 2014, \mnras, 442, 406

\bibitem[{{Hunter}(2007)}]{Hunter2007}
{Hunter}, J.~D. 2007, Computing in Science and Engineering, 9, 90

\bibitem[{{Ishiyama} {et~al.}(2021){Ishiyama}, {Prada}, {Klypin}, {Sinha},
  {Metcalf}, {Jullo}, {Altieri}, {Cora}, {Croton}, {de la Torre},
  {Mill{\'a}n-Calero}, {Oogi}, {Ruedas}, \&
  {Vega-Mart{\'\i}nez}}]{Ishiyama2021}
{Ishiyama}, T., {Prada}, F., {Klypin}, A.~A., {et~al.} 2021, \mnras, 506, 4210

\bibitem[{{Ivezi{\'c}} {et~al.}(2019){Ivezi{\'c}}, {Kahn}, {Tyson}, {Abel},
  {Acosta}, {Allsman}, {Alonso}, {AlSayyad}, {Anderson}, {Andrew}, {Angel},
  {Angeli}, {Ansari}, {Antilogus}, {Araujo}, {Armstrong}, {Arndt}, {Astier},
  {Aubourg}, {Auza}, {Axelrod}, {Bard}, {Barr}, {Barrau}, {Bartlett}, {Bauer},
  {Bauman}, {Baumont}, {Bechtol}, {Bechtol}, {Becker}, {Becla}, {Beldica},
  {Bellavia}, {Bianco}, {Biswas}, {Blanc}, {Blazek}, {Blandford}, {Bloom},
  {Bogart}, {Bond}, {Booth}, {Borgland}, {Borne}, {Bosch}, {Boutigny},
  {Brackett}, {Bradshaw}, {Brandt}, {Brown}, {Bullock}, {Burchat}, {Burke},
  {Cagnoli}, {Calabrese}, {Callahan}, {Callen}, {Carlin}, {Carlson},
  {Chandrasekharan}, {Charles-Emerson}, {Chesley}, {Cheu}, {Chiang}, {Chiang},
  {Chirino}, {Chow}, {Ciardi}, {Claver}, {Cohen-Tanugi}, {Cockrum}, {Coles},
  {Connolly}, {Cook}, {Cooray}, {Covey}, {Cribbs}, {Cui}, {Cutri}, {Daly},
  {Daniel}, {Daruich}, {Daubard}, {Daues}, {Dawson}, {Delgado}, {Dellapenna},
  {de Peyster}, {de Val-Borro}, {Digel}, {Doherty}, {Dubois},
  {Dubois-Felsmann}, {Durech}, {Economou}, {Eifler}, {Eracleous}, {Emmons},
  {Fausti Neto}, {Ferguson}, {Figueroa}, {Fisher-Levine}, {Focke}, {Foss},
  {Frank}, {Freemon}, {Gangler}, {Gawiser}, {Geary}, {Gee}, {Geha}, {Gessner},
  {Gibson}, {Gilmore}, {Glanzman}, {Glick}, {Goldina}, {Goldstein}, {Goodenow},
  {Graham}, {Gressler}, {Gris}, {Guy}, {Guyonnet}, {Haller}, {Harris},
  {Hascall}, {Haupt}, {Hernandez}, {Herrmann}, {Hileman}, {Hoblitt}, {Hodgson},
  {Hogan}, {Howard}, {Huang}, {Huffer}, {Ingraham}, {Innes}, {Jacoby}, {Jain},
  {Jammes}, {Jee}, {Jenness}, {Jernigan}, {Jevremovi{\'c}}, {Johns}, {Johnson},
  {Johnson}, {Jones}, {Juramy-Gilles}, {Juri{\'c}}, {Kalirai}, {Kallivayalil},
  {Kalmbach}, {Kantor}, {Karst}, {Kasliwal}, {Kelly}, {Kessler}, {Kinnison},
  {Kirkby}, {Knox}, {Kotov}, {Krabbendam}, {Krughoff}, {Kub{\'a}nek},
  {Kuczewski}, {Kulkarni}, {Ku}, {Kurita}, {Lage}, {Lambert}, {Lange},
  {Langton}, {Le Guillou}, {Levine}, {Liang}, {Lim}, {Lintott}, {Long},
  {Lopez}, {Lotz}, {Lupton}, {Lust}, {MacArthur}, {Mahabal}, {Mandelbaum},
  {Markiewicz}, {Marsh}, {Marshall}, {Marshall}, {May}, {McKercher}, {McQueen},
  {Meyers}, {Migliore}, {Miller}, {Mills}, {Miraval}, {Moeyens}, {Moolekamp},
  {Monet}, {Moniez}, {Monkewitz}, {Montgomery}, {Morrison}, {Mueller},
  {Muller}, {Mu{\~n}oz Arancibia}, {Neill}, {Newbry}, {Nief}, {Nomerotski},
  {Nordby}, {O'Connor}, {Oliver}, {Olivier}, {Olsen}, {O'Mullane}, {Ortiz},
  {Osier}, {Owen}, {Pain}, {Palecek}, {Parejko}, {Parsons}, {Pease},
  {Peterson}, {Peterson}, {Petravick}, {Libby Petrick}, {Petry},
  {Pierfederici}, {Pietrowicz}, {Pike}, {Pinto}, {Plante}, {Plate}, {Plutchak},
  {Price}, {Prouza}, {Radeka}, {Rajagopal}, {Rasmussen}, {Regnault}, {Reil},
  {Reiss}, {Reuter}, {Ridgway}, {Riot}, {Ritz}, {Robinson}, {Roby}, {Roodman},
  {Rosing}, {Roucelle}, {Rumore}, {Russo}, {Saha}, {Sassolas}, {Schalk},
  {Schellart}, {Schindler}, {Schmidt}, {Schneider}, {Schneider}, {Schoening},
  {Schumacher}, {Schwamb}, {Sebag}, {Selvy}, {Sembroski}, {Seppala}, {Serio},
  {Serrano}, {Shaw}, {Shipsey}, {Sick}, {Silvestri}, {Slater}, {Smith},
  {Smith}, {Sobhani}, {Soldahl}, {Storrie-Lombardi}, {Stover}, {Strauss},
  {Street}, {Stubbs}, {Sullivan}, {Sweeney}, {Swinbank}, {Szalay}, {Takacs},
  {Tether}, {Thaler}, {Thayer}, {Thomas}, {Thornton}, {Thukral}, {Tice},
  {Trilling}, {Turri}, {Van Berg}, {Vanden Berk}, {Vetter}, {Virieux},
  {Vucina}, {Wahl}, {Walkowicz}, {Walsh}, {Walter}, {Wang}, {Wang}, {Warner},
  {Wiecha}, {Willman}, {Winters}, {Wittman}, {Wolff}, {Wood-Vasey}, {Wu},
  {Xin}, {Yoachim}, \& {Zhan}}]{Ivezic2019}
{Ivezi{\'c}}, {\v{Z}}., {Kahn}, S.~M., {Tyson}, J.~A., {et~al.} 2019, \apj,
  873, 111

\bibitem[{{Jaff{\'e}} {et~al.}(2015){Jaff{\'e}}, {Smith}, {Candlish},
  {Poggianti}, {Sheen}, \& {Verheijen}}]{Jaffe2015}
{Jaff{\'e}}, Y.~L., {Smith}, R., {Candlish}, G.~N., {et~al.} 2015, \mnras, 448,
  1715

\bibitem[{{Jaff{\'e}} {et~al.}(2016){Jaff{\'e}}, {Verheijen}, {Haines}, {Yoon},
  {Cybulski}, {Montero-Casta{\~n}o}, {Smith}, {Chung}, {Deshev},
  {Fern{\'a}ndez}, {van Gorkom}, {Poggianti}, {Yun}, {Finoguenov}, {Smith}, \&
  {Okabe}}]{Jaffe2016}
{Jaff{\'e}}, Y.~L., {Verheijen}, M. A.~W., {Haines}, C.~P., {et~al.} 2016,
  \mnras, 461, 1202

\bibitem[{{Jin} {et~al.}(2024){Jin}, {Trager}, {Dalton}, {Aguerri}, {Drew},
  {Falc{\'o}n-Barroso}, {G{\"a}nsicke}, {Hill}, {Iovino}, {Pieri}, {Poggianti},
  {Smith}, {Vallenari}, {Abrams}, {Aguado}, {Antoja}, {Arag{\'o}n-Salamanca},
  {Ascasibar}, {Babusiaux}, {Balcells}, {Barrena}, {Battaglia}, {Belokurov},
  {Bensby}, {Bonifacio}, {Bragaglia}, {Carrasco}, {Carrera}, {Cornwell},
  {Dom{\'\i}nguez-Palmero}, {Duncan}, {Famaey}, {Fari{\~n}a}, {Gonzalez},
  {Guest}, {Hatch}, {Hess}, {Hoskin}, {Irwin}, {Knapen}, {Koposov}, {Kuchner},
  {Laigle}, {Lewis}, {Longhetti}, {Lucatello}, {M{\'e}ndez-Abreu}, {Mercurio},
  {Molaeinezhad}, {Mongui{\'o}}, {Morrison}, {Murphy}, {Peralta de Arriba},
  {P{\'e}rez}, {P{\'e}rez-R{\`a}fols}, {Pic{\'o}}, {Raddi}, {Romero-G{\'o}mez},
  {Royer}, {Siebert}, {Seabroke}, {Som}, {Terrett}, {Thomas}, {Wesson},
  {Worley}, {Alfaro}, {Allende Prieto}, {Alonso-Santiago}, {Amos}, {Ashley},
  {Balaguer-N{\'u}{\~n}ez}, {Balbinot}, {Bellazzini}, {Benn}, {Berlanas},
  {Bernard}, {Best}, {Bettoni}, {Bianco}, {Bishop}, {Blomqvist}, {Boeche},
  {Bolzonella}, {Bonoli}, {Bosma}, {Britavskiy}, {Busarello}, {Caffau},
  {Cantat-Gaudin}, {Castro-Ginard}, {Couto}, {Carbajo-Hijarrubia}, {Carter},
  {Casamiquela}, {Conrado}, {Corcho-Caballero}, {Costantin}, {Deason}, {de
  Burgos}, {De Grandi}, {Di Matteo}, {Dom{\'\i}nguez-G{\'o}mez}, {Dorda},
  {Drake}, {Dutta}, {Erkal}, {Feltzing}, {Ferr{\'e}-Mateu}, {Feuillet},
  {Figueras}, {Fossati}, {Franciosini}, {Frasca}, {Fumagalli}, {Gallazzi},
  {Garc{\'\i}a-Benito}, {Gentile Fusillo}, {Gebran}, {Gilbert}, {Gledhill},
  {Gonz{\'a}lez Delgado}, {Greimel}, {Guarcello}, {Guerra}, {Gullieuszik},
  {Haines}, {Hardcastle}, {Harris}, {Haywood}, {Helmi}, {Hernandez}, {Herrero},
  {Hughes}, {Ir{\v{s}}i{\v{c}}}, {Jablonka}, {Jarvis}, {Jordi}, {Kondapally},
  {Kordopatis}, {Krogager}, {La Barbera}, {Lam}, {Larsen}, {Lemasle}, {Lewis},
  {Lhom{\'e}}, {Lind}, {Lodi}, {Longobardi}, {Lonoce}, {Magrini}, {Ma{\'\i}z
  Apell{\'a}niz}, {Marchal}, {Marco}, {Martin}, {Matsuno}, {Maurogordato},
  {Merluzzi}, {Miralda-Escud{\'e}}, {Molinari}, {Monari}, {Morelli}, {Mottram},
  {Naylor}, {Negueruela}, {O{\~n}orbe}, {Pancino}, {Peirani}, {Peletier},
  {Pozzetti}, {Rainer}, {Ramos}, {Read}, {Rossi}, {R{\"o}ttgering},
  {Rubi{\~n}o-Mart{\'\i}n}, {Sabater}, {San Juan}, {Sanna}, {Schallig},
  {Schiavon}, {Schultheis}, {Serra}, {Shimwell}, {Sim{\'o}n-D{\'\i}az},
  {Smith}, {Sordo}, {Sorini}, {Soubiran}, {Starkenburg}, {Steele}, {Stott},
  {Stuik}, {Tolstoy}, {Tortora}, {Tsantaki}, {Van der Swaelmen}, {van Weeren},
  {Vergani}, {Verheijen}, {Verro}, {Vink}, {Vioque}, {Walcher}, {Walton},
  {Wegg}, {Weijmans}, {Williams}, {Wilson}, {Wright}, {Xylakis-Dornbusch},
  {Youakim}, {Zibetti}, \& {Zurita}}]{Jin2024}
{Jin}, S., {Trager}, S.~C., {Dalton}, G.~B., {et~al.} 2024, \mnras, 530, 2688

\bibitem[{{Kesebonye} {et~al.}(2023){Kesebonye}, {Hilton}, {Knowles}, {Cotton},
  {Clarke}, {Loubser}, {Moodley}, \& {Sikhosana}}]{Kesebonye2023}
{Kesebonye}, K.~C., {Hilton}, M., {Knowles}, K., {et~al.} 2023, \mnras, 518,
  3004

\bibitem[{{Khalil} {et~al.}(2024){Khalil}, {Finoguenov}, {Tempel}, \&
  {Mamon}}]{Khalil2024}
{Khalil}, H., {Finoguenov}, A., {Tempel}, E., \& {Mamon}, G.~A. 2024, \aap,
  690, A212

\bibitem[{{Kluge} {et~al.}(2024){Kluge}, {Comparat}, {Liu}, {Balzer}, {Bulbul},
  {Ider Chitham}, {Ghirardini}, {Garrel}, {Bahar}, {Artis}, {Bender}, {Clerc},
  {Dwelly}, {Fabricius}, {Grandis}, {Hern{\'a}ndez-Lang}, {Hill}, {Joshi},
  {Lamer}, {Merloni}, {Nandra}, {Pacaud}, {Predehl}, {Ramos-Ceja}, {Reiprich},
  {Salvato}, {Sanders}, {Schrabback}, {Seppi}, {Zelmer}, {Zenteno}, \&
  {Zhang}}]{Kluge2024}
{Kluge}, M., {Comparat}, J., {Liu}, A., {et~al.} 2024, \aap, 688, A210

\bibitem[{{Knuth}(2006)}]{Knuth2006}
{Knuth}, K.~H. 2006, arXiv e-prints, physics/0605197

\bibitem[{{Kriek} {et~al.}(2009){Kriek}, {van Dokkum}, {Labb{\'e}}, {Franx},
  {Illingworth}, {Marchesini}, \& {Quadri}}]{Kriek2009}
{Kriek}, M., {van Dokkum}, P.~G., {Labb{\'e}}, I., {et~al.} 2009, \apj, 700,
  221

\bibitem[{{Kuchner} {et~al.}(2022){Kuchner}, {Haggar}, {Arag{\'o}n-Salamanca},
  {Pearce}, {Gray}, {Rost}, {Cui}, {Knebe}, \& {Yepes}}]{Kuchner2022}
{Kuchner}, U., {Haggar}, R., {Arag{\'o}n-Salamanca}, A., {et~al.} 2022, \mnras,
  510, 581

\bibitem[{{Kugel} {et~al.}(2025){Kugel}, {Schaye}, {Schaller}, {Forouhar
  Moreno}, \& {McGibbon}}]{Kugel2025}
{Kugel}, R., {Schaye}, J., {Schaller}, M., {Forouhar Moreno}, V.~J., \&
  {McGibbon}, R.~J. 2025, \mnras, 537, 2179

\bibitem[{{Kulier} {et~al.}(2023){Kulier}, {Poggianti}, {Tonnesen}, {Smith},
  {Ignesti}, {Akerman}, {Marasco}, {Vulcani}, {Moretti}, \&
  {Wolter}}]{Kulier2023}
{Kulier}, A., {Poggianti}, B., {Tonnesen}, S., {et~al.} 2023, \apj, 954, 177

\bibitem[{{Lima} {et~al.}(2022){Lima}, {Sodr{\'e}}, {Bom}, {Teixeira},
  {Nakazono}, {Buzzo}, {Queiroz}, {Herpich}, {Castellon}, {Dantas}, {Dors},
  {Souza}, {Akras}, {Jim{\'e}nez-Teja}, {Kanaan}, {Ribeiro}, \&
  {Schoennell}}]{Lima2022}
{Lima}, E.~V.~R., {Sodr{\'e}}, L., {Bom}, C.~R., {et~al.} 2022, Astronomy and
  Computing, 38, 100510

\bibitem[{{Lopes} {et~al.}(2024){Lopes}, {Ribeiro}, \& {Brambila}}]{Lopes2024}
{Lopes}, P. A.~A., {Ribeiro}, A. L.~B., \& {Brambila}, D. 2024, \mnras, 527,
  L19

\bibitem[{{Lopez} {et~al.}(2008){Lopez}, {Barrientos}, {Lira}, {Padilla},
  {Gilbank}, {Gladders}, {Maza}, {Tejos}, {Vidal}, \& {Yee}}]{Lopez2008}
{Lopez}, S., {Barrientos}, L.~F., {Lira}, P., {et~al.} 2008, \apj, 679, 1144

\bibitem[{{Lucey} {et~al.}(1983){Lucey}, {Dickens}, {Mitchell}, \&
  {Dawe}}]{Lucey1983}
{Lucey}, J.~R., {Dickens}, R.~J., {Mitchell}, R.~J., \& {Dawe}, J.~A. 1983,
  \mnras, 203, 545

\bibitem[{{Lyke} {et~al.}(2020){Lyke}, {Higley}, {McLane}, {Schurhammer},
  {Myers}, {Ross}, {Dawson}, {Chabanier}, {Martini}, {Busca}, {Mas des
  Bourboux}, {Salvato}, {Streblyanska}, {Zarrouk}, {Burtin}, {Anderson},
  {Bautista}, {Bizyaev}, {Brandt}, {Brinkmann}, {Brownstein}, {Comparat},
  {Green}, {de la Macorra}, {Mu{\~n}oz Guti{\'e}rrez}, {Hou}, {Newman},
  {Palanque-Delabrouille}, {P{\^a}ris}, {Percival}, {Petitjean}, {Rich},
  {Rossi}, {Schneider}, {Smith}, {Vivek}, \& {Weaver}}]{Lyke2020}
{Lyke}, B.~W., {Higley}, A.~N., {McLane}, J.~N., {et~al.} 2020, \apjs, 250, 8

\bibitem[{{Mamon} {et~al.}(2004){Mamon}, {Sanchis}, {Salvador-Sol{\'e}}, \&
  {Solanes}}]{Mamon2004}
{Mamon}, G.~A., {Sanchis}, T., {Salvador-Sol{\'e}}, E., \& {Solanes}, J.~M.
  2004, \aap, 414, 445

\bibitem[{{Mart{\'\i}nez} {et~al.}(2016){Mart{\'\i}nez}, {Muriel}, \&
  {Coenda}}]{Martinez2016}
{Mart{\'\i}nez}, H.~J., {Muriel}, H., \& {Coenda}, V. 2016, \mnras, 455, 127

\bibitem[{{McClintock} {et~al.}(2019){McClintock}, {Varga}, {Gruen}, {Rozo},
  {Rykoff}, {Shin}, {Melchior}, {DeRose}, {Seitz}, {Dietrich}, {Sheldon},
  {Zhang}, {von der Linden}, {Jeltema}, {Mantz}, {Romer}, {Allen}, {Becker},
  {Bermeo}, {Bhargava}, {Costanzi}, {Everett}, {Farahi}, {Hamaus}, {Hartley},
  {Hollowood}, {Hoyle}, {Israel}, {Li}, {MacCrann}, {Morris}, {Palmese},
  {Plazas}, {Pollina}, {Rau}, {Simet}, {Soares-Santos}, {Troxel}, {Vergara
  Cervantes}, {Wechsler}, {Zuntz}, {Abbott}, {Abdalla}, {Allam}, {Annis},
  {Avila}, {Bridle}, {Brooks}, {Burke}, {Carnero Rosell}, {Carrasco Kind},
  {Carretero}, {Castander}, {Crocce}, {Cunha}, {D'Andrea}, {da Costa}, {Davis},
  {De Vicente}, {Diehl}, {Doel}, {Drlica-Wagner}, {Evrard}, {Flaugher},
  {Fosalba}, {Frieman}, {Garc{\'\i}a-Bellido}, {Gaztanaga}, {Gerdes},
  {Giannantonio}, {Gruendl}, {Gutierrez}, {Honscheid}, {James}, {Kirk},
  {Krause}, {Kuehn}, {Lahav}, {Li}, {Lima}, {March}, {Marshall}, {Menanteau},
  {Miquel}, {Mohr}, {Nord}, {Ogando}, {Roodman}, {Sanchez}, {Scarpine},
  {Schindler}, {Sevilla-Noarbe}, {Smith}, {Smith}, {Sobreira}, {Suchyta},
  {Swanson}, {Tarle}, {Tucker}, {Vikram}, {Walker}, {Weller}, \& {DES
  Collaboration}}]{McClintock2019}
{McClintock}, T., {Varga}, T.~N., {Gruen}, D., {et~al.} 2019, \mnras, 482, 1352

\bibitem[{{McGee} {et~al.}(2009){McGee}, {Balogh}, {Bower}, {Font}, \&
  {McCarthy}}]{McGee2009}
{McGee}, S.~L., {Balogh}, M.~L., {Bower}, R.~G., {Font}, A.~S., \& {McCarthy},
  I.~G. 2009, \mnras, 400, 937

\bibitem[{{Mendes de Oliveira} {et~al.}(2019){Mendes de Oliveira}, {Ribeiro},
  {Schoenell}, {Kanaan}, {Overzier}, {Molino}, {Sampedro}, {Coelho}, {Barbosa},
  {Cortesi}, {Costa-Duarte}, {Herpich}, {Hernandez-Jimenez}, {Placco},
  {Xavier}, {Abramo}, {Saito}, {Chies-Santos}, {Ederoclite}, {Lopes de
  Oliveira}, {Gon{\c{c}}alves}, {Akras}, {Almeida}, {Almeida-Fernandes},
  {Beers}, {Bonatto}, {Bonoli}, {Cypriano}, {Vinicius-Lima}, {de Souza},
  {Fabiano de Souza}, {Ferrari}, {Gon{\c{c}}alves}, {Gonzalez},
  {Guti{\'e}rrez-Soto}, {Hartmann}, {Jaffe}, {Kerber}, {Lima-Dias}, {Lopes},
  {Menendez-Delmestre}, {Nakazono}, {Novais}, {Ortega-Minakata}, {Pereira},
  {Perottoni}, {Queiroz}, {Reis}, {Santos}, {Santos-Silva}, {Santucci},
  {Barbosa}, {Siffert}, {Sodr{\'e}}, {Torres-Flores}, {Westera}, {Whitten},
  {Alcaniz}, {Alonso-Garc{\'\i}a}, {Alencar}, {Alvarez-Candal}, {Amram},
  {Azanha}, {Barb{\'a}}, {Bernardinelli}, {Borges Fernandes}, {Branco},
  {Brito-Silva}, {Buzzo}, {Caffer}, {Campillay}, {Cano}, {Carvano}, {Castejon},
  {Cid Fernandes}, {Dantas}, {Daflon}, {Damke}, {de la Reza}, {de Melo de
  Azevedo}, {De Paula}, {Diem}, {Donnerstein}, {Dors}, {Dupke}, {Eikenberry},
  {Escudero}, {Faifer}, {Far{\'\i}as}, {Fernandes}, {Fernandes}, {Fontes},
  {Galarza}, {Hirata}, {Katena}, {Gregorio-Hetem},
  {Hern{\'a}ndez-Fern{\'a}ndez}, {Izzo}, {Jaque Arancibia}, {Jatenco-Pereira},
  {Jim{\'e}nez-Teja}, {Kann}, {Krabbe}, {Labayru}, {Lazzaro}, {Lima Neto},
  {Lopes}, {Magalh{\~a}es}, {Makler}, {de Menezes}, {Miralda-Escud{\'e}},
  {Monteiro-Oliveira}, {Montero-Dorta}, {Mu{\~n}oz-Elgueta}, {Nemmen}, {Nilo
  Castell{\'o}n}, {Oliveira}, {Ort{\'\i}z}, {Pattaro}, {Pereira}, {Quint},
  {Riguccini}, {Rocha Pinto}, {Rodrigues}, {Roig}, {Rossi}, {Saha}, {Santos},
  {Schnorr M{\"u}ller}, {Sesto}, {Silva}, {Smith Castelli}, {Teixeira},
  {Telles}, {Thom de Souza}, {Th{\"o}ne}, {Trevisan}, {de Ugarte Postigo},
  {Urrutia-Viscarra}, {Veiga}, {Vika}, {Vitorelli}, {Werle}, {Werner}, \&
  {Zaritsky}}]{MendesdeOliveira2019}
{Mendes de Oliveira}, C., {Ribeiro}, T., {Schoenell}, W., {et~al.} 2019,
  \mnras, 489, 241

\bibitem[{{Merloni} {et~al.}(2019){Merloni}, {Alexander}, {Banerji}, {Boller},
  {Comparat}, {Dwelly}, {Fotopoulou}, {McMahon}, {Nandra}, {Salvato}, {Croom},
  {Finoguenov}, {Krumpe}, {Lamer}, {Rosario}, {Schwope}, {Shanks}, {Steinmetz},
  {Wisotzki}, \& {Worseck}}]{Merloni2019}
{Merloni}, A., {Alexander}, D.~A., {Banerji}, M., {et~al.} 2019, The Messenger,
  175, 42

\bibitem[{{Merloni} {et~al.}(2024){Merloni}, {Lamer}, {Liu}, {Ramos-Ceja},
  {Brunner}, {Bulbul}, {Dennerl}, {Doroshenko}, {Freyberg}, {Friedrich},
  {Gatuzz}, {Georgakakis}, {Haberl}, {Igo}, {Kreykenbohm}, {Liu}, {Maitra},
  {Malyali}, {Mayer}, {Nandra}, {Predehl}, {Robrade}, {Salvato}, {Sanders},
  {Stewart}, {Tub{\'\i}n-Arenas}, {Weber}, {Wilms}, {Arcodia}, {Artis},
  {Aschersleben}, {Avakyan}, {Aydar}, {Bahar}, {Balzer}, {Becker}, {Berger},
  {Boller}, {Bornemann}, {Br{\"u}ggen}, {Brusa}, {Buchner}, {Burwitz},
  {Camilloni}, {Clerc}, {Comparat}, {Coutinho}, {Czesla}, {Dannhauer},
  {Dauner}, {Dauser}, {Dietl}, {Dolag}, {Dwelly}, {Egg}, {Ehl}, {Freund},
  {Friedrich}, {Gaida}, {Garrel}, {Ghirardini}, {Gokus}, {Gr{\"u}nwald},
  {Grandis}, {Grotova}, {Gruen}, {Gueguen}, {H{\"a}mmerich}, {Hamaus},
  {Hasinger}, {Haubner}, {Homan}, {Ider Chitham}, {Joseph}, {Joyce},
  {K{\"o}nig}, {Kaltenbrunner}, {Khokhriakova}, {Kink}, {Kirsch}, {Kluge},
  {Knies}, {Krippendorf}, {Krumpe}, {Kurpas}, {Li}, {Liu}, {Locatelli},
  {Lorenz}, {M{\"u}ller}, {Magaudda}, {Mannes}, {McCall}, {Meidinger},
  {Michailidis}, {Migkas}, {Mu{\~n}oz-Giraldo}, {Musiimenta}, {Nguyen-Dang},
  {Ni}, {Olechowska}, {Ota}, {Pacaud}, {Pasini}, {Perinati}, {Pires},
  {Pommranz}, {Ponti}, {Poppenhaeger}, {P{\"u}hlhofer}, {Rau}, {Reh},
  {Reiprich}, {Roster}, {Saeedi}, {Santangelo}, {Sasaki}, {Schmitt},
  {Schneider}, {Schrabback}, {Schuster}, {Schwope}, {Seppi}, {Serim},
  {Shreeram}, {Sokolova-Lapa}, {Starck}, {Stelzer}, {Stierhof}, {Suleimanov},
  {Tenzer}, {Traulsen}, {Tr{\"u}mper}, {Tsuge}, {Urrutia}, {Veronica},
  {Waddell}, {Willer}, {Wolf}, {Yeung}, {Zainab}, {Zangrandi}, {Zhang},
  {Zhang}, \& {Zheng}}]{Merloni2024}
{Merloni}, A., {Lamer}, G., {Liu}, T., {et~al.} 2024, \aap, 682, A34

\bibitem[{{Merluzzi} {et~al.}(2015){Merluzzi}, {Busarello}, {Haines},
  {Mercurio}, {Okabe}, {Pimbblet}, {Dopita}, {Grado}, {Limatola}, {Bourdin},
  {Mazzotta}, {Capaccioli}, {Napolitano}, \& {Schipani}}]{Merluzzi2015}
{Merluzzi}, P., {Busarello}, G., {Haines}, C.~P., {et~al.} 2015, \mnras, 446,
  803

\bibitem[{{Miller} {et~al.}(2016){Miller}, {Stark}, {Gifford}, \&
  {Kern}}]{Miller2016}
{Miller}, C.~J., {Stark}, A., {Gifford}, D., \& {Kern}, N. 2016, \apj, 822, 41

\bibitem[{{Montaguth} {et~al.}(2024){Montaguth}, {Monachesi}, {Torres-Flores},
  {G{\'o}mez}, {Lima-Dias}, {Cortesi}, {Mendes de Oliveira}, {Telles}, {Panda},
  {Grossi}, {Lopes}, {O'Mill}, {Hernandez-Jimenez}, {Olave-Rojas}, {Demarco},
  {Kanaan}, {Ribeiro}, \& {Schoenell}}]{Montaguth2024}
{Montaguth}, G.~P., {Monachesi}, A., {Torres-Flores}, S., {et~al.} 2024, \aap\
  in press, arXiv:2406.14671

\bibitem[{{Moretti} {et~al.}(2017){Moretti}, {Gullieuszik}, {Poggianti},
  {Paccagnella}, {Couch}, {Vulcani}, {Bettoni}, {Fritz}, {Cava}, {Fasano},
  {D'Onofrio}, \& {Omizzolo}}]{Moretti2017}
{Moretti}, A., {Gullieuszik}, M., {Poggianti}, B., {et~al.} 2017, \aap, 599,
  A81

\bibitem[{{Moretti} {et~al.}(2014){Moretti}, {Poggianti}, {Fasano}, {Bettoni},
  {D'Onofrio}, {Fritz}, {Cava}, {Varela}, {Vulcani}, {Gullieuszik}, {Couch},
  {Omizzolo}, {Valentinuzzi}, {Dressler}, {Moles}, {Kj{\ae}rgaard},
  {Smareglia}, \& {Molinaro}}]{Moretti2014}
{Moretti}, A., {Poggianti}, B.~M., {Fasano}, G., {et~al.} 2014, \aap, 564, A138

\bibitem[{{Munari} {et~al.}(2013){Munari}, {Biviano}, {Borgani}, {Murante}, \&
  {Fabjan}}]{Munari2013}
{Munari}, E., {Biviano}, A., {Borgani}, S., {Murante}, G., \& {Fabjan}, D.
  2013, \mnras, 430, 2638

\bibitem[{{Muzzin} {et~al.}(2014){Muzzin}, {van der Burg}, {McGee}, {Balogh},
  {Franx}, {Hoekstra}, {Hudson}, {Noble}, {Taranu}, {Webb}, {Wilson}, \&
  {Yee}}]{Muzzin2014}
{Muzzin}, A., {van der Burg}, R.~F.~J., {McGee}, S.~L., {et~al.} 2014, \apj,
  796, 65

\bibitem[{{Natarajan} {et~al.}(2002){Natarajan}, {Kneib}, \&
  {Smail}}]{Natarajan2002}
{Natarajan}, P., {Kneib}, J.-P., \& {Smail}, I. 2002, \apjl, 580, L11

\bibitem[{{Navarro} {et~al.}(1996){Navarro}, {Frenk}, \& {White}}]{Navarro1996}
{Navarro}, J.~F., {Frenk}, C.~S., \& {White}, S. D.~M. 1996, \apj, 462, 563

\bibitem[{{Okabe} \& {Smith}(2016)}]{Okabe2016}
{Okabe}, N. \& {Smith}, G.~P. 2016, \mnras, 461, 3794

\bibitem[{{O'Kane} {et~al.}(2024){O'Kane}, {Kuchner}, {Gray}, \&
  {Arag{\'o}n-Salamanca}}]{OKane2024}
{O'Kane}, C.~J., {Kuchner}, U., {Gray}, M.~E., \& {Arag{\'o}n-Salamanca}, A.
  2024, \mnras, 534, 1682

\bibitem[{{Oman} {et~al.}(2013){Oman}, {Hudson}, \& {Behroozi}}]{Oman2013}
{Oman}, K.~A., {Hudson}, M.~J., \& {Behroozi}, P.~S. 2013, \mnras, 431, 2307

\bibitem[{{Pallero} {et~al.}(2022){Pallero}, {G{\'o}mez}, {Padilla},
  {Bah{\'e}}, {Vega-Mart{\'\i}nez}, \& {Torres-Flores}}]{Pallero2022}
{Pallero}, D., {G{\'o}mez}, F.~A., {Padilla}, N.~D., {et~al.} 2022, \mnras,
  511, 3210

\bibitem[{{Piffaretti} {et~al.}(2011){Piffaretti}, {Arnaud}, {Pratt},
  {Pointecouteau}, \& {Melin}}]{Piffaretti2011}
{Piffaretti}, R., {Arnaud}, M., {Pratt}, G.~W., {Pointecouteau}, E., \&
  {Melin}, J.~B. 2011, \aap, 534, A109

\bibitem[{{Piraino-Cerda} {et~al.}(2024){Piraino-Cerda}, {Jaff{\'e}},
  {Louren{\c{c}}o}, {Crossett}, {Salinas}, {Kim}, {Sheen}, {Kelkar}, {Pallero},
  \& {Bravo-Alfaro}}]{PirainoCerda2024}
{Piraino-Cerda}, F., {Jaff{\'e}}, Y.~L., {Louren{\c{c}}o}, A.~C., {et~al.}
  2024, \mnras, 528, 919

\bibitem[{{Pizzardo} {et~al.}(2024){Pizzardo}, {Geller}, {Kenyon}, \&
  {Damjanov}}]{Pizzardo2024}
{Pizzardo}, M., {Geller}, M.~J., {Kenyon}, S.~J., \& {Damjanov}, I. 2024, \aap,
  683, A82

\bibitem[{{Planck Collaboration} {et~al.}(2016{\natexlab{a}}){Planck
  Collaboration}, {Ade}, {Aghanim}, {Arnaud}, {Ashdown}, {Aumont},
  {Baccigalupi}, {Banday}, {Barreiro}, {Barrena}, {Bartlett}, {Bartolo},
  {Battaner}, {Battye}, {Benabed}, {Beno{\^\i}t}, {Benoit-L{\'e}vy}, {Bernard},
  {Bersanelli}, {Bielewicz}, {Bikmaev}, {B{\"o}hringer}, {Bonaldi}, {Bonavera},
  {Bond}, {Borrill}, {Bouchet}, {Bucher}, {Burenin}, {Burigana}, {Butler},
  {Calabrese}, {Cardoso}, {Carvalho}, {Catalano}, {Challinor}, {Chamballu},
  {Chary}, {Chiang}, {Chon}, {Christensen}, {Clements}, {Colombi}, {Colombo},
  {Combet}, {Comis}, {Couchot}, {Coulais}, {Crill}, {Curto}, {Cuttaia},
  {Dahle}, {Danese}, {Davies}, {Davis}, {de Bernardis}, {de Rosa}, {de Zotti},
  {Delabrouille}, {D{\'e}sert}, {Dickinson}, {Diego}, {Dolag}, {Dole},
  {Donzelli}, {Dor{\'e}}, {Douspis}, {Ducout}, {Dupac}, {Efstathiou},
  {Eisenhardt}, {Elsner}, {En{\ss}lin}, {Eriksen}, {Falgarone}, {Fergusson},
  {Feroz}, {Ferragamo}, {Finelli}, {Forni}, {Frailis}, {Fraisse}, {Franceschi},
  {Frejsel}, {Galeotta}, {Galli}, {Ganga}, {G{\'e}nova-Santos}, {Giard},
  {Giraud-H{\'e}raud}, {Gjerl{\o}w}, {Gonz{\'a}lez-Nuevo}, {G{\'o}rski},
  {Grainge}, {Gratton}, {Gregorio}, {Gruppuso}, {Gudmundsson}, {Hansen},
  {Hanson}, {Harrison}, {Hempel}, {Henrot-Versill{\'e}},
  {Hern{\'a}ndez-Monteagudo}, {Herranz}, {Hildebrandt}, {Hivon}, {Hobson},
  {Holmes}, {Hornstrup}, {Hovest}, {Huffenberger}, {Hurier}, {Jaffe}, {Jaffe},
  {Jin}, {Jones}, {Juvela}, {Keih{\"a}nen}, {Keskitalo}, {Khamitov}, {Kisner},
  {Kneissl}, {Knoche}, {Kunz}, {Kurki-Suonio}, {Lagache}, {Lamarre}, {Lasenby},
  {Lattanzi}, {Lawrence}, {Leonardi}, {Lesgourgues}, {Levrier}, {Liguori},
  {Lilje}, {Linden-V{\o}rnle}, {L{\'o}pez-Caniego}, {Lubin},
  {Mac{\'\i}as-P{\'e}rez}, {Maggio}, {Maino}, {Mak}, {Mandolesi}, {Mangilli},
  {Martin}, {Mart{\'\i}nez-Gonz{\'a}lez}, {Masi}, {Matarrese}, {Mazzotta},
  {McGehee}, {Mei}, {Melchiorri}, {Melin}, {Mendes}, {Mennella}, {Migliaccio},
  {Mitra}, {Miville-Desch{\^e}nes}, {Moneti}, {Montier}, {Morgante},
  {Mortlock}, {Moss}, {Munshi}, {Murphy}, {Naselsky}, {Nastasi}, {Nati},
  {Natoli}, {Netterfield}, {N{\o}rgaard-Nielsen}, {Noviello}, {Novikov},
  {Novikov}, {Olamaie}, {Oxborrow}, {Paci}, {Pagano}, {Pajot}, {Paoletti},
  {Pasian}, {Patanchon}, {Pearson}, {Perdereau}, {Perotto}, {Perrott},
  {Perrotta}, {Pettorino}, {Piacentini}, {Piat}, {Pierpaoli}, {Pietrobon},
  {Plaszczynski}, {Pointecouteau}, {Polenta}, {Pratt}, {Pr{\'e}zeau}, {Prunet},
  {Puget}, {Rachen}, {Reach}, {Rebolo}, {Reinecke}, {Remazeilles}, {Renault},
  {Renzi}, {Ristorcelli}, {Rocha}, {Rosset}, {Rossetti}, {Roudier}, {Rozo},
  {Rubi{\~n}o-Mart{\'\i}n}, {Rumsey}, {Rusholme}, {Rykoff}, {Sandri}, {Santos},
  {Saunders}, {Savelainen}, {Savini}, {Schammel}, {Scott}, {Seiffert},
  {Shellard}, {Shimwell}, {Spencer}, {Stanford}, {Stern}, {Stolyarov},
  {Stompor}, {Streblyanska}, {Sudiwala}, {Sunyaev}, {Sutton}, {Suur-Uski},
  {Sygnet}, {Tauber}, {Terenzi}, {Toffolatti}, {Tomasi}, {Tramonte},
  {Tristram}, {Tucci}, {Tuovinen}, {Umana}, {Valenziano}, {Valiviita}, {Van
  Tent}, {Vielva}, {Villa}, {Wade}, {Wandelt}, {Wehus}, {White}, {Wright},
  {Yvon}, {Zacchei}, \& {Zonca}}]{PlanckCollaboration2016}
{Planck Collaboration}, {Ade}, P.~A.~R., {Aghanim}, N., {et~al.}
  2016{\natexlab{a}}, \aap, 594, A27

\bibitem[{{Planck Collaboration} {et~al.}(2016{\natexlab{b}}){Planck
  Collaboration}, {Ade}, {Aghanim}, {Arnaud}, {Ashdown}, {Aumont},
  {Baccigalupi}, {Banday}, {Barreiro}, {Bartlett}, {Bartolo}, {Battaner},
  {Battye}, {Benabed}, {Beno{\^\i}t}, {Benoit-L{\'e}vy}, {Bernard},
  {Bersanelli}, {Bielewicz}, {Bock}, {Bonaldi}, {Bonavera}, {Bond}, {Borrill},
  {Bouchet}, {Bucher}, {Burigana}, {Butler}, {Calabrese}, {Cardoso},
  {Catalano}, {Challinor}, {Chamballu}, {Chary}, {Chiang}, {Christensen},
  {Church}, {Clements}, {Colombi}, {Colombo}, {Combet}, {Comis}, {Couchot},
  {Coulais}, {Crill}, {Curto}, {Cuttaia}, {Danese}, {Davies}, {Davis}, {de
  Bernardis}, {de Rosa}, {de Zotti}, {Delabrouille}, {D{\'e}sert}, {Diego},
  {Dolag}, {Dole}, {Donzelli}, {Dor{\'e}}, {Douspis}, {Ducout}, {Dupac},
  {Efstathiou}, {Elsner}, {En{\ss}lin}, {Eriksen}, {Falgarone}, {Fergusson},
  {Finelli}, {Forni}, {Frailis}, {Fraisse}, {Franceschi}, {Frejsel},
  {Galeotta}, {Galli}, {Ganga}, {Giard}, {Giraud-H{\'e}raud}, {Gjerl{\o}w},
  {Gonz{\'a}lez-Nuevo}, {G{\'o}rski}, {Gratton}, {Gregorio}, {Gruppuso},
  {Gudmundsson}, {Hansen}, {Hanson}, {Harrison}, {Henrot-Versill{\'e}},
  {Hern{\'a}ndez-Monteagudo}, {Herranz}, {Hildebrandt}, {Hivon}, {Hobson},
  {Holmes}, {Hornstrup}, {Hovest}, {Huffenberger}, {Hurier}, {Jaffe}, {Jaffe},
  {Jones}, {Juvela}, {Keih{\"a}nen}, {Keskitalo}, {Kisner}, {Kneissl},
  {Knoche}, {Kunz}, {Kurki-Suonio}, {Lagache}, {L{\"a}hteenm{\"a}ki},
  {Lamarre}, {Lasenby}, {Lattanzi}, {Lawrence}, {Leonardi}, {Lesgourgues},
  {Levrier}, {Liguori}, {Lilje}, {Linden-V{\o}rnle}, {L{\'o}pez-Caniego},
  {Lubin}, {Mac{\'\i}as-P{\'e}rez}, {Maggio}, {Maino}, {Mandolesi}, {Mangilli},
  {Maris}, {Martin}, {Mart{\'\i}nez-Gonz{\'a}lez}, {Masi}, {Matarrese},
  {McGehee}, {Meinhold}, {Melchiorri}, {Melin}, {Mendes}, {Mennella},
  {Migliaccio}, {Mitra}, {Miville-Desch{\^e}nes}, {Moneti}, {Montier},
  {Morgante}, {Mortlock}, {Moss}, {Munshi}, {Murphy}, {Naselsky}, {Nati},
  {Natoli}, {Netterfield}, {N{\o}rgaard-Nielsen}, {Noviello}, {Novikov},
  {Novikov}, {Oxborrow}, {Paci}, {Pagano}, {Pajot}, {Paoletti}, {Partridge},
  {Pasian}, {Patanchon}, {Pearson}, {Perdereau}, {Perotto}, {Perrotta},
  {Pettorino}, {Piacentini}, {Piat}, {Pierpaoli}, {Pietrobon}, {Plaszczynski},
  {Pointecouteau}, {Polenta}, {Popa}, {Pratt}, {Pr{\'e}zeau}, {Prunet},
  {Puget}, {Rachen}, {Rebolo}, {Reinecke}, {Remazeilles}, {Renault}, {Renzi},
  {Ristorcelli}, {Rocha}, {Roman}, {Rosset}, {Rossetti}, {Roudier},
  {Rubi{\~n}o-Mart{\'\i}n}, {Rusholme}, \& {Sandri}}]{PlanckCollaboration2016a}
{Planck Collaboration}, {Ade}, P.~A.~R., {Aghanim}, N., {et~al.}
  2016{\natexlab{b}}, \aap, 594, A24

\bibitem[{{Planck Collaboration} {et~al.}(2020){Planck Collaboration},
  {Aghanim}, {Akrami}, {Ashdown}, {Aumont}, {Baccigalupi}, {Ballardini},
  {Banday}, {Barreiro}, {Bartolo}, {Basak}, {Battye}, {Benabed}, {Bernard},
  {Bersanelli}, {Bielewicz}, {Bock}, {Bond}, {Borrill}, {Bouchet}, {Boulanger},
  {Bucher}, {Burigana}, {Butler}, {Calabrese}, {Cardoso}, {Carron},
  {Challinor}, {Chiang}, {Chluba}, {Colombo}, {Combet}, {Contreras}, {Crill},
  {Cuttaia}, {de Bernardis}, {de Zotti}, {Delabrouille}, {Delouis}, {Di
  Valentino}, {Diego}, {Dor{\'e}}, {Douspis}, {Ducout}, {Dupac}, {Dusini},
  {Efstathiou}, {Elsner}, {En{\ss}lin}, {Eriksen}, {Fantaye}, {Farhang},
  {Fergusson}, {Fernandez-Cobos}, {Finelli}, {Forastieri}, {Frailis},
  {Fraisse}, {Franceschi}, {Frolov}, {Galeotta}, {Galli}, {Ganga},
  {G{\'e}nova-Santos}, {Gerbino}, {Ghosh}, {Gonz{\'a}lez-Nuevo}, {G{\'o}rski},
  {Gratton}, {Gruppuso}, {Gudmundsson}, {Hamann}, {Handley}, {Hansen},
  {Herranz}, {Hildebrandt}, {Hivon}, {Huang}, {Jaffe}, {Jones}, {Karakci},
  {Keih{\"a}nen}, {Keskitalo}, {Kiiveri}, {Kim}, {Kisner}, {Knox},
  {Krachmalnicoff}, {Kunz}, {Kurki-Suonio}, {Lagache}, {Lamarre}, {Lasenby},
  {Lattanzi}, {Lawrence}, {Le Jeune}, {Lemos}, {Lesgourgues}, {Levrier},
  {Lewis}, {Liguori}, {Lilje}, {Lilley}, {Lindholm}, {L{\'o}pez-Caniego},
  {Lubin}, {Ma}, {Mac{\'\i}as-P{\'e}rez}, {Maggio}, {Maino}, {Mandolesi},
  {Mangilli}, {Marcos-Caballero}, {Maris}, {Martin}, {Martinelli},
  {Mart{\'\i}nez-Gonz{\'a}lez}, {Matarrese}, {Mauri}, {McEwen}, {Meinhold},
  {Melchiorri}, {Mennella}, {Migliaccio}, {Millea}, {Mitra},
  {Miville-Desch{\^e}nes}, {Molinari}, {Montier}, {Morgante}, {Moss}, {Natoli},
  {N{\o}rgaard-Nielsen}, {Pagano}, {Paoletti}, {Partridge}, {Patanchon},
  {Peiris}, {Perrotta}, {Pettorino}, {Piacentini}, {Polastri}, {Polenta},
  {Puget}, {Rachen}, {Reinecke}, {Remazeilles}, {Renzi}, {Rocha}, {Rosset},
  {Roudier}, {Rubi{\~n}o-Mart{\'\i}n}, {Ruiz-Granados}, {Salvati}, {Sandri},
  {Savelainen}, {Scott}, {Shellard}, {Sirignano}, {Sirri}, {Spencer},
  {Sunyaev}, {Suur-Uski}, {Tauber}, {Tavagnacco}, {Tenti}, {Toffolatti},
  {Tomasi}, {Trombetti}, {Valenziano}, {Valiviita}, {Van Tent}, {Vibert},
  {Vielva}, {Villa}, {Vittorio}, {Wandelt}, {Wehus}, {White}, {White},
  {Zacchei}, \& {Zonca}}]{PlanckCollaboration2020}
{Planck Collaboration}, {Aghanim}, N., {Akrami}, Y., {et~al.} 2020, \aap, 641,
  A6

\bibitem[{{Proust} {et~al.}(2006){Proust}, {Quintana}, {Carrasco},
  {Reisenegger}, {Slezak}, {Muriel}, {D{\"u}nner}, {Sodr{\'e}}, {Drinkwater},
  {Parker}, \& {Ragone}}]{Proust2006}
{Proust}, D., {Quintana}, H., {Carrasco}, E.~R., {et~al.} 2006, \aap, 447, 133

\bibitem[{{Quilis} {et~al.}(2017){Quilis}, {Planelles}, \&
  {Ricciardelli}}]{Quilis2017}
{Quilis}, V., {Planelles}, S., \& {Ricciardelli}, E. 2017, \mnras, 469, 80

\bibitem[{{Raj} {et~al.}(2024){Raj}, {Awad}, {Peletier}, {Smith}, {Kuchner},
  {van de Weygaert}, {Libeskind}, {Canducci}, {Ti{\v{n}}o}, \&
  {Bunte}}]{Raj2024}
{Raj}, M.~A., {Awad}, P., {Peletier}, R.~F., {et~al.} 2024, \aap, 690, A92

\bibitem[{{Reiprich} {et~al.}(2021){Reiprich}, {Veronica}, {Pacaud},
  {Ramos-Ceja}, {Ota}, {Sanders}, {Kara}, {Erben}, {Klein}, {Erler}, {Kerp},
  {Hoang}, {Br{\"u}ggen}, {Marvil}, {Rudnick}, {Biffi}, {Dolag},
  {Aschersleben}, {Basu}, {Brunner}, {Bulbul}, {Dennerl}, {Eckert}, {Freyberg},
  {Gatuzz}, {Ghirardini}, {K{\"a}fer}, {Merloni}, {Migkas}, {Nandra},
  {Predehl}, {Robrade}, {Salvato}, {Whelan}, {Diaz-Ocampo}, {Hernandez-Lang},
  {Zenteno}, {Brown}, {Collier}, {Diego}, {Hopkins}, {Kapinska}, {Koribalski},
  {Mroczkowski}, {Norris}, {O'Brien}, \& {Vardoulaki}}]{Reiprich2021}
{Reiprich}, T.~H., {Veronica}, A., {Pacaud}, F., {et~al.} 2021, \aap, 647, A2

\bibitem[{{Reisenegger} {et~al.}(2000){Reisenegger}, {Quintana}, {Carrasco}, \&
  {Maze}}]{Reisenegger2000}
{Reisenegger}, A., {Quintana}, H., {Carrasco}, E.~R., \& {Maze}, J. 2000, \aj,
  120, 523

\bibitem[{{Rines} \& {Diaferio}(2006)}]{Rines2006}
{Rines}, K. \& {Diaferio}, A. 2006, \aj, 132, 1275

\bibitem[{{Rossetti} {et~al.}(2017){Rossetti}, {Gastaldello}, {Eckert}, {Della
  Torre}, {Pantiri}, {Cazzoletti}, \& {Molendi}}]{Rossetti2017}
{Rossetti}, M., {Gastaldello}, F., {Eckert}, D., {et~al.} 2017, \mnras, 468,
  1917

\bibitem[{{Rozo} {et~al.}(2014){Rozo}, {Bartlett}, {Evrard}, \&
  {Rykoff}}]{Rozo2014}
{Rozo}, E., {Bartlett}, J.~G., {Evrard}, A.~E., \& {Rykoff}, E.~S. 2014,
  \mnras, 438, 78

\bibitem[{{Rozo} {et~al.}(2016){Rozo}, {Rykoff}, {Abate}, {Bonnett}, {Crocce},
  {Davis}, {Hoyle}, {Leistedt}, {Peiris}, {Wechsler}, {Abbott}, {Abdalla},
  {Banerji}, {Bauer}, {Benoit-L{\'e}vy}, {Bernstein}, {Bertin}, {Brooks},
  {Buckley-Geer}, {Burke}, {Capozzi}, {Rosell}, {Carollo}, {Kind}, {Carretero},
  {Castander}, {Childress}, {Cunha}, {D'Andrea}, {Davis}, {DePoy}, {Desai},
  {Diehl}, {Dietrich}, {Doel}, {Eifler}, {Evrard}, {Neto}, {Flaugher},
  {Fosalba}, {Frieman}, {Gaztanaga}, {Gerdes}, {Glazebrook}, {Gruen},
  {Gruendl}, {Honscheid}, {James}, {Jarvis}, {Kim}, {Kuehn}, {Kuropatkin},
  {Lahav}, {Lidman}, {Lima}, {Maia}, {March}, {Martini}, {Melchior}, {Miller},
  {Miquel}, {Mohr}, {Nichol}, {Nord}, {O'Neill}, {Ogando}, {Plazas}, {Romer},
  {Roodman}, {Sako}, {Sanchez}, {Santiago}, {Schubnell}, {Sevilla-Noarbe},
  {Smith}, {Soares-Santos}, {Sobreira}, {Suchyta}, {Swanson}, {Thaler},
  {Thomas}, {Uddin}, {Vikram}, {Walker}, {Wester}, {Zhang}, \& {da
  Costa}}]{Rozo2016}
{Rozo}, E., {Rykoff}, E.~S., {Abate}, A., {et~al.} 2016, \mnras, 461, 1431

\bibitem[{{Rykoff} {et~al.}(2014){Rykoff}, {Rozo}, {Busha}, {Cunha},
  {Finoguenov}, {Evrard}, {Hao}, {Koester}, {Leauthaud}, {Nord}, {Pierre},
  {Reddick}, {Sadibekova}, {Sheldon}, \& {Wechsler}}]{Rykoff2014}
{Rykoff}, E.~S., {Rozo}, E., {Busha}, M.~T., {et~al.} 2014, \apj, 785, 104

\bibitem[{{Sereno}(2015)}]{Sereno2015}
{Sereno}, M. 2015, \mnras, 450, 3665

\bibitem[{{Sif{\'o}n} \& {Han}(2024)}]{Sifon2024}
{Sif{\'o}n}, C. \& {Han}, J. 2024, \aap, 686, A163

\bibitem[{{Smith} {et~al.}(2022){Smith}, {Calder{\'o}n-Castillo}, {Shin},
  {Raouf}, \& {Ko}}]{Smith2022}
{Smith}, R., {Calder{\'o}n-Castillo}, P., {Shin}, J., {Raouf}, M., \& {Ko}, J.
  2022, \aj, 164, 95

\bibitem[{{Smith} {et~al.}(2016){Smith}, {Choi}, {Lee}, {Rhee},
  {Sanchez-Janssen}, \& {Yi}}]{Smith2016}
{Smith}, R., {Choi}, H., {Lee}, J., {et~al.} 2016, \apj, 833, 109

\bibitem[{{Stroe} {et~al.}(2017){Stroe}, {Sobral}, {Paulino-Afonso}, {Alegre},
  {Calhau}, {Santos}, \& {van Weeren}}]{Stroe2017}
{Stroe}, A., {Sobral}, D., {Paulino-Afonso}, A., {et~al.} 2017, \mnras, 465,
  2916

\bibitem[{{Taylor} {et~al.}(2011){Taylor}, {Hopkins}, {Baldry}, {Brown},
  {Driver}, {Kelvin}, {Hill}, {Robotham}, {Bland-Hawthorn}, {Jones}, {Sharp},
  {Thomas}, {Liske}, {Loveday}, {Norberg}, {Peacock}, {Bamford}, {Brough},
  {Colless}, {Cameron}, {Conselice}, {Croom}, {Frenk}, {Gunawardhana},
  {Kuijken}, {Nichol}, {Parkinson}, {Phillipps}, {Pimbblet}, {Popescu},
  {Prescott}, {Sutherland}, {Tuffs}, {van Kampen}, \&
  {Wijesinghe}}]{Taylor2011}
{Taylor}, E.~N., {Hopkins}, A.~M., {Baldry}, I.~K., {et~al.} 2011, \mnras, 418,
  1587

\bibitem[{{Tempel} {et~al.}(2018){Tempel}, {Kruuse}, {Kipper}, {Tuvikene},
  {Sorce}, \& {Stoica}}]{Tempel2018}
{Tempel}, E., {Kruuse}, M., {Kipper}, R., {et~al.} 2018, \aap, 618, A81

\bibitem[{{Tempel} {et~al.}(2020{\natexlab{a}}){Tempel}, {Norberg}, {Tuvikene},
  {Bensby}, {Chiappini}, {Christlieb}, {Cioni}, {Comparat}, {Davies},
  {Guiglion}, {Koch}, {Kordopatis}, {Krumpe}, {Loveday}, {Merloni}, {Micheva},
  {Minchev}, {Roukema}, {Sorce}, {Starkenburg}, {Storm}, {Swann}, {Thi},
  {Traven}, \& {de Jong}}]{Tempel2020a}
{Tempel}, E., {Norberg}, P., {Tuvikene}, T., {et~al.} 2020{\natexlab{a}}, \aap,
  635, A101

\bibitem[{{Tempel} {et~al.}(2017){Tempel}, {Tuvikene}, {Kipper}, \&
  {Libeskind}}]{Tempel2017}
{Tempel}, E., {Tuvikene}, T., {Kipper}, R., \& {Libeskind}, N.~I. 2017, \aap,
  602, A100

\bibitem[{{Tempel} {et~al.}(2020{\natexlab{b}}){Tempel}, {Tuvikene}, {Muru},
  {Stoica}, {Bensby}, {Chiappini}, {Christlieb}, {Cioni}, {Comparat},
  {Feltzing}, {Hook}, {Koch}, {Kordopatis}, {Krumpe}, {Loveday}, {Minchev},
  {Norberg}, {Roukema}, {Sorce}, {Storm}, {Swann}, {Taylor}, {Traven},
  {Walcher}, \& {de Jong}}]{Tempel2020}
{Tempel}, E., {Tuvikene}, T., {Muru}, M.~M., {et~al.} 2020{\natexlab{b}},
  \mnras, 497, 4626

\bibitem[{{The Pandas Development Team}(2024)}]{Team2024}
{The Pandas Development Team}. 2024, {pandas-dev/pandas: Pandas}

\bibitem[{{Tollet} {et~al.}(2017){Tollet}, {Cattaneo}, {Mamon}, {Moutard}, \&
  {van den Bosch}}]{Tollet2017}
{Tollet}, {\'E}., {Cattaneo}, A., {Mamon}, G.~A., {Moutard}, T., \& {van den
  Bosch}, F.~C. 2017, \mnras, 471, 4170

\bibitem[{{Treu} {et~al.}(2003){Treu}, {Ellis}, {Kneib}, {Dressler}, {Smail},
  {Czoske}, {Oemler}, \& {Natarajan}}]{Treu2003}
{Treu}, T., {Ellis}, R.~S., {Kneib}, J.-P., {et~al.} 2003, \apj, 591, 53

\bibitem[{{Veronica} {et~al.}(2024){Veronica}, {Reiprich}, {Pacaud}, {Ota},
  {Aschersleben}, {Biffi}, {Bulbul}, {Clerc}, {Dolag}, {Erben}, {Gatuzz},
  {Ghirardini}, {Kerp}, {Klein}, {Liu}, {Liu}, {Migkas}, {Ramos-Ceja},
  {Sanders}, \& {Spinelli}}]{Veronica2024}
{Veronica}, A., {Reiprich}, T.~H., {Pacaud}, F., {et~al.} 2024, \aap, 681, A108

\bibitem[{{Virtanen} {et~al.}(2020){Virtanen}, {Gommers}, {Oliphant},
  {Haberland}, {Reddy}, {Cournapeau}, {Burovski}, {Peterson}, {Weckesser},
  {Bright}, {van der Walt}, {Brett}, {Wilson}, {Millman}, {Mayorov}, {Nelson},
  {Jones}, {Kern}, {Larson}, {Carey}, {Polat}, {Feng}, {Moore}, {VanderPlas},
  {Laxalde}, {Perktold}, {Cimrman}, {Henriksen}, {Quintero}, {Harris},
  {Archibald}, {Ribeiro}, {Pedregosa}, {van Mulbregt}, \& {SciPy 1. 0
  Contributors}}]{Virtanen2020}
{Virtanen}, P., {Gommers}, R., {Oliphant}, T.~E., {et~al.} 2020, Nature
  Methods, 17, 261

\bibitem[{{Wilman} {et~al.}(2009){Wilman}, {Oemler}, {Mulchaey}, {McGee},
  {Balogh}, \& {Bower}}]{Wilman2009}
{Wilman}, D.~J., {Oemler}, A., J., {Mulchaey}, J.~S., {et~al.} 2009, \apj, 692,
  298

\bibitem[{{Zabludoff} {et~al.}(1996){Zabludoff}, {Zaritsky}, {Lin}, {Tucker},
  {Hashimoto}, {Shectman}, {Oemler}, \& {Kirshner}}]{Zabludoff1996}
{Zabludoff}, A.~I., {Zaritsky}, D., {Lin}, H., {et~al.} 1996, \apj, 466, 104

\end{thebibliography}

\begin{appendix}

\onecolumn

\section{CHANCES cluster samples}

\FloatBarrier

\Cref{t:lowz,t:evolution} list the CHANCES cluster samples for the Low-z and Evolution subsurveys, respectively, along with the mass estimates and corresponding $r_{200}$. All of these clusters will be observed out to $5r_{200}$ with the 4MOST spectrograph through the five-year duration of the CHANCES survey.

\begin{table*}[h!]
\centering\small
\caption{CHANCES Low-z cluster sample.}
\label{t:lowz}
\begin{tabular}{lcccccccl}
\hline
(1) & (2) & (3) & (4) & (5) & (6) & (7) & (8) \\
Cluster name & 
RA & 
Dec & 
Redshift & 
\multicolumn{1}{c}{$M_{200}$} & 
$r_{200}$ & 
\multicolumn{1}{c}{$\theta_{200}$} & 
Mass source \\
 & 
 hh:mm:ss & 
 dd:mm:ss & 
 & 
 \multicolumn{1}{c}{$10^{14}\,\Msun$} & 
 $\mathrm{Mpc}$ & 
 \multicolumn{1}{c}{arcmin} & 
 \\[0.5ex]
 \hline
Abell 85 & 00:41:49.9 & $-$09:18:07.2 & 0.056 & 8.4 & 1.96 & 29.00 & MENeaCS \\
Abell 119 & 00:56:16.1 & $-$01:15:18.0 & 0.044 & 7.8 & 1.92 & 35.63 & MENeaCS \\
Abell 133 & 01:02:41.8 & $-$21:52:55.2 & 0.057 & 4.1 & 1.54 & 22.45 & MENeaCS \\
Abell 147 & 01:08:11.5 & $+$02:10:33.6 & 0.044 & 1.4 & 1.08 & 20.10 & MCXC \\
Abell 151 & 01:08:50.9 & $-$15:24:25.2 & 0.053 & 2.5 & 1.30 & 20.28 & MCXC \\
Abell 168 & 01:15:02.4 & $+$00:18:54.0 & 0.044 & 2.4 & 1.29 & 24.04 & MCXC \\
Abell 194 & 01:25:50.4 & $-$01:24:07.2 & 0.017 & 0.6 & 0.83 & 38.79 & AXES-2MRS \\
Abell 496 & 04:33:38.4 & $-$13:15:32.4 & 0.034 & 5.7 & 1.73 & 41.61 & AXES-2MRS \\
Abell 500 & 04:38:51.8 & $-$22:06:00.0 & 0.067 & 3.0 & 1.37 & 17.26 & MCXC \\
Abell 548 & 05:48:29.0 & $-$25:28:58.8 & 0.041 & 2.7 & 1.34 & 26.63 & MCXC \\
Abell 754 & 09:08:31.9 & $-$09:36:57.6 & 0.054 & 14.9 & 2.37 & 36.33 & MENeaCS \\
Abell 780 & 09:18:06.0 & $-$12:04:58.8 & 0.057 & 6.5 & 1.80 & 26.20 & MENeaCS \\
Abell 957 & 10:13:37.9 & $-$00:54:57.6 & 0.045 & 2.2 & 1.25 & 22.83 & MCXC \\
Abell 970 & 10:17:33.8 & $-$10:39:57.6 & 0.059 & 3.6 & 1.48 & 20.85 & MCXC \\
Abell 1069 & 10:39:43.0 & $-$08:40:58.8 & 0.062 & 1.5 & 1.10 & 14.79 & WINGS \\
Abell 1520 & 12:19:19.7 & $-$13:15:36.0 & 0.068 & 3.7 & 1.48 & 18.29 & MCXC \\
Abell 1631 & 12:52:51.8 & $-$15:24:00.0 & 0.047 & 2.4 & 1.29 & 22.59 & WINGS \\
Abell 1644 & 12:57:10.8 & $-$17:24:00.0 & 0.048 & 6.0 & 1.76 & 30.11 & WINGS \\
Abell 2399 & 21:57:22.1 & $-$07:48:39.6 & 0.058 & 3.4 & 1.44 & 20.67 & WINGS \\
Abell 2415 & 22:05:39.4 & $-$05:35:38.4 & 0.058 & 1.5 & 1.11 & 15.87 & WINGS \\
Abell 2457 & 22:35:40.8 & $+$01:30:21.6 & 0.058 & 2.0 & 1.22 & 17.47 & WINGS \\
Abell 2717 & 00:03:13.0 & $-$35:55:58.8 & 0.050 & 1.4 & 1.09 & 17.93 & WINGS \\
Abell 2734 & 00:11:20.6 & $-$28:51:18.0 & 0.061 & 2.4 & 1.28 & 17.56 & WINGS \\
Abell 2870 & 01:07:43.9 & $-$46:54:00.0 & 0.023 & 4.5 & 1.59 & 55.14 & CODEX \\
Abell 2877 & 01:10:00.2 & $-$45:55:22.8 & 0.024 & 1.2 & 1.05 & 34.83 & AXES-2MRS \\
Abell 3223 & 04:08:16.1 & $-$30:53:38.4 & 0.060 & 3.0 & 1.38 & 19.25 & MCXC \\
Abell 3266 & 04:31:13.0 & $-$61:27:00.0 & 0.054 & 7.0 & 1.84 & 28.21 & SPT-SZ \\
Abell 3301 & 05:00:46.6 & $-$38:40:40.8 & 0.054 & 2.4 & 1.30 & 19.89 & MCXC \\
Abell 3341 & 05:25:34.1 & $-$31:35:42.0 & 0.037 & 1.9 & 1.20 & 25.96 & AXES-2MRS \\
Abell 3376 & 06:00:40.8 & $-$40:01:58.8 & 0.046 & 4.0 & 1.53 & 27.36 & WINGS \\
Abell 3391 & 06:26:22.8 & $-$53:41:49.2 & 0.051 & 4.3 & 1.56 & 25.32 & MCXC \\
Abell 3395 & 06:27:36.0 & $-$54:25:58.8 & 0.050 & 5.1 & 1.65 & 27.28 & PSZ2 \\
Abell 3490 & 11:45:19.9 & $-$34:19:58.8 & 0.069 & 3.0 & 1.38 & 16.91 & AXES-LEGACY \\
Abell 3497 & 12:00:03.8 & $-$31:22:58.8 & 0.068 & 3.1 & 1.39 & 17.22 & WINGS \\
Abell 3565 & 13:36:39.1 & $-$33:57:31.2 & 0.013 & 0.2 & 0.60 & 36.67 & MCXC \\
Abell 3571 & 13:47:28.3 & $-$32:50:56.4 & 0.039 & 8.7 & 1.99 & 41.49 & AXES-2MRS \\
Abell 3574 & 13:49:06.7 & $-$30:19:33.6 & 0.016 & 2.7 & 1.36 & 67.25 & AXES-2MRS \\
Abell 3581 & 14:07:28.1 & $-$27:00:54.0 & 0.023 & 2.0 & 1.23 & 42.74 & AXES-2MRS \\
Abell 3651 & 19:52:16.3 & $-$55:03:43.2 & 0.060 & 2.9 & 1.37 & 19.01 & MCXC \\
Abell 3667 & 20:12:26.9 & $-$56:48:57.6 & 0.053 & 8.2 & 1.94 & 30.28 & ACT-DR5 \\
Abell 3716 & 20:51:30.0 & $-$52:42:57.6 & 0.045 & 4.5 & 1.60 & 29.12 & WINGS \\
Abell 3809 & 21:46:58.8 & $-$43:52:58.8 & 0.063 & 1.0 & 0.97 & 12.85 & WINGS \\
Abell 4059 & 23:57:00.0 & $-$34:45:32.4 & 0.048 & 6.2 & 1.77 & 30.28 & ACT-DR5 \\
Abell S560 & 06:00:48.2 & $-$58:35:13.2 & 0.037 & 0.8 & 0.91 & 19.90 & AXES-2MRS \\
Antlia & 10:30:03.4 & $-$35:19:22.8 & 0.009 & 1.3 & 1.06 & 92.08 & PSZ2 \\
Fornax & 03:38:27.8 & $-$35:26:52.8 & 0.005 & 0.3 & 0.69 & 107.25 & MCXC \\
Hydra (A1060) & 10:36:41.8 & $-$27:31:26.4 & 0.012 & 2.2 & 1.26 & 82.82 & AXES-2MRS \\
IIZw108 & 21:13:55.9 & $+$02:33:54.0 & 0.048 & 1.5 & 1.11 & 19.09 & WINGS \\
MKW4 & 12:04:27.6 & $+$01:53:42.0 & 0.020 & 1.4 & 1.08 & 42.82 & AXES-2MRS \\
MKW8 & 14:40:42.2 & $+$03:28:19.2 & 0.027 & 1.4 & 1.09 & 32.46 & AXES-2MRS \\
\hline
\end{tabular}
\tablefoot{ Clusters which are part of the Shapely and Horologium-Reticulum superclusters, which are also part of the Low-z survey, are not included in this table. Columns are: (1) literature cluster name; (2) and (3): right ascension and declination corresponding to the adopted cluster centre; (4): literature cluster redshift; (5) and (6): mass and corresponding radius enclosing a mean density of 200 times the critical mass density of the Universe at the cluster redshift; (7) cluster angular size, $\theta_{200}=r_{200}/D_\mathrm{A}$, where $D_\mathrm{A}$ is the angular diameter distance; and (8) catalogue from which we take the cluster mass, after normalizing the published masses as discussed in Sect. \ref{s:masses}.}
\end{table*}

\begin{table*}[h!]
\centering\small
\caption{CHANCES Evolution sample.}
\label{t:evolution}
\begin{tabular}{lccccccc}
\hline
Cluster name & 
RA & 
Dec & 
Redshift & 
\multicolumn{1}{c}{$M_{200}$} & 
$r_{200}$ & 
\multicolumn{1}{c}{$\theta_{200}$} & 
Mass source \\
 & 
 hh:mm:ss & 
 dd:mm:ss & 
 & 
 \multicolumn{1}{c}{$10^{14}\,\Msun$} & 
 $\mathrm{Mpc}$ & 
 \multicolumn{1}{c}{arcmin} & 
 \\[0.5ex]
 \hline
Abell 209 & 01:31:53.5 & $-$13:36:46.8 & 0.209 & 8.5 & 1.86 & 8.80 & MENeaCS \\
Abell 370 & 02:39:50.4 & $-$01:35:06.0 & 0.373 & 24.1 & 2.47 & 7.76 & MENeaCS \\
Abell 520 & 04:54:06.5 & $+$02:57:43.2 & 0.203 & 11.5 & 2.07 & 9.99 & MENeaCS \\
Abell 521 & 04:54:09.1 & $-$10:14:20.4 & 0.247 & 5.8 & 1.62 & 6.74 & LoCuSS \\
Abell 1300 & 11:31:54.5 & $-$19:55:40.8 & 0.306 & 12.5 & 2.04 & 7.29 & SPT-ECS \\
Abell 1437 & 12:00:26.2 & $+$03:20:52.8 & 0.134 & 3.9 & 1.48 & 10.03 & CoMaLit \\
Abell 1650 & 12:58:42.0 & $-$01:45:32.4 & 0.085 & 10.5 & 2.09 & 21.21 & MENeaCS \\
Abell 1651 & 12:59:28.1 & $-$04:12:03.6 & 0.085 & 8.3 & 1.93 & 19.60 & MENeaCS \\
Abell 1689 & 13:11:30.0 & $-$01:20:06.0 & 0.183 & 11.4 & 2.07 & 10.86 & LoCuSS \\
Abell 1835 & 14:00:52.3 & $+$02:52:40.8 & 0.252 & 10.5 & 1.96 & 8.06 & LoCuSS \\
Abell 2163 & 16:15:49.2 & $-$06:09:07.2 & 0.206 & 13.1 & 2.15 & 10.30 & MENeaCS \\
Abell 2420 & 22:10:16.6 & $-$12:10:37.2 & 0.085 & 8.4 & 1.94 & 19.58 & MENeaCS \\
Abell 2744 & 00:14:18.7 & $-$30:23:20.4 & 0.308 & 20.6 & 2.41 & 8.58 & CoMaLit \\
Abell 2811 & 00:42:08.6 & $-$28:32:09.6 & 0.108 & 6.2 & 1.74 & 14.22 & CoMaLit \\
Abell 2813 & 00:43:27.8 & $-$20:37:01.2 & 0.292 & 8.5 & 1.80 & 6.66 & LoCuSS \\
Abell 3048 & 02:46:27.6 & $-$20:32:06.0 & 0.310 & 8.2 & 1.77 & 6.27 & ACT-DR5 \\
Abell 3186 & 03:52:14.6 & $-$74:00:28.8 & 0.127 & 10.2 & 2.03 & 14.44 & PSZ2 \\
Abell 3378 & 06:05:52.6 & $-$35:18:32.4 & 0.139 & 8.8 & 1.93 & 12.70 & PSZ2 \\
Abell 3404 & 06:45:29.3 & $-$54:13:08.4 & 0.164 & 16.0 & 2.33 & 13.34 & ACT-DR5 \\
Abell 3444 & 10:23:50.9 & $-$27:15:32.4 & 0.254 & 11.5 & 2.02 & 8.25 & SPT-ECS \\
Abell 3695 & 20:34:49.4 & $-$35:49:30.0 & 0.089 & 7.5 & 1.87 & 18.10 & CoMaLit \\
Abell 3822 & 21:54:06.7 & $-$57:51:46.8 & 0.076 & 5.3 & 1.67 & 18.68 & ACT-DR5 \\
Abell 3827 & 22:01:52.6 & $-$59:56:20.4 & 0.098 & 12.5 & 2.20 & 19.62 & ACT-DR5 \\
Abell 3911 & 22:46:17.0 & $-$52:43:19.2 & 0.097 & 6.7 & 1.79 & 16.11 & PSZ2 \\
Abell 3921 & 22:49:49.4 & $-$64:24:54.0 & 0.094 & 7.0 & 1.82 & 16.77 & SPT-SZ \\
Abell S780 & 14:59:29.3 & $-$18:11:13.2 & 0.236 & 12.3 & 2.08 & 8.98 & PSZ2 \\
Bullet & 06:58:31.0 & $-$55:56:49.2 & 0.296 & 14.3 & 2.14 & 7.83 & CoMaLit \\
MACS J0329.7$-$0211 & 03:29:41.5 & $-$02:11:45.6 & 0.450 & 8.6 & 1.70 & 4.78 & CoMaLit \\
MACS J0416.1$-$2403 & 04:16:09.8 & $-$24:03:57.6 & 0.397 & 10.7 & 1.87 & 5.65 & CoMaLit \\
MACS J0553.4$-$3342 & 05:53:27.1 & $-$33:42:54.0 & 0.430 & 21.4 & 2.34 & 6.75 & ACT-DR5 \\
MACS J1206.2$-$0847 & 12:06:12.2 & $-$08:48:00.0 & 0.441 & 18.1 & 2.19 & 6.22 & CoMaLit \\
PSZ2 G205.93$-$39.64 & 04:17:37.4 & $-$11:53:45.6 & 0.443 & 30.6 & 2.61 & 7.38 & CoMaLit \\
PSZ2 G208.60$-$26.00 & 05:10:47.8 & $-$08:01:44.4 & 0.219 & 13.0 & 2.14 & 9.74 & ACT-DR5 \\
PSZ2 G241.11$-$28.68 & 05:42:57.1 & $-$35:59:02.4 & 0.420 & 10.0 & 1.82 & 5.32 & ACT-DR5 \\
PSZ2 G241.76$-$30.88 & 05:32:56.0 & $-$37:01:34.0 & 0.275 & 14.7 & 2.18 & 8.40 & ACT-DR5 \\
PSZ2 G259.98$-$63.43 & 02:32:18.7 & $-$44:20:42.0 & 0.284 & 17.3 & 2.29 & 8.63 & ACT-DR5 \\
PSZ2 G262.27$-$35.38 & 05:16:36.7 & $-$54:31:12.0 & 0.295 & 9.5 & 1.87 & 6.86 & CoMaLit \\
PSZ2 G262.73$-$40.92 & 04:38:19.0 & $-$54:19:04.8 & 0.421 & 15.0 & 2.07 & 6.05 & ACT-DR5 \\
PSZ2 G271.18$-$30.95 & 05:49:18.2 & $-$62:04:58.8 & 0.376 & 16.4 & 2.18 & 6.80 & SPT-SZ \\
PSZ2 G277.76$-$51.74 & 02:54:23.0 & $-$58:57:50.4 & 0.438 & 11.2 & 1.87 & 5.33 & ACT-DR5 \\
PSZ2 G286.98+32.90 & 11:50:49.2 & $-$28:04:37.2 & 0.390 & 31.0 & 2.67 & 8.17 & CoMaLit \\
PSZ2 G348.90$-$67.37 & 23:25:13.0 & $-$41:12:28.8 & 0.358 & 10.9 & 1.91 & 6.15 & ACT-DR5 \\
RXC J0528.9$-$3827 & 05:28:53.0 & $-$39:28:15.5 & 0.284 & 12.8 & 2.07 & 7.81 & ACT-DR5 \\
RXC J1314.4$-$2515 & 13:14:28.1 & $-$25:15:39.6 & 0.244 & 13.8 & 2.16 & 9.06 & SPT-ECS \\
RXC J1347.5$-$1144 & 13:47:30.5 & $-$11:45:10.8 & 0.452 & 16.1 & 2.10 & 5.88 & MENeaCS \\
RXC J1514.9$-$1523 & 15:14:58.1 & $-$15:23:09.6 & 0.223 & 13.8 & 2.18 & 9.81 & PSZ2 \\
RXC J2031.8$-$4037 & 20:31:51.6 & $-$40:37:15.6 & 0.341 & 14.7 & 2.13 & 7.07 & ACT-DR5 \\
RXC J2211.7$-$0350 & 22:11:43.4 & $-$03:49:44.4 & 0.397 & 19.4 & 2.28 & 6.89 & CoMaLit \\
RXC J2248.7$-$4431 & 22:48:43.4 & $-$44:31:44.4 & 0.347 & 18.7 & 2.30 & 7.55 & CoMaLit \\
SMACS J0723.3$-$7327 & 07:23:21.4 & $-$73:26:20.4 & 0.390 & 12.8 & 1.99 & 6.09 & PSZ2 \\
\hline
\end{tabular}
\tablefoot{Columns are as in \Cref{t:lowz}.}
\end{table*}

\FloatBarrier

\section{Properties of \redmapper\ groups}
\label{s:redmapper}

Here we describe the calculation of the properties of \redmapper\ groups used in Sect. \ref{s:environment}.

The \redmapper\ catalogue constructed by \cite{Kluge2024} reports the redshift for each group as derived from one of three techniques:
\texttt{cg\_spec\_z}, when the cluster redshift is equal to the central galaxy's spectroscopic redshift; \texttt{spec\_z\_boot}, bootstrapped from multiple (at least three) available spectroscopic redshifts; and \texttt{photo\_z}, photometric redshifts from the multi-colour red sequence as derived by \cite{Rykoff2014}. Uncertainties for the latter two are taken from the \redmapper\ catalogue directly. Uncertainties $\delta z$ for groups with \texttt{cg\_spec\_z} redshifts are calculated as
\begin{equation}
    \delta z = \left[(\delta z_\mathrm{cg})^2 + (\sigma_v)^2\right]^{1/2},
\end{equation}
where $\delta z_\mathrm{cg}$ is the uncertainty in the central galaxy redshift (equal to the reported group redshift uncertainty) and $\sigma_v$ is the velocity dispersion.

Of the 320 \redmapper\ groups within $5r_{200}$ of CHANCES Low-z clusters, there are 90 (28\%) with \texttt{cg\_spec\_z}, 161 (50\%)  with \texttt{spec\_z\_boot}), and 69 (22\%) \texttt{photo\_z}. Typical uncertainties  for each set are (250--400, 100--1200, 2800--4600) km~s$^{-1}$, respectively. In the Evolution sample there are a total of 1075 groups within $5r_{200}$ of CHANCES clusters: (86, 157, 832) groups, corresponding to (8\%, 15\%, 77\%), with (\texttt{cg\_spec\_z}, \texttt{spec\_z\_boot}, \texttt{photo\_z}) and typical uncertainties (300--500, 150--900, 1700--4000) km~s$^{-1}$, respectively. These all correspond to 16th--84th percentile ranges on the uncertainties.

We calculate the peculiar velocity of each \redmapper\ group as
\begin{equation}
    v_\mathrm{pec} = \frac{c(z-z_\mathrm{cl})}{1+z_\mathrm{cl}},
\end{equation}
where $z$ is the redshift reported in the \redmapper\ catalogue and $z_\mathrm{cl}$ is the redshift of the associated CHANCES cluster, as listed in \Cref{t:lowz,t:evolution}. The uncertainty on the peculiar velocity is then $\delta v_\mathrm{pec}=c[\delta z/(1+z_\mathrm{cl})]$.

In order to estimate group masses, $M_{200}$, we multiply the reported richness, $\lambda$, by 1.21 to match the DES richness scale as described by \cite{Kluge2024} and use the relation between $\lambda$ and weak lensing mass derived by \cite{McClintock2019} for DES clusters. We convert from $M_\mathrm{200m}$ (with respect to the mean matter density), estimated from the \cite{McClintock2019} relation, to $M_\mathrm{200}$ (with respect to the critical matter density) using the mass-concentration relation of \cite{Ishiyama2021} as implemented in \texttt{colossus} \citep{Diemer2018}.

Finally, we compare the richness-derived masses with the masses listed in \Cref{t:lowz,t:evolution}, since we combine the two to calculate the infall mass function (Fig. \ref{f:imf}). For the Low-z clusters we find a median logarithmic difference $\langle \log M_{200}^\mathrm{group} - \log M_{200}^\mathrm{main}\rangle=0.02$ dex and a 16--84th percentile range $[-0.47, 0.26]$ dex; that is, typically richness-derived masses are between 34\% and 182\% of the nominal CHANCES masses. Similarly, for Evolution clusters we find a median of $-0.07$ dex and a 16--84th percentile range $[-0.23, 0.18]$ dex. That is, richness-derived masses are typically between 66\% and 151\% the nominal CHANCES masses. Although the scatter between the two mass estimates is large, particularly in the case of Low-z, they are statistically consistent with each other.

\end{appendix}

\end{document}